\documentclass{aa}  

\usepackage{graphicx}
\usepackage{txfonts}

\usepackage{lscape,longtable}
\usepackage{natbib}
\usepackage{amsfonts,amsmath,subfigure,ulem}
\usepackage{color,xcolor,colortbl}
\usepackage{xtab,booktabs}
\usepackage{hyperref}
\hypersetup{breaklinks,colorlinks,citecolor=blue,linkcolor=blue,urlcolor=red}
\bibpunct{(}{)}{;}{a}{}{,} 

\definecolor{Gray}{gray}{0.9}
\definecolor{LCyan}{rgb}{0.88,1,1}
\definecolor{maroon}{cmyk}{0,0.87,0.68,0.32}
\definecolor{forestgreen}{cmyk}{0.76,0,0.76,0.45}

\begin{document} 

\title{Study of solar brightness profiles in the $18$ -- $26$~GHz frequency range with INAF radio telescopes I: solar radius}

%
\author{M.~Marongiu\thanks{marco.marongiu@inaf.it}\inst{1}
\and A.~Pellizzoni\inst{1}
\and S.~Mulas\inst{1}
\and S.~Righini\inst{2}
\and R.~Nesti\inst{3}
\and G.~Murtas\inst{4}
\and E.~Egron\inst{1}
\and M.~N.~Iacolina\inst{5}
\and A.~Melis~\inst{1}
\and G.~Valente\inst{5}
\and G.~Serra\inst{5}
\and S.~L.~Guglielmino\inst{6}
\and A.~Zanichelli\inst{2}
\and P.~Romano\inst{6}
\and S.~Loru\inst{6}
\and M.~Bachetti\inst{1}
\and A.~Bemporad\inst{7}
\and F.~Buffa\inst{1}
\and R.~Concu\inst{1}
\and G.~L.~Deiana\inst{1}
\and C.~Karakotia\inst{2}
\and A.~Ladu\inst{1}
\and A.~Maccaferri\inst{2}
\and P.~Marongiu~\inst{1}
\and M.~Messerotti\inst{8,9}
\and A.~Navarrini~\inst{1,10}
\and A.~Orfei\inst{2}
\and P.~Ortu\inst{1}
\and M.~Pili\inst{1}
\and T.~Pisanu\inst{1}
\and G.~Pupillo\inst{2}
\and A.~Saba\inst{5}
\and L.~Schirru\inst{1}
\and C.~Tiburzi\inst{1}
\and P.~Zucca\inst{11}
}

%
\institute{INAF - Cagliari Astronomical Observatory, Via della Scienza 5, I--09047 Selargius (CA), Italy                
\and INAF - Institute of Radio Astronomy, Via Gobetti 101, I--40129 Bologna, Italy                                      
\and Istituto Nazionale di Astrofisica (INAF/OAA), Largo Enrico Fermi 5, I--50125 Firenze, Italy                        
\and Los Alamos National Laboratory, Bikini Atoll Rd, Los Alamos, NM 87545, USA                                         
\and ASI - c/o Cagliari Astronomical Observatory, Via della Scienza 5, I--09047 Selargius (CA), Italy                   
\and INAF - Catania Astrophysical Observatory, Via Santa Sofia 78, I--95123 Catania, Italy                              
\and INAF - Turin Astrophysical Observatory, Via Osservatorio 20, I--10025 Pino Torinese (TO), Italy                    
\and INAF – Trieste Astronomical Observatory, Via Giambattista Tiepolo 11, I--34131 Trieste, Italy                      
\and Department of Physics, University of Trieste, Via Alfonso Valerio 2, I--34127 Trieste, Italy                       
\and NRAO -- Central Development Laboratory, 1180 Boxwood Estate Rd, Charlottesville, VA 22903, USA                     
\and ASTRON – The Netherlands Institute for Radio Astronomy, Oude Hoogeveensedijk 4, 7991 PD Dwingeloo, The Netherlands 
}

\date{Received November 28, 2023; accepted January 20, 2024}

\abstract
{
The Sun is an extraordinary workbench, from which several fundamental astronomical parameters can be measured with high precision.
Among these parameters, the solar radius $R_{\odot}$ plays an important role in several aspects, such as in evolutionary models.
Moreover, it conveys information about the structure of the different layers that compose the solar interior and its atmosphere.
Despite the efforts in obtaining accurate measurements of $R_{\odot}$, the subject is still debated and measurements are puzzling and/or lacking in many frequency ranges.
}
%
{
We aimed to determine the mean, equatorial, and polar radii of the Sun ($R_c$, $R_{eq}$, and $R_{pol}$) in the frequency range $18.1$ -- $26.1$~GHz.
We employed single-dish observations from the newly-appointed Medicina "Gavril Grueff" Radio Telescope and the Sardinia Radio Telescope (SRT) throughout 5 years, from $2018$ to mid-$2023$, in the framework of the SunDish project for solar monitoring.
}
%
{
Two methods to calculate the radius at radio frequencies -- the half-power and the inflection-point -- are considered and compared.
To assess the quality of our radius determinations, we also analysed the possible degrading effects of the antenna beam pattern on our solar maps, using two 2D-models (ECB and 2GECB).
We carried out a correlation analysis with the evolution of the solar cycle through the calculation of Pearson's correlation coefficient $\rho$ in the 13-month running means.
}
%
{
We obtained several values for the solar radius -- ranging between $959$ and $994$~arcsec -- and $\rho$, with typical errors of a few arcsec.
These values constrain the correlation between the solar radius and the solar activity, and allow us to estimate the level of the Sun prolatness in the centimetric frequency range.
}
%
{
Our $R_{\odot}$ measurements are consistent with values reported in literature, and provide refined estimations in the centimetric range.
The results suggest a weak prolatness of the solar limb ($R_{eq}$ > $R_{pol}$), although $R_{eq}$ and $R_{pol}$ are statistically compatible within 3$\sigma$ errors.
The correlation analysis using the solar images from the Grueff Radio Telescope shows (1) a positive correlation between the solar activity and the temporal variation of $R_c$ (and $R_{eq}$) at all observing frequencies, and (2) a weak anti-correlation between the temporal variation of $R_{pol}$ and the solar activity at $25.8$~GHz.
}


\titlerunning{Solar radius}
\authorrunning{Marongiu et al.}

\maketitle

%

\section{Introduction}
\label{par:intro}

The Sun is a mildly-active star with an 11-year activity cycle \citep{Gleissberg66}, emitting radiation in the whole electromagnetic spectrum, from radio to gamma-ray frequencies (e.g., \citealp{Aschwanden04,deglInnocenti07}).
This star is the closest for which highly-precise fundamental astronomical parameters are known.
Among these parameters, the solar radius $R_{\odot}$ -- usually normalised to the unit distance (1 AU) -- has been the subject of increasingly accurate measurements and investigations (e.g. \citealp{Gilliland81,Vaquero16,Rozelot18}).

The value of $R_{\odot}$ allows us to obtain several physical information, such as (1) eclipse computations, (2) the prolatness of the solar limb and (3) insights on the structure of the different layers in the solar atmosphere.
Moreover, determining the changes in $R_{\odot}$ provides insights into the physical mechanisms that may be responsible for its variation (e.g., \citealp{Gilliland81,Ribes91}), enabling a better understanding of (1) the variation of the luminosity and its possible climate effects on the Earth and its environment (e.g., \citealp{Dansgaard75,Stuiver80,Hiremath04,Chapman08,Hiremath15}), and (2) the structure of the solar interior, and the role of the underlying magnetism (e.g., \citealp{Emilio2000,Kilic11,Rozelot15,Kosovichev18}).
Until a few decades ago, only optical imaging observations of the Sun were available.
The canonical value at optical frequencies of the solar photospheric radius $R_{\odot, opt}$ -- accepted by the International Astronomical Union (IAU, \citealp{Mamajek15,Prsa16}) -- is $695.66 \pm 0.14$~Mm (corresponding to $959.16 \pm 0.19$~arcsec; \citealp{Haberreiter08}), and it is widely used in literature.
Observations at radio frequencies started in 1950, with the development of different techniques to measure $R_{\odot}$ in this band (e.g., \citealp{Coates58,Wrixon70,Swanson73,Pelyushenko83,Costa86,Costa99,Selhorst04,Alissandrakis17,Menezes17,Selhorst19a,Selhorst19b}).

Despite the importance of having an accurate estimate of $R_{\odot}$, its measurement is still a matter of debate, especially for the radio band.
Discrepancies in the values reported for $R_{\odot}$ -- and their uncertainties, that over time decreased below the arcsec level (e.g., \citealp{Ribes87,Ribes89}; Table 1 in \citealp{Menezes17}) -- could be explained as effects, difficult to quantify and/or not fully included in the measurement procedures.
These effects arise from (1) employing instruments with different performances (both ground-based and space-based, \citealp{Gough01}), in turn characterised by different angular resolutions, 
(2) different definitions of the solar limb in the intensity profile (e.g., \citealp{Meftah14}) and (3) the presence of a particularly-variable phenomenology close to the limb, such as the limb brightening (e.g., \citealp{Withbroe70,Lindsey81,Horne81,Selhorst19b}), whose properties are still debated (e.g., \citealp{Simon69,Fuerst79,Kosugi86,Belkora92}).
Limb brightening can provide important information on the temperature gradient of the solar atmosphere (e.g., \citealp{Vernazza81,Fontenla93,Selhorst05}).
The presence of the limb brightening -- if confirmed -- will be crucial for the measurement of $R_{\odot}$, especially in the radio domain.
Several authors reported limb brightening only in the polar regions of the solar disk -- known as polar brightening -- at radio frequencies,  in particular at $17$~GHz (e.g., \citealp{Shibasaki98,Nindos99,Selhorst03}) with the Nobeyama Radioheliograph (NoRH; \citealp{Nakajima94,Nakajima95}) and at $100$ and $230$~GHz \citep{Selhorst19b} with the Atacama Large Millimeter/submillimeter Array (ALMA; \citealp{alma04}).
The analysis of \citet{Selhorst03} showed that at $17$~GHz the trend of the polar brightening is strongly anti-correlated with solar activity, and \citet{Selhorst19b} found that at $100$ and $230$~GHz the intensity of the polar brightening is more pronounced at the southern pole.

The increasingly-accurate value of $R_{\odot}$ obtained in recent years suggests to analyse the Sun sphericity.
This analysis is deemed to be controversial because of the use of different measurement techniques, instrument calibration (if any), frequency domain of measurements, and -- for ground-based instruments --  atmospheric effects.
At extreme-UV (EUV) frequencies ($\sim 10^{16}$~GHz), using the solar images obtained by the Extreme Ultraviolet Imager (EIT) aboard the Solar and Heliospheric Observatory (SoHO) satellite \citep{soho95}, different authors showed that the equatorial radius $R_{eq}$ is larger than the polar radius $R_{pol}$ \citep{Auchere98,GimenezDeCastro07}.
Recently, the same conclusion was shown by \citet{Zhang22} at very low radio frequencies ($20$--$80$~MHz) using the radio interferometer LOw Frequency ARray (LOFAR; \citealp{vanHaarlem13}), in accordance with other similar works (e.g., \citealp{Ramesh06,Mercier15,Melnik18}).
On the other hand, \citet{Menezes21} showed a negligible difference between $R_{eq}$ and $R_{pol}$, obtained from observation with the Solar Submillimeter-wave Telescope (SST, \citealp{Kaufmann94}) and ALMA at high radio frequencies ($100$--$405$~GHz).

The radius dependence on the solar cycle is also a good indicator of the changes that occur in the solar atmosphere, but this dependence is still a subject of debate (e.g., \citealp{ReisNeto03,Kuhn04}).
Anti-correlated variations with other solar cycle indicators, such as sunspots, were found by several authors (e.g., \citealp{Secchi1872,Gilliland81,Sofia83,Wittmann93,Laclare96,Dziembowski01,Egidi06,Qu13}), whereas other works find the opposite behaviour (e.g., \citealp{Ulrich95,Rozelot98,Emilio2000,Delmas02,Noel04,Chapman08}). 
Furthermore, other authors (e.g., \citealp{Neckel95,Antia98,Kuhn04,Bush10}) report no -- or very small -- variations in $R_{\odot}$ correlated with the solar cycle in their observations.
At radio frequencies, variations of $R_{\odot}$ in phase with the solar cycle are reported by several authors (e.g., \citealp{Bachurin83} at $8$~GHz and $13$~GHz; \citealp{Selhorst19a} at $37$~GHz; \citealp{Costa99} at $48$~GHz).
\citet{Selhorst04} studied the variation of $R_{\odot}$ at $17$~GHz using NoRH maps over a solar cycle (1992--2003), revealing a good positive correlation between the mean radius $R_{c}$ and the solar cycle, but an anti-correlation when the polar radius $R_{pol}$ is considered.
This anti-correlation suggests a possible increase of polar brightening during the solar minima.
On the other hand, \citet{Menezes17} reported a strong anti-correlation between the solar activity and the variation of $R_{\odot}$ as measured at 212 and 405 GHz, using $\sim 16600$ solar maps throughout 18 years (from 1999 to 2017), obtained through SST.

Multi-frequency observations -- from radio to X-EUV domain -- of the Sun suggest a typical relationship between $R_{\odot}$ and the observing frequency, but today there is no unique theoretical model to reproduce this dependence.
\citet{Rozelot15} showed that this relationship can be modelled by a parabola, with a minimum at around $45$~THz.
At different radio frequencies, several authors \citep{Menezes17,Menezes21} show that $R_{\odot}$ follows an exponential trend, as the frequency decreases, where the curve seems to get flatter at higher frequencies ($\gtrsim 200$ GHz).
This trend is in contrast with other analyses (e.g., \citealp{Meftah18,Quaglia21}) that claimed the extreme weakness of the correlation between $R_{\odot}$ and the observing frequency.

In this complex and continually-evolving landscape of solar science, the present paper focuses on the measures of the solar radius ($R_{c}$, E-W direction $R_{eq}$, and N-S direction $R_{pol}$) and its behaviour over time (also with respect to the solar activity).
We use single-dish observations performed with the INAF Medicina "Gavril Grueff" Radio Telescope (hereafter Grueff Radio Telescope) and Sardinia Radio Telescope (SRT) from 2018 to mid-2023 in K-band ($18.1$ -- $26.1$~GHz) in the frame of the SunDish project \citep{Pellizzoni22}.
Our analysis allows us to enhance the data available in a frequency range -- especially at $20$ -- $25$~GHz -- characterised by few and often dated radii measurements (e.g., \citealp{Fuerst79,Costa86}), and to reveal a different behaviour of $R_{eq}$ and $R_{pol}$ with respect to the activity level of the Sun.
Thanks to this project, the external layers of the solar atmosphere can be deeply analysed in this radio domain, gaining insights into the properties of the corona in terms of temperature and density distributions \citep{Marongiu23b}.
The coverage of the entire solar disk with the suitable resolution, the low noise, the accurate absolute calibration, and the great sensitivity of INAF's radio telescopes make these data crucial for our purpose.

We organise this paper as follows.
A brief description of the INAF radio telescopes used in this work, with the relative implementation of the instrumental configurations and the observing techniques adopted for radio-continuum solar imaging is reported in Section~\ref{par:obs_datared}.
The techniques employed for determining $R_{\odot}$, and for analysing the role of the antenna beam pattern on our solar maps are described in Section~\ref{par:data_an}.
We present our results in Section~\ref{par:results}, where we also include an investigation of the correlation between (1) our solar radius time series and the solar activity cycle, and (2) the equatorial and the polar radius time series.
After the discussion of our results in Section~\ref{par:disc_concl}, we give our conclusions in Section~\ref{par:concl_svil}.

\section{Observations and data reduction}
\label{par:obs_datared}

In this work the solar data set was obtained by single-dish observations, using the INAF single-dish radio telescopes network\footnote{\url{https://www.radiotelescopes.inaf.it}} in the radio K-band ($18$ -- $26$~GHz).
This network includes the Grueff Radio Telescope (previously-known as the Medicina radio telescope), the Sardinia Radio Telescope (SRT), and the Noto Radio Telescope\footnote{We did not carry out any solar observations with the Noto radio telescope, since it is currently under maintenance operations.}.
The solar data set covers more than 5 years of observations (from 2018 to mid-2023), corresponding to about half a solar cycle.
These solar campaigns are a cornerstone of the "SunDish Project" (PI: A. Pellizzoni) \footnote{\url{https://sites.google.com/inaf.it/sundish}}, a collaboration between INAF and ASI \citep{Pellizzoni19,Plainaki20,Pellizzoni22} involved in deeply probing the solar atmosphere.

\begin{figure*} 
\centering
{\includegraphics[width=89.5mm]{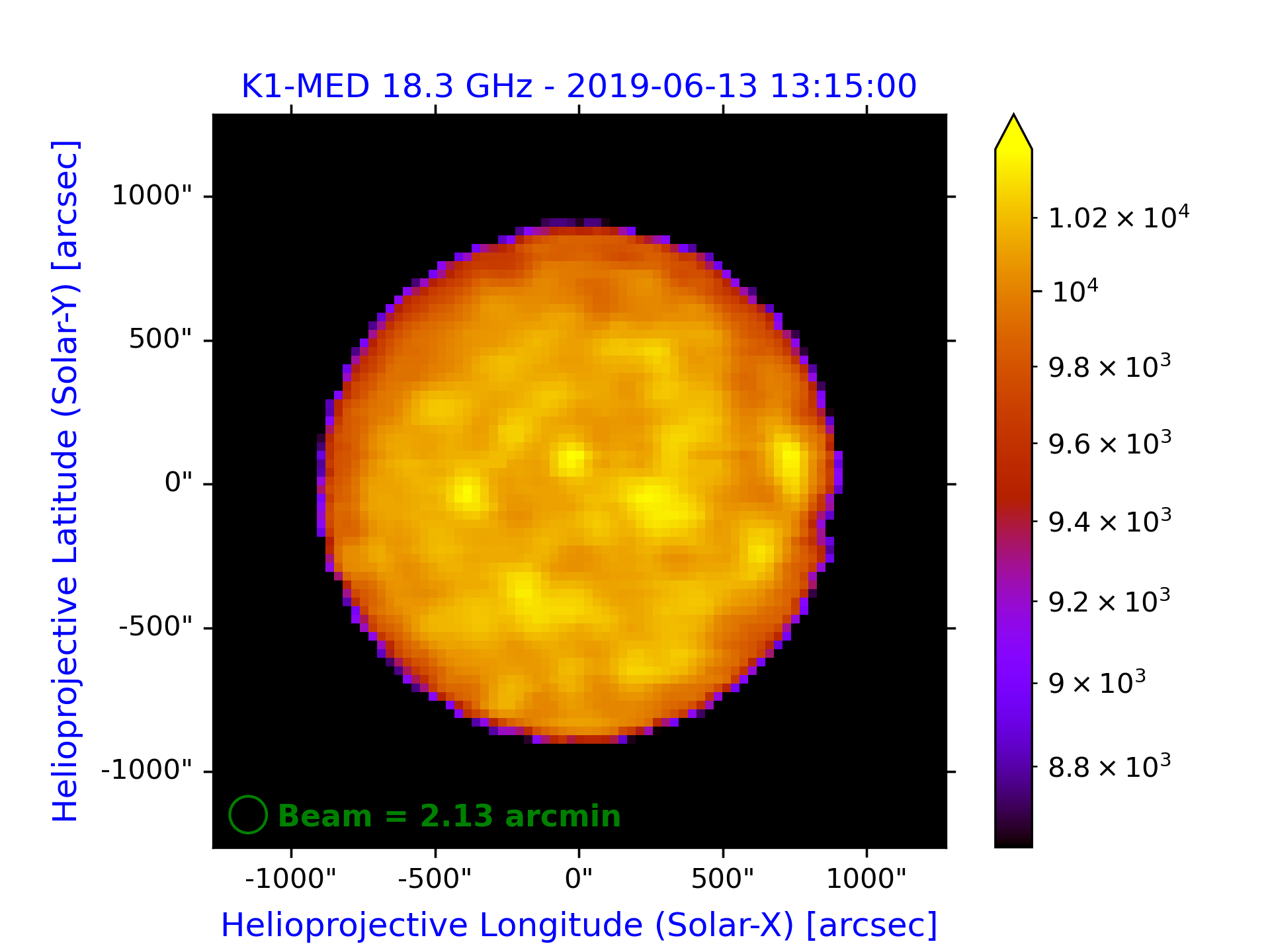}} \quad
{\includegraphics[width=89.5mm]{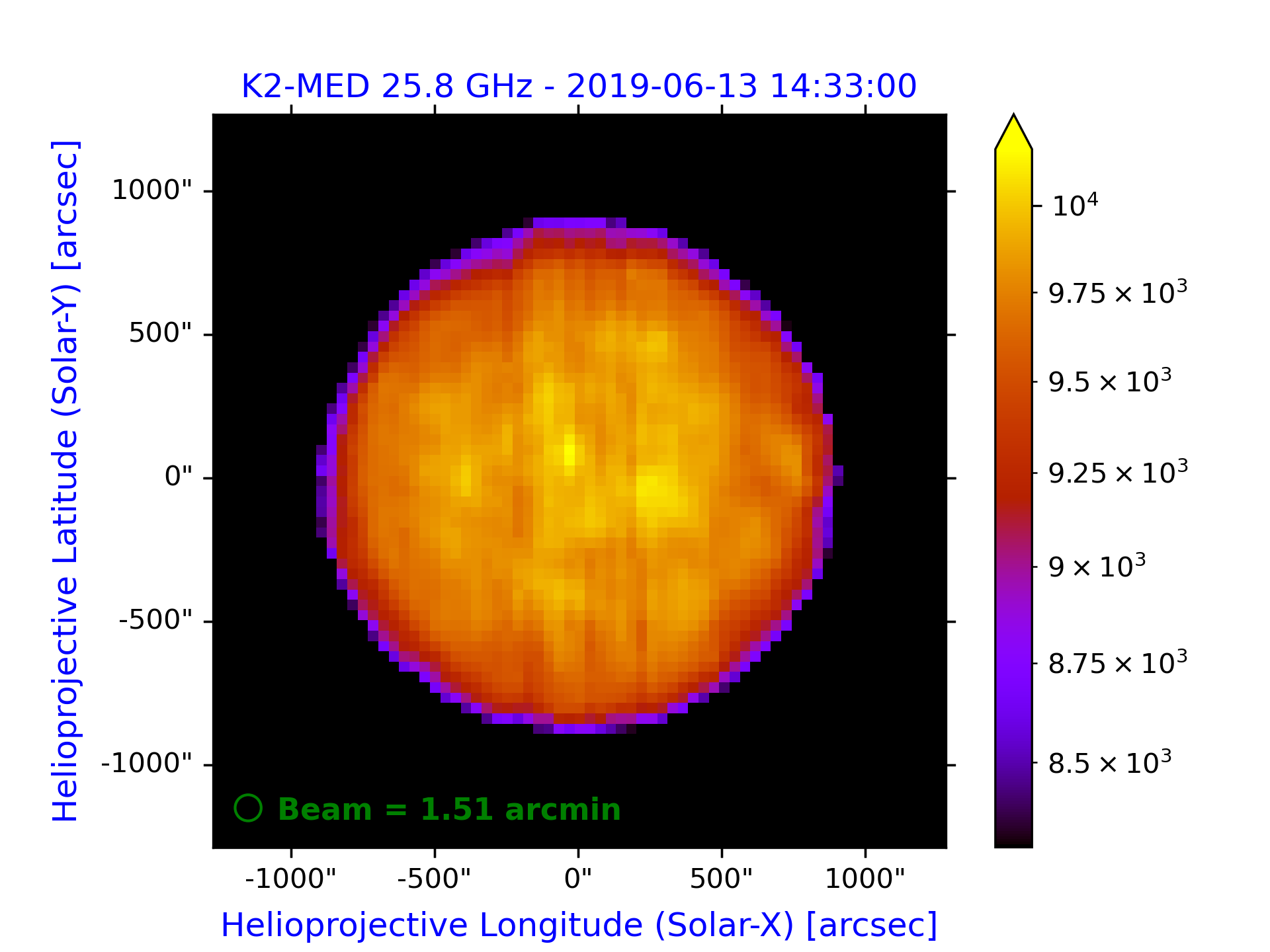}} \\
\caption{
Solar disk image at $18.3$~GHz (left) and $25.8$~GHz (right) obtained with the Grueff Radio Telescope on June 13$^{th}$ 2019, processed with the {\sc SUNDARA} package \citep{Marongiu21}.
Colorbars indicate $T_B$ of the solar maps in units of Kelvin.
The green circles on the lower left corner of each map mark the Beam Width Half Maximum (BWHM) at the observed frequencies.
}
\label{fig:sun_map_med}
\end{figure*}

\begin{figure*} 
\centering
{\includegraphics[width=89.5mm]{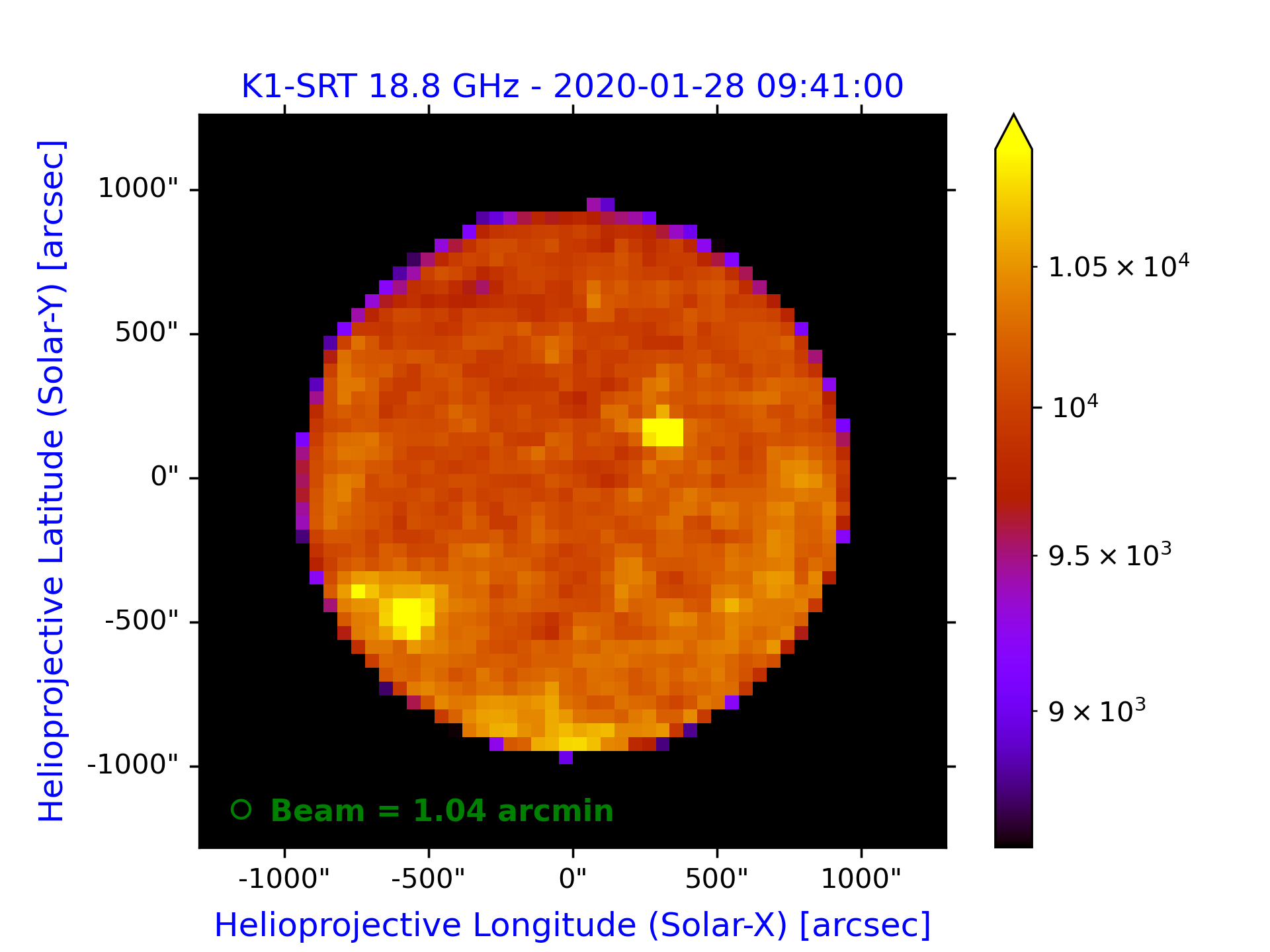}} \quad
{\includegraphics[width=89.5mm]{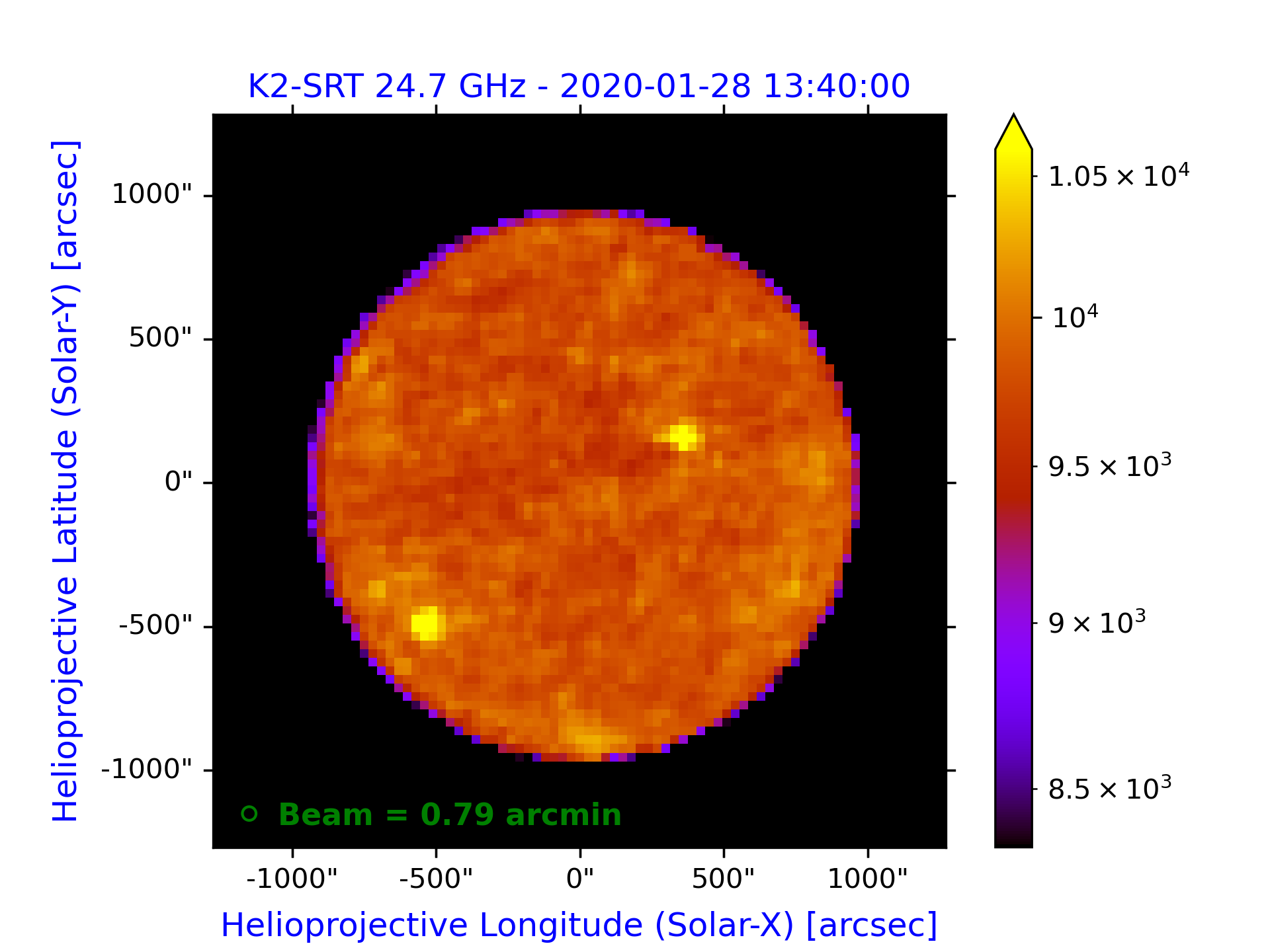}} \\
\caption{
Solar disk image at $18.8$~GHz (left) and $24.7$~GHz (right) obtained with SRT on January 28$^{th}$ 2020, processed with the {\sc SUNDARA} package \citep{Marongiu21}.
See the caption of Figure \ref{fig:sun_map_med} for a full description of the colorbars and the symbols.
}
\label{fig:sun_map_srt}
\end{figure*}

The 32-m Grueff Radio Telescope has been observing the full solar disk almost once a week since February 2018.
The instrument is located in Medicina (near Bologna, Italy) at 25~m elevation in the heart of the Po Valley.
Among multiple receivers covering the range $1.3$ -- $26.5$~GHz, we observed the Sun through the K-band dual-feed receiver, primarily using $18.3$ and $25.8$~GHz as central frequencies.
The corresponding beam sizes are $2.1$ and $1.5$~arcmin, respectively (green circles in Fig.~\ref{fig:sun_map_med}).
The 64-m Sardinia Radio Telescope (SRT) is located on an isolated plateau at $650$~m elevation in Sardinia (Italy).
Up to date, the solar campaigns with SRT have taken place mostly once a month at primarily $18.8$ and $24.7$~GHz through a $7$~feeds dual polarization K-band receiver \citep{Bolli15,Prandoni17}.
This receiver, customised for solar observations \citep{Pellizzoni22}, is characterised by a beam size of $1.0$ and $0.8$~arcmin, respectively (green circles, Fig.~\ref{fig:sun_map_srt}).
SRT is presently upgrading its capabilities with new receivers -- exploitable also to observe the Sun -- operating up to $116$~GHz, thanks to the financial support of the Italian Ministry of University and
Research in the context of the National Operative Programme (Programma Operativo Nazionale-PON; \citealp{Govoni21})\footnote{\url{https://sites.google.com/a/inaf.it/pon-srt/home}}.

In the time frame $2018$ -- mid-$2023$, our observations of the Sun in K-band were mostly obtained at Medicina, with a $\sim$ 5\% of the total number of sessions provided by the larger SRT.
In the frequency range $18.1$ -- $26.1$~GHz we obtained an extensive data set of $327$ maps.
Among this data set, $145$ maps were obtained at $18.3$~GHz, and $128$ maps were obtained at $25.8$~GHz.
However, due to high atmospheric opacity and occasional instrumental failures, some maps ($\sim 10\%$) had to be discarded, thus reducing the final amount to $287$ maps.
From 2019 to 2021, the need for a manual solar setup for each observation has limited the use of SRT in the frequency range of $18.3$ -- $25.5$~GHz to a few solar sessions/year.
This configuration resulted in $19$ solar maps: among these maps, $10$ were observed at $18.8$~GHz, and $7$ were observed at $24.7$~GHz. $2$ maps at $18.8$~GHz were discarded as they were acquired under unsuitable weather conditions or were interested by instrumental failures.
As shown in Fig.~\ref{fig:sun_map_srt}, maps with SRT were mainly performed at $18.8$ and $24.7$~GHz to exploit the best receiver performance and avoid the Radio-Frequency Interference (RFI).

The solar maps are composed by On-The-Fly (OTF) scans \citep{Prandoni17}, commanded using the {\sc DISCOS} antenna control system, in Right Ascension (RA, at the lower frequency) and Declination (DEC, at the higher frequency).
These maps cover an area of about $1.3^{\circ} \times 1.3^{\circ}$ in the sky at Medicina (Fig.~\ref{fig:sun_map_med}), and $1.5^{\circ} \times 1.5^{\circ}$ at SRT (Fig.~\ref{fig:sun_map_srt}).
The radio signal is processed with two different back-ends: (1) the full-stokes spectral-polarimeter SARDARA ($1.5$~GHz theoretical bandwidth, \citealp{Melis18}) at SRT, and (2) the Total-power/intensity at the Grueff Radio Telescope\footnote{From July 2021 this radio telescope is also equipped by the SARDARA system \citep{Mulas22}, characterised by a theoretical bandwidth of $1.2$~GHz. The scientific community can use this back-end since the observing semester 2023A.}, characterised by a theoretical bandwidth of $0.3$~GHz.
The high dynamic range of SARDARA allows us to detect in the same image the emission coming both from the bright solar disk and the faint tails near the limb.
The Supernova Remnant Cas~A is selected for the procedure of absolute brightness temperature calibration (for details see \citealp{Pellizzoni22, Mulas22}).
The solar images from the Grueff Radio Telescope and SRT were extracted from the FITS files produced by the INAF solar pipeline SUNPIT \citep{Marongiu22}.
Details about the OTF mapping techniques, the setup configurations used for the receiver and back-end, the observing strategy, and the data processing (RFI rejection, baseline background subtraction, and image production) are available -- both for the Grueff Radio Telescope and SRT -- in the first solar paper of the SunDish collaboration (\citealp{Pellizzoni22}, and references therein).
Moreover, the SunDish Archive\footnote{\url{https://sites.google.com/inaf.it/sundish/sundish-images-archive/sundish-archive-summary}} includes the data set -- regularly updated -- of the Grueff Radio Telescope and SRT.

\section{Data analysis}
\label{par:data_an}

In this work, we analysed and measured the size of the Sun, in order to obtain information on the value of $R_{c}$, $R_{eq}$, and $R_{pol}$, and to study the temporal evolution of these radii.
For the radius calculations we used a specific procedure, described in Sect.~\ref{par:det_radius}.
To obtain more robust measurements of the solar radii, and to corroborate the presence of the coronal emission -- the analysis of which will be described in detail in a further paper \citep{Marongiu23b} -- we also analysed the contribution of the antenna beam pattern on the solar signal (Section~\ref{par:ant_beam}).

\subsection{Prescription for calculating the solar radius}
\label{par:det_radius}

The measurement of $R_{\odot}$ at radio frequencies requires an unambiguous definition of the solar limb, a parameter strongly influenced by the instability of the solar atmosphere.
These instabilities are characterised by ever-changing small structures -- prominent in the observed radio Sun -- such as active regions (ARs), sunspots, spicules, and faculae.
Two widely-used methods for measuring the radio $R_{\odot}$ in the literature are the so-called "half-power method" (hereafter HP-method, \citealp{Costa99,Selhorst11,Menezes17,Selhorst19a}), and the "inflection-point of the limb darkening function" (hereafter IP-method, e.g., \citealp{Emilio12,Emilio15,Alissandrakis17,Menezes21}).
In our analysis both methods are applied and compared.

The HP-method calculates $R_{\odot}$ at the points where the brightness temperature $T_B$ is half of its quiet-Sun Level (QSL) $T_{qS}$ (Fig.~\ref{fig:raggio_example}, top left), the most common mean temperature in the distribution of the solar disk intensity (for further details, see \citealp{Landi08,Pellizzoni22}).
In the IP-method, the solar radius is determined by the position of the inflection-point of the solar limb-darkening profile (Fig.~\ref{fig:raggio_example}, top right); this method is less susceptible to the irregularities of the telescope beams and the variations of the $T_B$ profiles of the Sun, such as the limb-brightening level and the presence of ARs \citep{Menezes22}.

To determine $R_{\odot}$ (with the methods described above) we follow a prescription similar to others described in the literature \citep{Costa99,Selhorst19a,Menezes17,Menezes21,Menezes22}.
The first step is the extraction of the solar limb coordinates from each map.
These coordinates were determined considering scans, corresponding to brightness profiles (for example, see the green curve in Fig.~\ref{fig:raggio_example} on top left), in both RA and DEC orientation.
During the limb point extraction, some criteria were adopted to avoid extracting limb points associated with ARs, or affected by weather and seasonal effects, instrumental errors, or high atmospheric opacity, which may increase the calculated local radius in that region and hence the average radius \citep{Menezes17,Menezes21}:
\begin{itemize}
    \item in the HP-method, we considered only coordinates ranging in a solar ring composed of points with a center-to-limb distance corresponding to a brightness level between $0.9 T_{qS}$ and $1.1 T_{qS}$;
    \item in the IP-method we considered only scans characterised by a signal exceeding the RMS of the solar images\footnote{These scans are characterised by at least a conservative 15\% of the pixels with $T_B \geq 0.15 T_{qS}$} ($\sim 10$~K); the limb coordinates are defined as the maximum and minimum points of the numerical derivative (blue curve in Fig.~\ref{fig:raggio_example}, top right) of each scan across the solar disk.
\end{itemize}

The second step consists in modelling the solar limb -- composed by the limb coordinates extracted in the first step -- for each solar map.
This limb is modelled following the generic parametric equation of the ellipse:
\begin{equation}
\left[\frac{X \cos{\theta} + Y \sin{\theta}}{R_{eq}}\right]^2 +
\left[\frac{X \sin{\theta} + Y \cos{\theta}}{R_{pol}}\right]^2 = 1
\label{eq:model_radius}
\end{equation}
where $\theta$ is the ellipse orientation (set to 0 in this analysis), $X = x - x_0$ and $Y = y - y_0$ ($x_0$ and $y_0$ are the centre coordinates), $R_{eq}$ and $R_{pol}$ indicate the semiaxes of the ellipse. 
Eq.~\ref{eq:model_radius} comes down to the case of the circle, assuming $R_{c} = R_{eq} = R_{pol}$.
The modelling is performed using a least-squares method, to determine $x_0$, $y_0$, and the solar radii ($R_{c}$ in the circular case, $R_{eq}$ and $R_{pol}$ in the elliptical case).
The calculation of $R_{\odot}$ is performed through two alternative approaches:
\begin{enumerate}
    \item Modelling procedure $\rightarrow$ the best-fit parameters are obtained from the modelling (Eq.~\ref{eq:model_radius} and Fig.~\ref{fig:raggio_example}, bottom left), using either the circle or the ellipse case.
    \item Statistical procedure $\rightarrow$ the radii are obtained through the median values of the centre-to-limb distances, calculated thanks to the distance between each limb point and the fitted centre position (both in the circular and in the elliptical case).
    The uncertainties in the median values are described by the third and first quartiles of the sample.
    We calculated three average radii: (a) the average radius $\bar R_{stat}$ considering all the distances, (b) the equatorial radius $\bar R_{eq,stat}$ considering only equatorial latitudes of the solar disk-points between $30^{\circ}$ N and $30^{\circ}$ S, and (c) the polar radius $\bar R_{pol,stat}$ considering only points above $60^{\circ}$ N and below $60^{\circ}$ S (Fig.~\ref{fig:raggio_example}, bottom right).
\end{enumerate}
For both approaches, we filtered the final solar limb coordinates thanks to the following iterative process:
\begin{figure*} 
\centering
{\includegraphics[width=87mm]{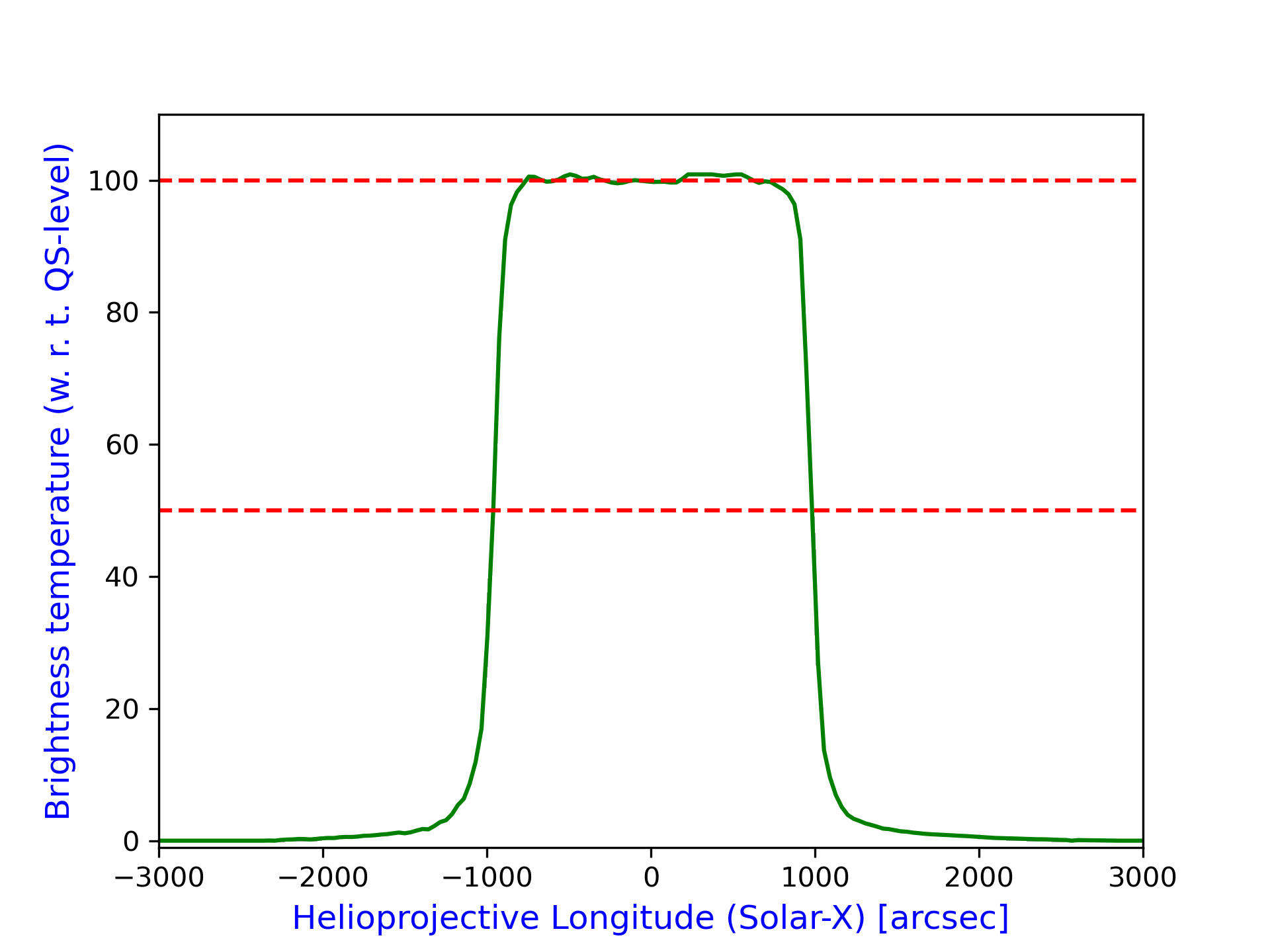}} \quad
{\includegraphics[width=86mm]{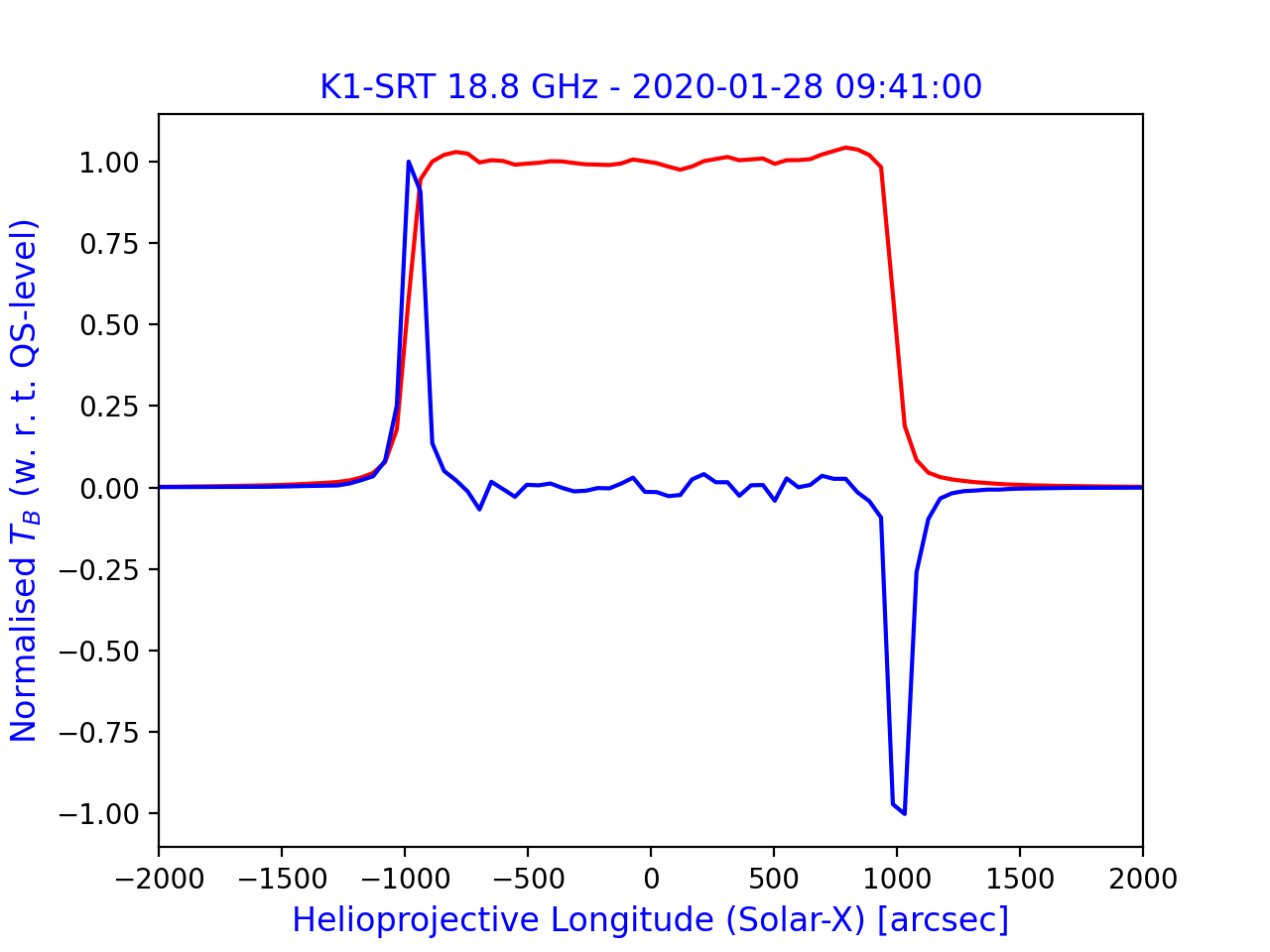}} \\
{\includegraphics[width=86mm]{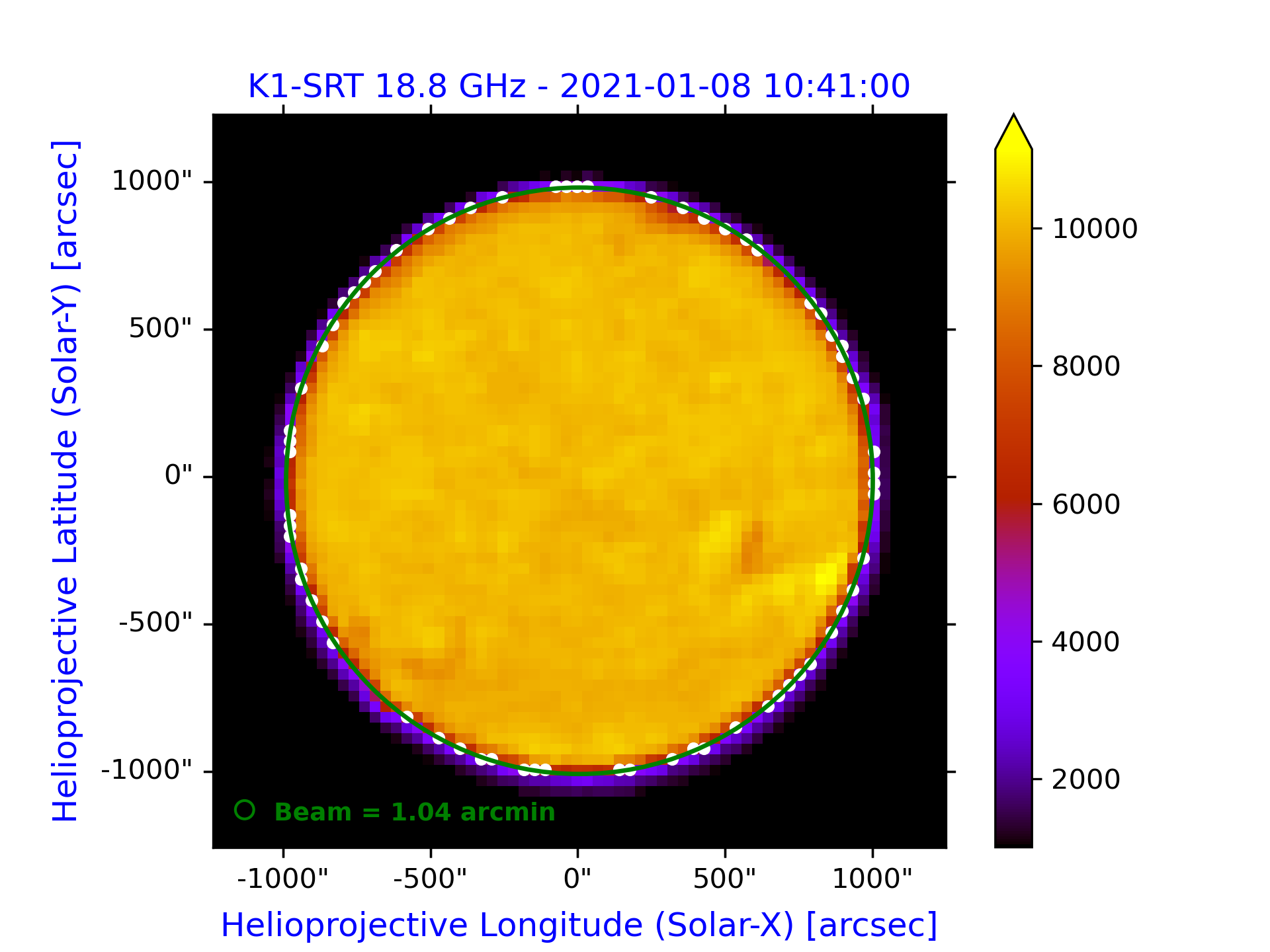}} \quad
{\includegraphics[width=88mm]{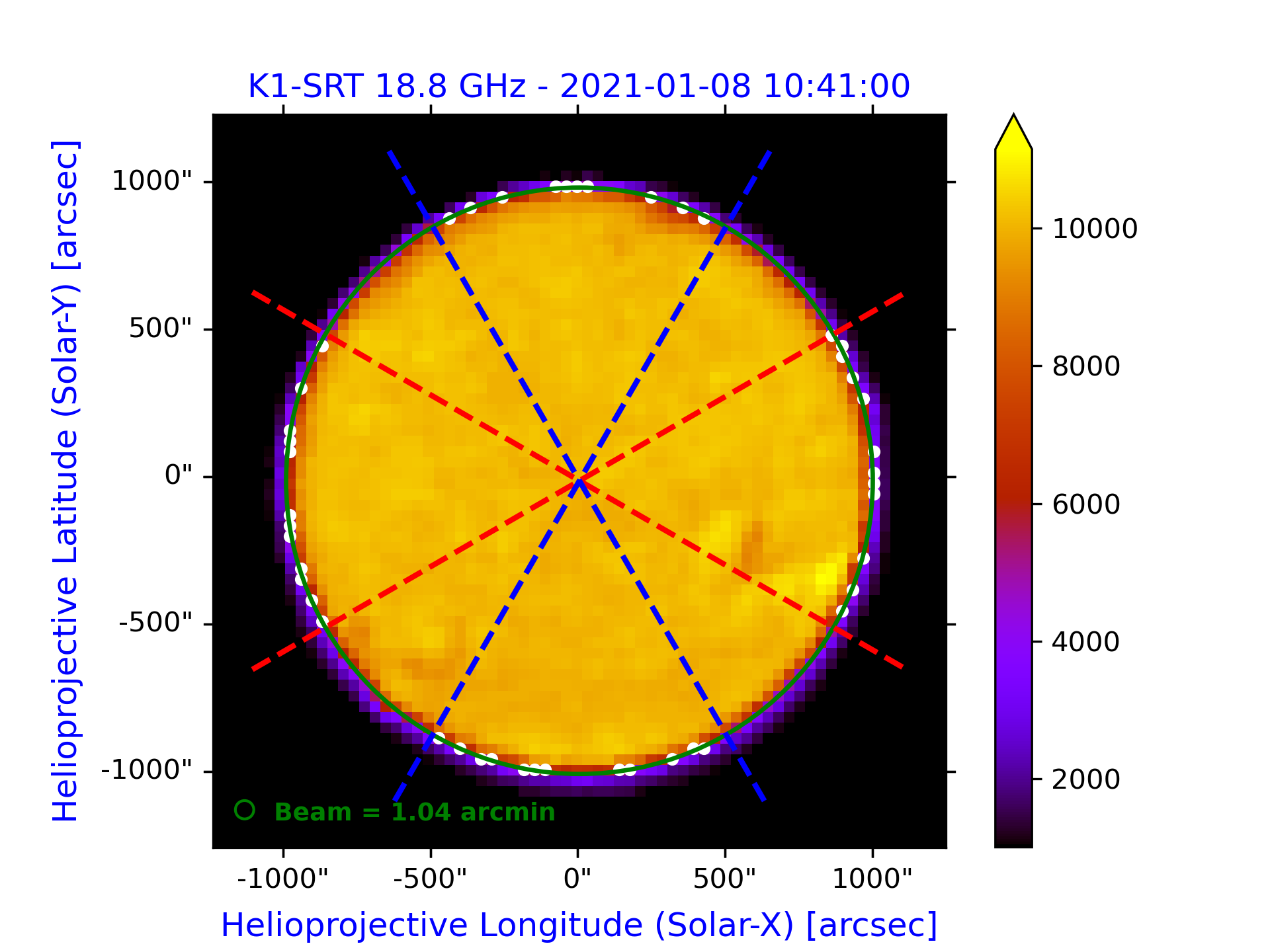}} \\
\caption{
Steps of solar radius measurement: (top left) brightness profile with the $T_B$ levels (dashed red lines) of $T_{qS}$ (top line) and the half value of $T_{qS}$ (bottom line), used in the HP-method; (top right) numerical differentiation of the brightness profile (blue curve), whose minimum and maximum points correspond to the limb coordinates of the equatorial scan (red curve) in the IP-method; (bottom left) limb coordinates (white points) extracted from a solar map with a fit (circle or ellipse, green solid line); (bottom right) limb coordinates (white points) extracted from a solar map with the statistical procedure to estimate $\bar R_{eq,stat}$ and $\bar R_{pol,stat}$ (dashed red and blue lines are the constraints of the equatorial and the polar regions, respectively).
}
\label{fig:raggio_example}
\end{figure*}
\begin{itemize}
    \item The average radius $\bar R$ is calculated thanks to the distance between each limb point and the fitted centre position.
    For each fit, the points with centre-to-limb distance (obtained from the modelling) between $\bar R - d$~arcsec and $\bar R + d$~arcsec are considered (where $d = 10$~arcsec for the circular fit, and $d = 20$~arcsec for the elliptical fit), and then a new fit is performed with the remaining points.
    This process is repeated until no other points are discarded.
    \item If there are less than $25$ points remaining, the entire map is discarded, otherwise the radius is calculated.
    In the case of the statistical procedure, for the calculation of $R_{eq}$ and $R_{pol}$ we assume a threshold of $10$ remaining points for each side.
    \item If the standard deviation of the solar limb coordinates is less than $20$~arcsec, then the calculated radius is stored and the next map is submitted to this process.
\end{itemize}
The same method is applied to both INAF radio telescopes, in order to compare these solar radii.
The resulting radius values is further normalised at the distance of 1 AU, through the correction for the orbit eccentricity of the Earth.
This correction makes the apparent $R_{\odot}$ ranging between $\sim 950$~arcsec (aphelion) and $\sim 1005$~arcsec (perihelion) during the year.

To calculate the monthly and annual medians of our calculated $R_{\odot}$, we apply a further criterion, similar to other prescriptions in the literature (e.g., \citealp{Menezes21,Menezes22}).
The following criterion reduces the scattering in the distribution of the calculated $R_{\odot}$, caused by maps of low quality (affected by systematic effects of the instrument and/or bad weather conditions).
\begin{itemize}
    \item For each observing frequency, we calculate the average solar radius $\bar R$, considering only the values included in the range $900$ -- $1050$~arcsec (a conservative range of $R_{\odot}$ obtained thanks to our solar observations at K-band).
    \item With the remaining points (at least 3), we apply the statistical Chauvenet's criterion \citep{Chauvenet1863,Maples18,Konz23}, aimed to extract further outliers from our set of $R_{\odot}$.
    \item With the remaining points (at least 3), we calculate again $\bar R$, discarding the values that are outside of the range $\bar R \pm 60$~arcsec.
    \item We repeat the same step considering the range $\bar R \pm 30$~arcsec.
    \item Finally, with the remaining points (at least 3), we calculate again $\bar R$, discarding the values that are outside the more stringent $\bar R \pm 10$~arcsec range.
    The latter process is repeated until no other points are discarded.
\end{itemize}
The final value of $R_{\odot}$ is the median of the remaining points of the sample, and the uncertainties are described by the third and first quartiles of the sample, respectively.
\begin{table*}
\caption{Measured solar radii at several frequencies in the range $18$ -- $26$~GHz with the Grueff Radio Telescope and SRT radio telescopes, obtained through the modelling described in Sects.~\ref{par:det_radius} and \ref{par:ant_beam}. As specified at $18.3$ and $25.8$~GHz, we also report the measured values obtained from the averaged solar maps.}
\label{tab:raggio_resume}
\centering
\begin{tabular}{ccccc|cc}
\hline
\hline
Frequency & Telescope  & Type radius          & HP-method             & IP-method             & ECB-model             & 2GECB-model           \\
(GHz)     &            &                      & (arcsec)              & (arcsec)              & (arcsec)              & (arcsec)              \\
\hline
18.1      & Grueff   & $R_{c}$              & $981.8_{-1.0}^{+2.0}$ & $976.7_{-3.4}^{+2.9}$ & -                     & -                     \\
18.1      & Grueff   & $R_{eq}$             & $984.6_{-1.4}^{+0.8}$ & $977.5_{-2.7}^{+2.2}$ & -                     & -                     \\
18.1      & Grueff   & $R_{pol}$            & $980.5_{-3.4}^{+2.4}$ & $976.2_{-2.0}^{+2.9}$ & -                     & -                     \\
\hline
18.3      & Grueff   & $R_{c}$              & $984.7_{-2.1}^{+2.6}$ & $978.7_{-1.9}^{+1.7}$ & -                     & -                     \\
18.3      & Grueff   & $R_{eq}$             & $985.4_{-2.4}^{+2.9}$ & $978.8_{-2.2}^{+2.1}$ & $990.6_{-3.0}^{+5.5}$ & $975.6_{-4.5}^{+4.9}$ \\
18.3      & Grueff   & $R_{pol}$            & $983.2_{-2.3}^{+1.7}$ & $978.9_{-2.5}^{+2.2}$ & $985.8_{-2.0}^{+4.3}$ & $968.9_{-4.1}^{+4.6}$ \\
18.3      & Grueff   & $R_{c}$ (averaged)   & $981.7 \pm 0.3$       & $976.4 \pm 0.3$       & -                     & -                     \\
18.3      & Grueff   & $R_{eq}$ (averaged)  & $984.9 \pm 0.6$       & $977.8 \pm 0.8$       & -                     & -                     \\
18.3      & Grueff   & $R_{pol}$ (averaged) & $978.8 \pm 0.6$       & $974.0 \pm 0.8$       & -                     & -                     \\
\hline
18.8      & SRT        & $R_{c}$              & $978.7_{-0.7}^{+0.3}$ & $974.1_{-0.7}^{+1.5}$ & -                     & -                     \\
18.8      & SRT        & $R_{eq}$             & $979.6_{-0.1}^{+0.3}$ & $974.9_{-2.1}^{+1.1}$ & $981.6_{-1.8}^{+0.6}$ & $973.1_{-3.9}^{+1.0}$ \\
18.8      & SRT        & $R_{pol}$            & $978.2_{-0.7}^{+0.9}$ & $972.0_{-3.0}^{+1.6}$ & $993.5_{-6.3}^{+2.8}$ & $960.6_{-1.4}^{+1.5}$ \\
\hline
23.6      & Grueff   & $R_{c}$              & $980.7_{-0.7}^{+1.2}$ & $974.8_{-1.1}^{+2.1}$ & -                     & -                     \\
23.6      & Grueff   & $R_{eq}$             & $979.8_{-1.8}^{+1.7}$ & $974.0_{-0.9}^{+0.4}$ & -                     & -                     \\
23.6      & Grueff   & $R_{pol}$            & $981.4_{-1.4}^{+1.5}$ & $976.2_{-1.5}^{+2.7}$ & -                     & -                     \\
\hline
24.7      & SRT        & $R_{c}$              & $976.7_{-0.9}^{+0.5}$ & $972.4_{-0.6}^{+0.1}$ & -                     & -                     \\
24.7      & SRT        & $R_{eq}$             & $976.0_{-0.2}^{+0.7}$ & $973.7_{-1.8}^{+0.2}$ & $980.3_{-5.8}^{+1.4}$ & $965.1_{-3.0}^{+0.2}$ \\
24.7      & SRT        & $R_{pol}$            & $976.2_{-0.9}^{+1.3}$ & $972.9_{-1.4}^{+0.1}$ & $982.3_{-1.8}^{+5.3}$ & $959.3_{-7.1}^{+6.2}$ \\
\hline
25.8      & Grueff   & $R_{c}$              & $982.0_{-1.5}^{+1.3}$ & $975.3_{-2.0}^{+1.6}$ & -                     & -                     \\
25.8      & Grueff   & $R_{eq}$             & $982.2_{-2.0}^{+2.3}$ & $975.1_{-2.2}^{+1.9}$ & $987.3_{-2.9}^{+3.7}$ & $964.2_{-4.4}^{+5.8}$ \\
25.8      & Grueff   & $R_{pol}$            & $981.2_{-1.3}^{+1.4}$ & $975.5_{-1.7}^{+1.2}$ & $985.2_{-2.9}^{+2.1}$ & $961.1_{-9.5}^{+5.6}$ \\
25.8      & Grueff   & $R_{c}$ (averaged)   & $979.3 \pm 0.4$       & $972.1 \pm 0.4$       & -                     & -                     \\
25.8      & Grueff   & $R_{eq}$ (averaged)  & $979.9 \pm 0.9$       & $971.9 \pm 0.9$       & -                     & -                     \\
25.8      & Grueff   & $R_{pol}$ (averaged) & $977.9 \pm 0.9$       & $973.1 \pm 0.9$       & -                     & -                     \\
\hline
26.1      & Grueff   & $R_{c}$              & $981.0_{-1.4}^{+0.9}$ & $973.8_{-0.6}^{+1.0}$ & -                     & -                     \\
26.1      & Grueff   & $R_{eq}$             & $980.6_{-1.4}^{+3.2}$ & $972.4_{-0.6}^{+0.6}$ & -                     & -                     \\
26.1      & Grueff   & $R_{pol}$            & $981.1_{-1.5}^{+1.0}$ & $974.2_{-0.7}^{+1.3}$ & -                     & -                     \\
\hline
\end{tabular}
\end{table*}
\begin{table*}
\caption{Measured solar radii at several frequencies in the range $18$ -- $26$~GHz with the Grueff and SRT radio telescopes, obtained through the statistical approach described in Sect.~\ref{par:det_radius}. As specified at $18.3$ and $25.8$~GHz, we also report the measured values obtained from the averaged solar maps.}
\label{tab:raggio_resume2}
\centering
\begin{tabular}{ccccccc}
\hline
\hline
Frequency & Telescope  & Type radius          & Stat - HP-method      & Stat - HP-method      & Stat - IP-method      & Stat - IP-method      \\
(GHz)     &            &                      & Ellipse (arcsec)      & Circle (arcsec)       & Ellipse (arcsec)      & Circle (arcsec)       \\
\hline
18.1      & Grueff   & $R_{c}$              & $981.8_{-0.9}^{+1.5}$ & $981.8_{-1.0}^{+2.0}$ & $976.7_{-3.2}^{+2.4}$ & $976.6_{-3.4}^{+2.9}$ \\
18.1      & Grueff   & $R_{eq}$             & $983.8_{-0.8}^{+1.4}$ & $982.6_{-2.0}^{+1.4}$ & $977.3_{-3.1}^{+1.7}$ & $977.3_{-4.5}^{+2.5}$ \\
18.1      & Grueff   & $R_{pol}$            & $981.2_{-3.4}^{+1.2}$ & $981.5_{-1.5}^{+1.9}$ & $975.9_{-1.9}^{+2.2}$ & $976.4_{-3.6}^{+2.4}$ \\
\hline
18.3      & Grueff   & $R_{c}$              & $984.1_{-2.1}^{+2.2}$ & $984.6_{-2.1}^{+2.6}$ & $978.6_{-2.6}^{+2.1}$ & $978.6_{-1.9}^{+1.7}$ \\
18.3      & Grueff   & $R_{eq}$             & $984.2_{-1.7}^{+3.1}$ & $984.5_{-2.0}^{+3.2}$ & $978.2_{-2.2}^{+2.2}$ & $978.3_{-1.9}^{+1.9}$ \\
18.3      & Grueff   & $R_{pol}$            & $982.9_{-2.1}^{+1.8}$ & $984.2_{-2.3}^{+2.2}$ & $978.1_{-2.0}^{+2.6}$ & $978.2_{-1.3}^{+1.9}$ \\
18.3      & Grueff   & $R_{c}$ (averaged)   & $981.4_{-7.2}^{+8.3}$ & $981.3_{-4.7}^{+5.7}$ & $975.6_{-7.0}^{+7.0}$ & $976.4_{-5.0}^{+4.0}$ \\
18.3      & Grueff   & $R_{eq}$ (averaged)  & $983.7_{-7.8}^{+8.5}$ & $981.0_{-5.3}^{+5.6}$ & $977.2_{-7.5}^{+6.7}$ & $976.8_{-4.0}^{+3.7}$ \\
18.3      & Grueff   & $R_{pol}$ (averaged) & $978.4_{-6.4}^{+8.6}$ & $980.9_{-3.8}^{+6.3}$ & $974.4_{-6.6}^{+5.6}$ & $975.9_{-4.8}^{+4.2}$ \\
\hline
18.8      & SRT        & $R_{c}$              & $978.7_{-0.1}^{+0.3}$ & $978.6_{-0.7}^{+0.3}$ & $973.9_{-2.6}^{+1.3}$ & $974.0_{-0.7}^{+1.5}$ \\
18.8      & SRT        & $R_{eq}$             & $979.4_{-0.5}^{+0.3}$ & $978.6_{-0.2}^{+0.5}$ & $974.4_{-2.2}^{+0.8}$ & $974.3_{-1.5}^{+0.8}$ \\
18.8      & SRT        & $R_{pol}$            & $978.3_{-1.2}^{+1.1}$ & $978.4_{-0.7}^{+0.2}$ & $972.8_{-3.9}^{+1.4}$ & $973.9_{-1.1}^{+1.3}$ \\
\hline
23.6      & Grueff   & $R_{c}$              & $980.9_{-1.5}^{+0.7}$ & $980.7_{-0.7}^{+1.2}$ & $975.1_{-1.2}^{+1.1}$ & $974.7_{-1.1}^{+2.1}$ \\
23.6      & Grueff   & $R_{eq}$             & $980.4_{-2.5}^{+0.9}$ & $980.3_{-0.7}^{+1.7}$ & $973.8_{-0.8}^{+1.1}$ & $975.0_{-1.6}^{+1.5}$ \\
23.6      & Grueff   & $R_{pol}$            & $981.6_{-1.4}^{+0.9}$ & $980.3_{-0.5}^{+2.0}$ & $975.6_{-1.8}^{+2.1}$ & $976.2_{-2.8}^{+0.9}$ \\
\hline
24.7      & SRT        & $R_{c}$              & $976.9_{-0.4}^{+0.1}$ & $976.6_{-0.9}^{+0.5}$ & $972.1_{-0.6}^{+1.1}$ & $972.4_{-0.6}^{+0.1}$ \\
24.7      & SRT        & $R_{eq}$             & $976.6_{-0.4}^{+0.9}$ & $976.0_{-0.1}^{+0.7}$ & $973.6_{-1.4}^{+1.4}$ & $972.1_{-0.3}^{+0.5}$ \\
24.7      & SRT        & $R_{pol}$            & $977.0_{-1.0}^{+0.2}$ & $976.0_{-0.4}^{+1.0}$ & $972.3_{-0.6}^{+0.7}$ & $971.9_{-0.8}^{+0.7}$ \\
\hline
25.8      & Grueff   & $R_{c}$              & $981.5_{-1.5}^{+1.1}$ & $982.0_{-1.5}^{+1.3}$ & $975.1_{-1.9}^{+1.5}$ & $975.3_{-2.0}^{+1.6}$ \\
25.8      & Grueff   & $R_{eq}$             & $982.2_{-2.3}^{+1.9}$ & $981.9_{-1.4}^{+1.4}$ & $974.8_{-1.9}^{+1.5}$ & $975.4_{-2.2}^{+1.6}$ \\
25.8      & Grueff   & $R_{pol}$            & $980.9_{-1.3}^{+1.1}$ & $981.8_{-1.5}^{+1.7}$ & $975.0_{-1.3}^{+1.3}$ & $975.3_{-1.7}^{+2.0}$ \\
25.8      & Grueff   & $R_{c}$ (averaged)   & $978.5_{-6.4}^{+8.2}$ & $978.8_{-3.4}^{+5.2}$ & $971.6_{-5.7}^{+7.3}$ & $972.1_{-4.3}^{+3.8}$ \\
25.8      & Grueff   & $R_{eq}$ (averaged)  & $978.5_{-5.2}^{+6.8}$ & $979.0_{-2.8}^{+5.2}$ & $969.6_{-5.1}^{+6.4}$ & $971.8_{-4.6}^{+3.6}$ \\
25.8      & Grueff   & $R_{pol}$ (averaged) & $977.8_{-7.3}^{+6.4}$ & $978.9_{-2.7}^{+5.3}$ & $971.2_{-4.9}^{+6.7}$ & $971.0_{-4.2}^{+4.1}$ \\
\hline
26.1      & Grueff   & $R_{c}$              & $981.2_{-1.6}^{+0.4}$ & $980.9_{-1.4}^{+0.9}$ & $972.8_{-0.6}^{+0.7}$ & $973.7_{-0.6}^{+1.0}$ \\
26.1      & Grueff   & $R_{eq}$             & $981.0_{-2.0}^{+2.0}$ & $981.8_{-2.3}^{+0.5}$ & $973.0_{-1.4}^{+1.6}$ & $974.3_{-1.5}^{+0.5}$ \\
26.1      & Grueff   & $R_{pol}$            & $981.5_{-1.4}^{+1.4}$ & $980.3_{-0.4}^{+0.5}$ & $973.4_{-1.6}^{+0.9}$ & $975.2_{-2.2}^{+0.5}$ \\
\hline
\end{tabular}
\end{table*}

\subsection{The role of the antenna beam pattern on the solar maps}
\label{par:ant_beam}

One of the most important features that influence the measurement of $R_{\odot}$ is the degrading effect of the antenna beam pattern on the solar signal (e.g., \citealp{GimenezDeCastro20,Menezes22}).
The analysis of this effect in our solar images allows us to (1) assess the quality of our radius determinations, and (2) reveal the presence of the coronal emission in our maps.
In the present work we assess the quality of the radius determinations.
The presence of the external coronal emission observed in our solar maps and probing its physical nature (see the tails in the brightness profiles; green line in Fig.~\ref{fig:raggio_example}, top left) is presented in \citet{Marongiu23b}.
This phenomenology arises from the fact that an observed solar map results from the convolution between the beam radiation pattern of the antenna, hereafter named "beam pattern", and the true solar signal, hereafter named "solar signal" (e.g., \citealp{Wilson13}).
The solar signal can be obtained from physical and/or empirical models, both using the solar maps (2D-approach) and the extraction of the $T_B$ profiles as a function of the solar coordinates (1D-approach).
We can see an example of these slices of the solar maps in Fig.~\ref{fig:raggio_example} (green curve, top left).
To analyse the effects of the antenna beam pattern on the solar signal, we developed a specific empirical 2D-model based on the convolution between the beam pattern and the solar signal.
This 2D modelling procedure requires in input the observed solar map and the beam pattern (both at the same observing frequency for the given radio telescope).
In output, it produces the solar signal, whose convolution with the beam pattern results in the observed solar map (the input).
This output gives the empirical properties of the solar signal in terms of the size of the solar disk ($R_{eq}$ and $R_{pol}$), and of the geometrical parameters of the external coronal emission (modelled as a 2D-Gaussian function).

\begin{figure*} 
\centering
{\includegraphics[width=85.5mm]{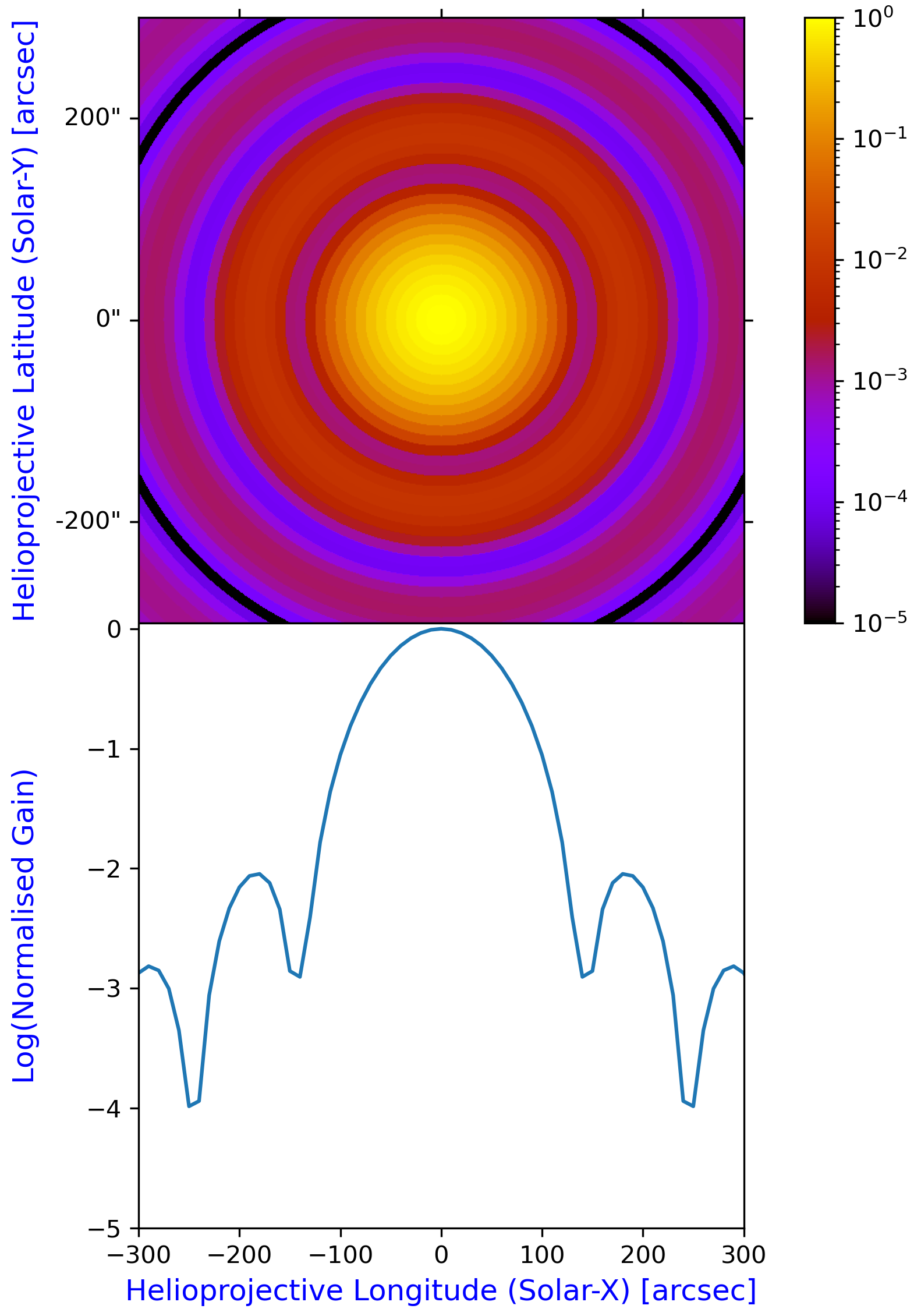}} \quad
{\includegraphics[width=86mm]{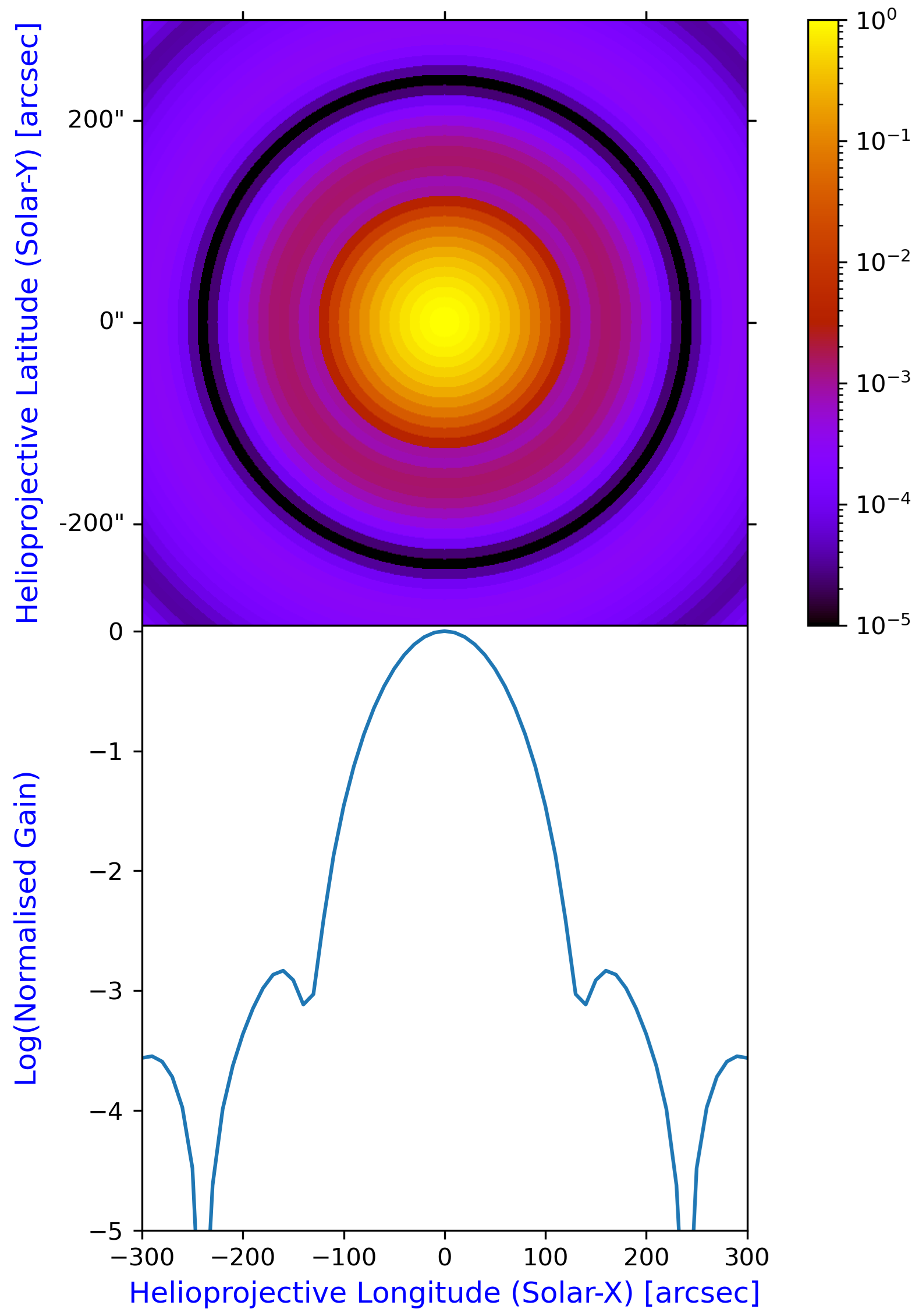}} \\
\caption{
GRASP nominal beam patterns of the Grueff Radio Telescope at $18.3$~GHz (left) and at $25.8$~GHz (right).
(Top) 2D-maps of the beam pattern; (Bottom) 1D-equatorial $T_B$ profiles of the 2D beam pattern.
All the maps are normalised in gain, and shown in logarithmic scale.
}
\label{fig:beam_pattern_med}
\end{figure*}

\begin{figure*} 
\centering
{\includegraphics[width=86mm]{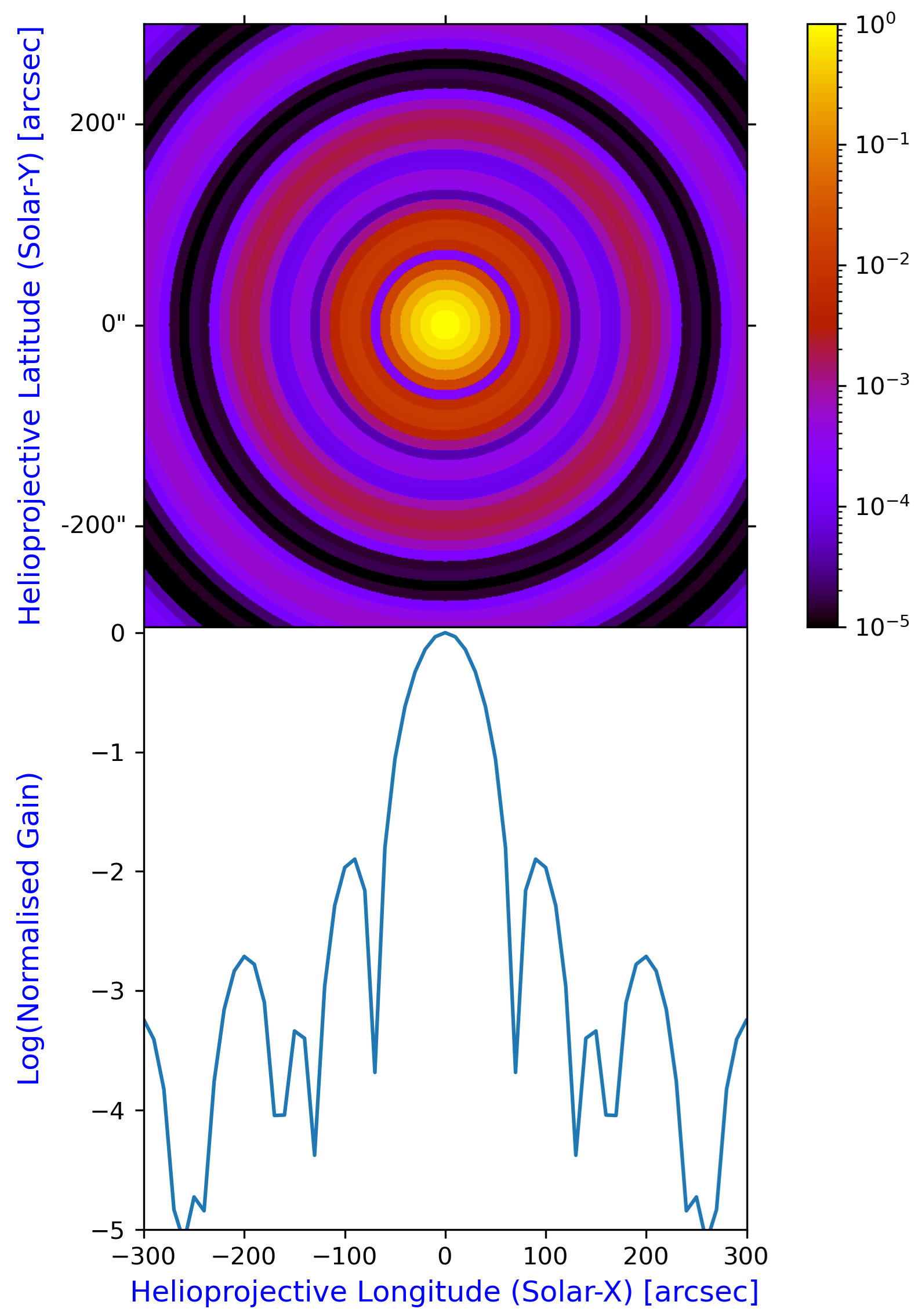}} \quad
{\includegraphics[width=86mm]{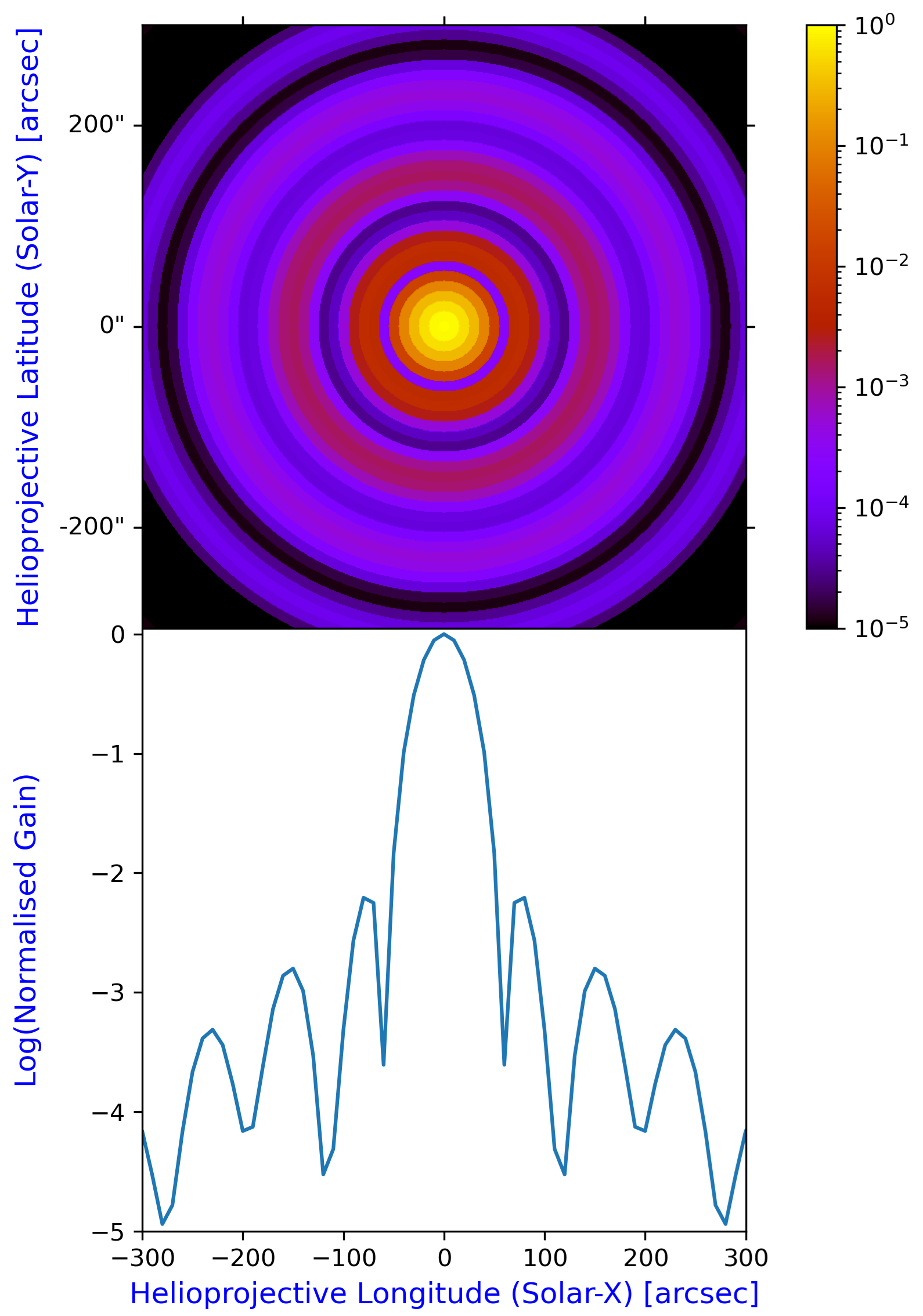}} \\
\caption{
GRASP nominal beam patterns of SRT at $18.8$~GHz (left) and at $24.7$~GHz (right).
(Top) 2D-maps of the beam pattern; (Bottom) 1D-equatorial $T_B$ profiles of the 2D beam pattern.
All the maps are normalised in gain, and shown in logarithmic scale.
}
\label{fig:beam_pattern_srt}
\end{figure*}

The Grueff and SRT beam patterns were reconstructed using GRASP, a dedicated software for electromagnetic analysis of reflector antenna systems integrated into the TICRA Tools software framework\footnote{\url{https://www.ticra.com/software/grasp/}}.
Excluding the measurement noise and the optical residual aberration, the measured beam patterns are in good accordance with the GRASP nominal beam patterns \citep{Prandoni17,Egron22}.
As shown in Figs.~\ref{fig:beam_pattern_med} (for Grueff) and \ref{fig:beam_pattern_srt} (for SRT), we used four 2D beam patterns ($18.3$ and $25.8$~GHz for Grueff; $18.8$ and $24.7$~GHz for SRT).
We selected these frequencies because they were employed in the vast majority of the solar maps.
The single-dish nominal half-power beam width (HPBW) of Grueff and SRT at our observing frequencies ranges between $0.8$ and $2.1$~arcmin (Sect.~\ref{par:obs_datared}).

The parametric model of the solar signal is defined as two normalised empirical 2D-models: the Elliptical-based Cylindrical Box (ECB-model), and the Combination between the ECB-model and a 2D-Gaussian function (2GECB-model).
The \textbf{ECB-model}, tailored only for the solar disk emission of our maps, is defined as:
\begin{equation}
    f_b(x,y,R_{eq},R_{pol}) = \left\{
    \begin{array}{l l}
        1 & {\rm if} \; U_b \leq 1 \\
        0 & {\rm otherwise}
    \end{array} \right.
    \label{eq:f_ellbox}
\end{equation}
where $U_b$ is the elliptical base of the cylindrical box, defined as:
\begin{equation}
    U_b(x,y,R_{eq},R_{pol}) = \left [\frac{(x - x_C)}{R_{eq}} \right]^2 + \left [\frac{(y - y_C)}{R_{pol}} \right]^2
    \label{eq:ellipse_base}
\end{equation}
where $x_C$ and $y_C$ indicate the coordinates of the ellipse center (set to 0, as defined for the solar maps at SRT and Medicina).
We used this model as a first approach to the beam pattern analysis.

\begin{figure} 
\centering
{\includegraphics[width=94mm]{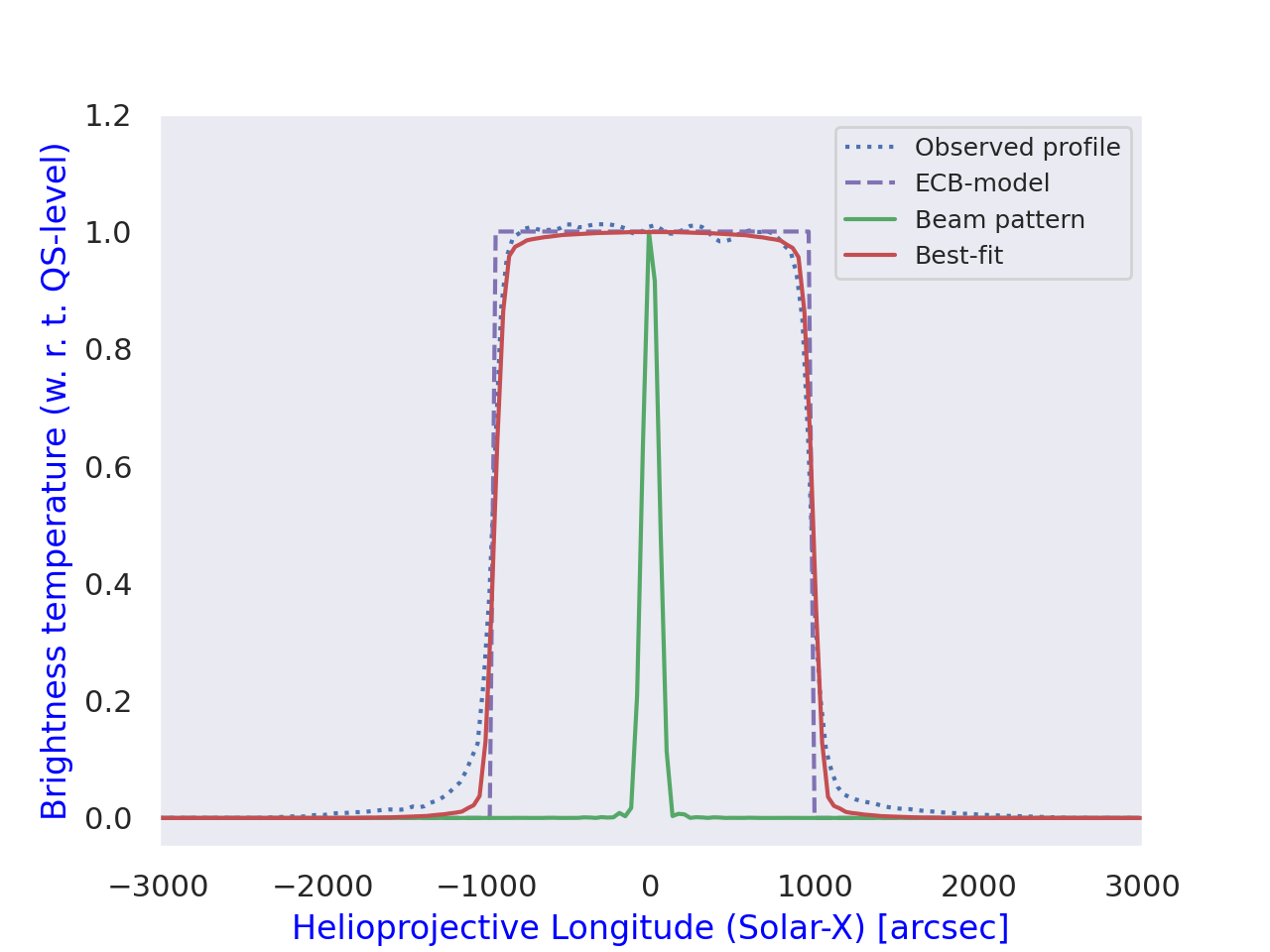}} \\
\caption{
Example of the analysis with the ECB-model using the solar signal in the Grueff Radio Telescope at $18.3$~GHz on September 2$^{nd}$ 2020.
Green solid line indicates the instrumental beam pattern, blue dashed line indicates the observed equatorial $T_B$ profile of the solar map, violet dotted line indicates the modelled equatorial $T_B$ profile of the ECB-model (Eq.~\ref{eq:ellipse_base}), and red solid line indicates the convolved best-fit $T_B$ profile using the instrumental beam pattern and the ECB-model as the solar signal.
}
\label{fig:fit_beam_ecb_simple}
\end{figure}

\begin{figure*} 
\centering
{\includegraphics[width=89.5mm]{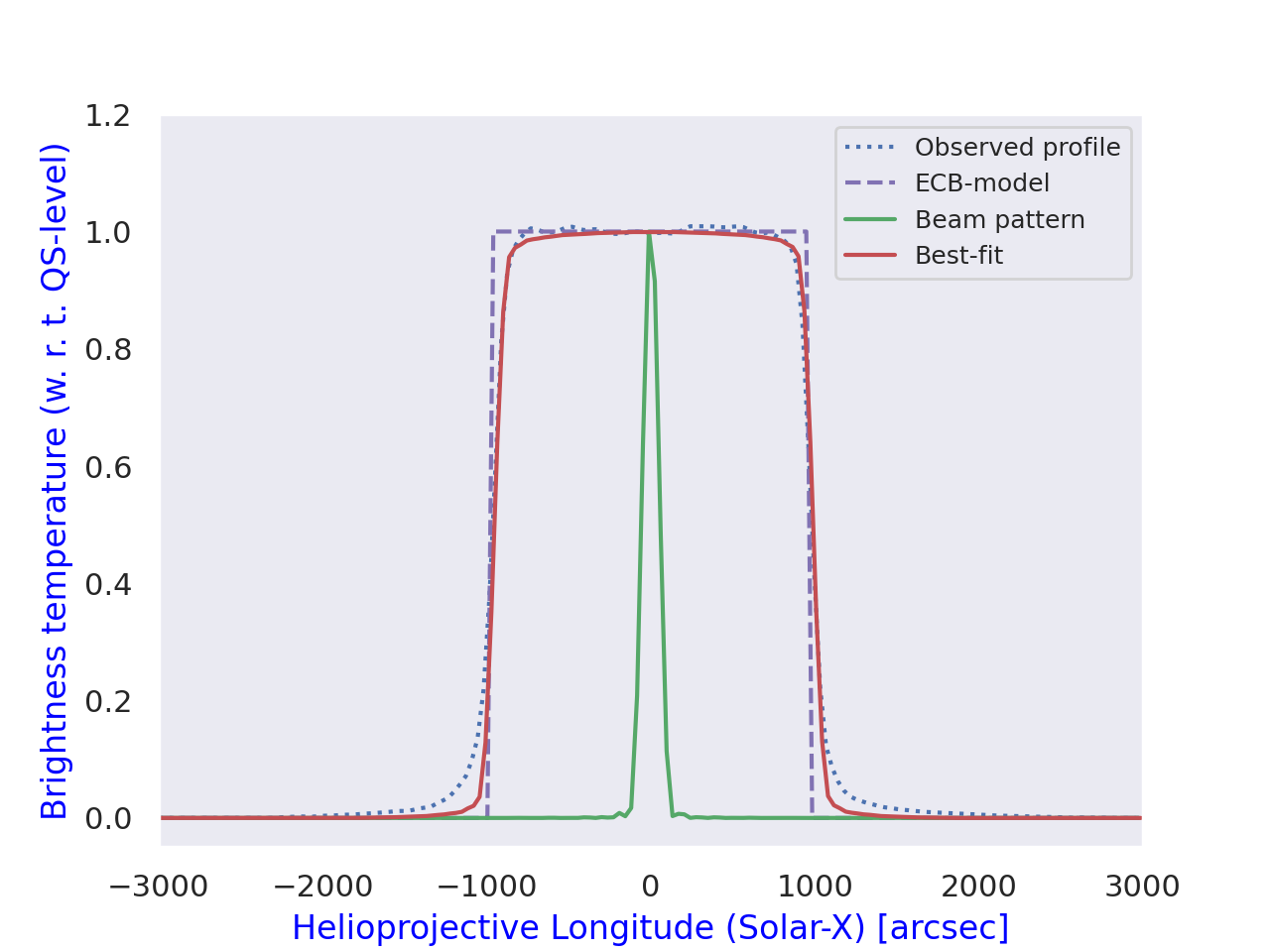}} \quad
{\includegraphics[width=89.5mm]{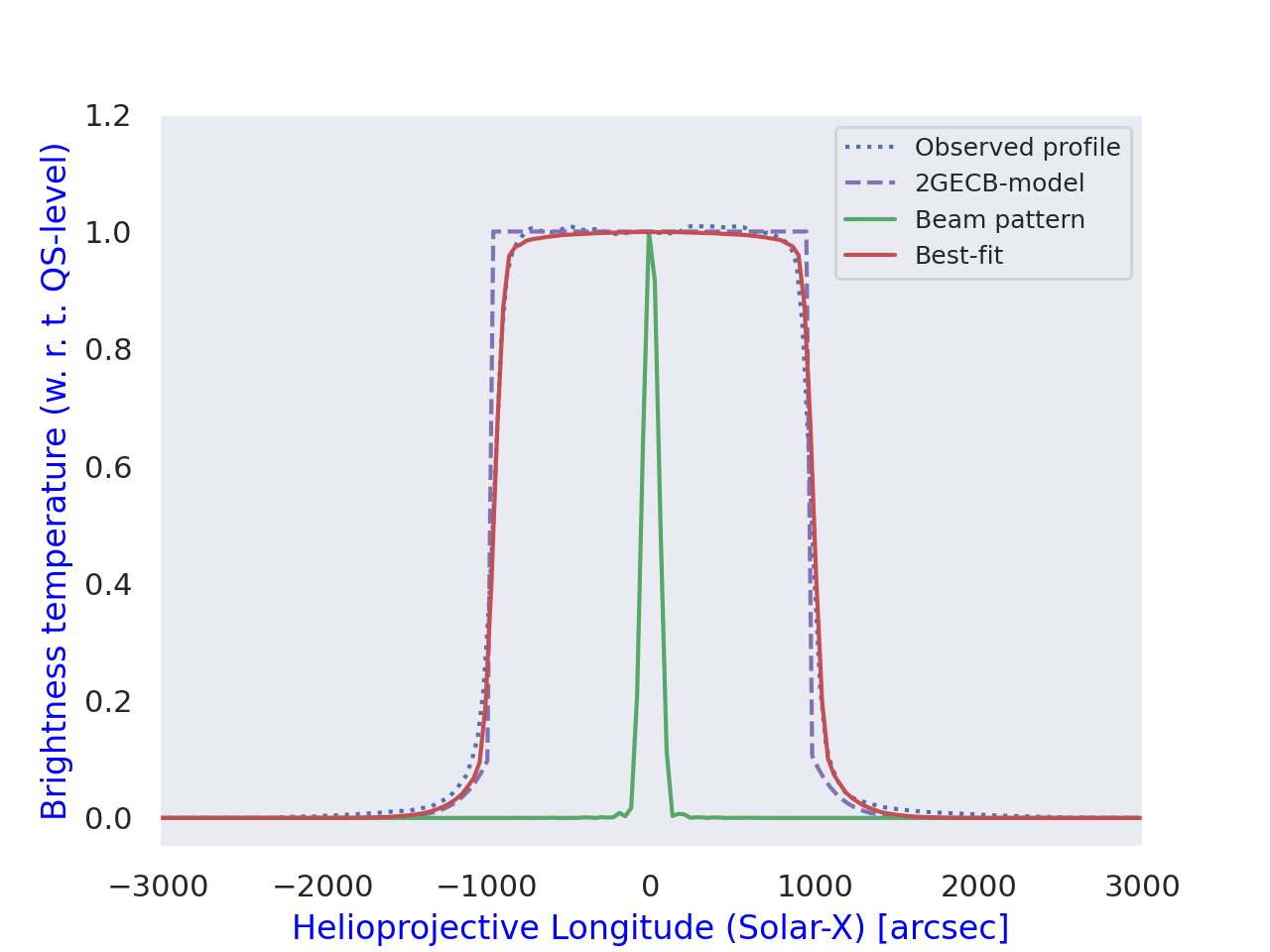}} \\
{\includegraphics[width=89.5mm]{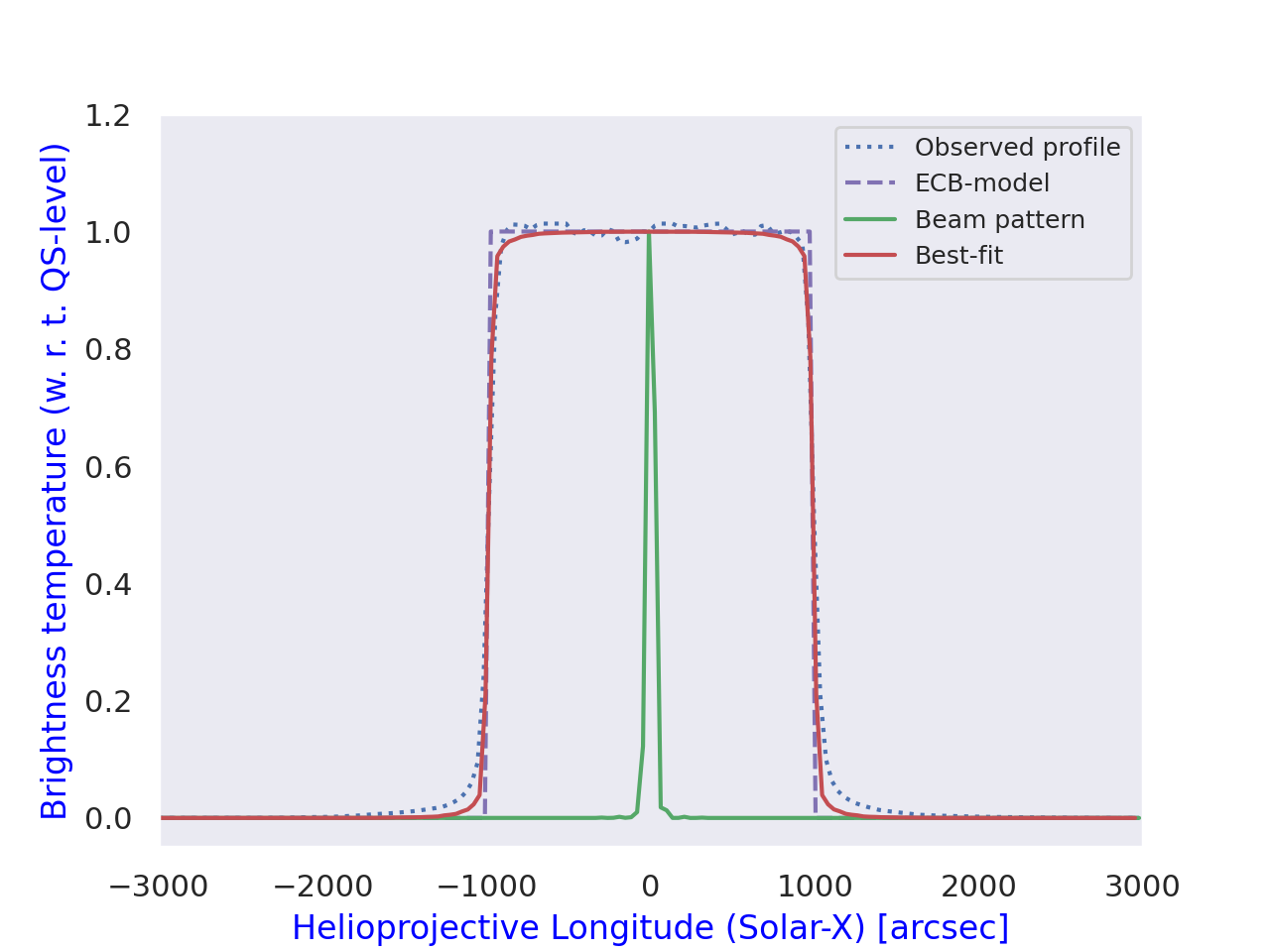}} \quad
{\includegraphics[width=89.5mm]{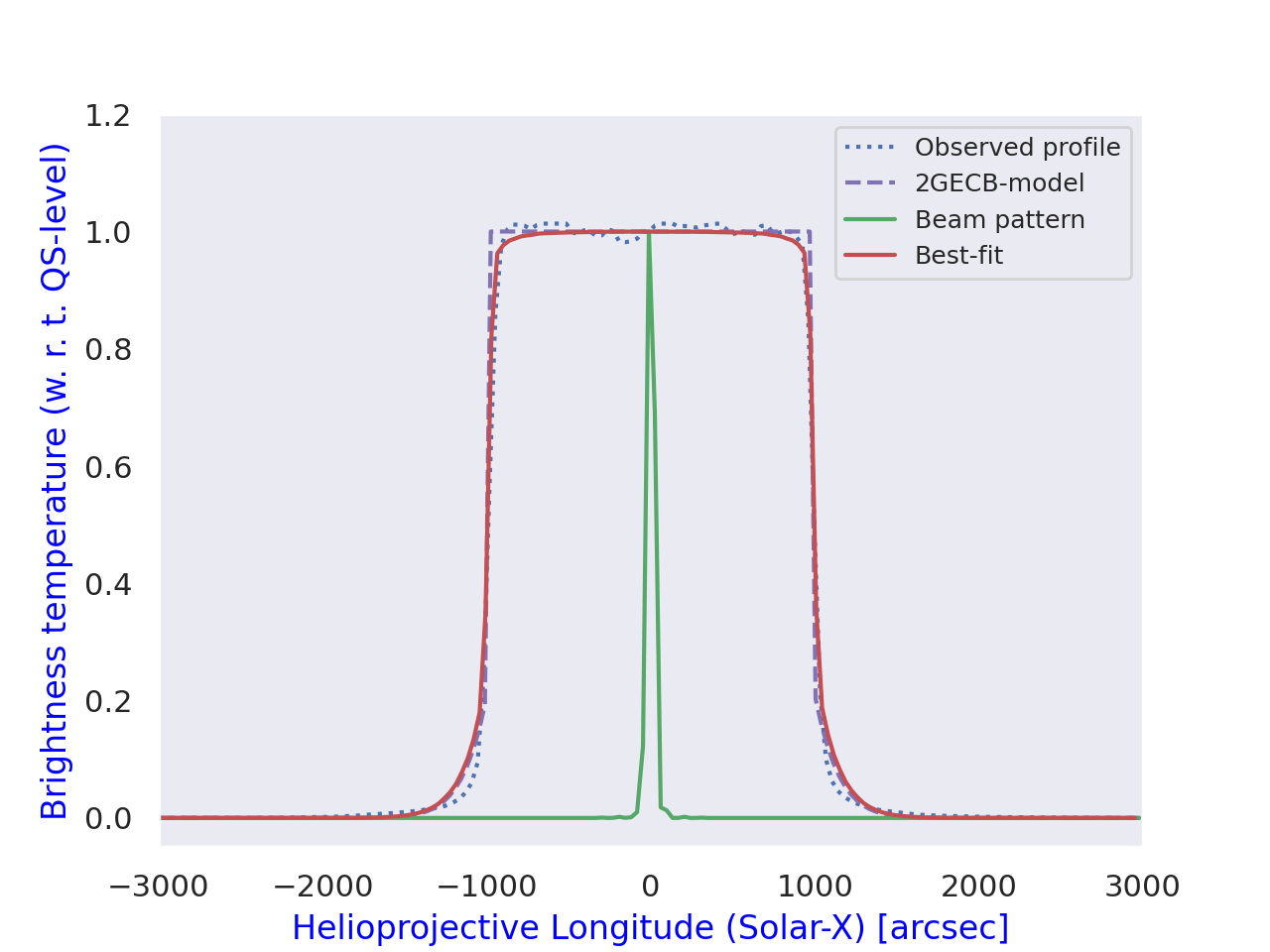}} \\
\caption{
Result of the analysis with the 2D-convolution model, in the context of the equatorial $T_B$ profile, using the ECB-model (Grueff at $18.3$~GHz, top left; SRT at $18.8$~GHz, bottom left) and the 2GECB-model (Grueff at $18.3$~GHz, top right; SRT at $18.8$~GHz, bottom right).
The $T_B$ profiles of the Grueff Radio Telescope are obtained on 6$^{th}$ September 2020; the SRT $T_B$ profiles are obtained on 8$^{th}$ January 2021.
See the caption of Fig.~\ref{fig:fit_beam_ecb_simple} for a full description of the profiles.
}
\label{fig:fit_beam_med}
\end{figure*}

The solar maps resulting from the ECB-model as the solar signal convolved with the beam pattern, showed that our observed solar maps display an external signal which cannot be explained simply by invoking systematic effects such the smoothing produced by the beam pattern on the solar signal.
This aspect is clearly shown in Fig.~\ref{fig:fit_beam_ecb_simple}, where the observed limb level of the solar disk (blue dashed line) is higher than the modelled limb level obtained through the ECB-model (red solid line).
We will focus on this aspect in a dedicated paper \citep{Marongiu23b}.
This incompatibility has led us to consider the \textbf{2GECB-model}, tailored both for the solar disk and the external coronal emission (Fig.~\ref{fig:fit_beam_med}).
This model is defined as:
\begin{equation}
    f_{bG}(x,y,R_{eq},R_{pol},R_{eq,G},R_{pol,G},A_G) = \left\{
    \begin{array}{l l}
        1 & {\rm if} \;  S > 1 \\
        S & {\rm otherwise}
    \end{array} \right.
    \label{eq:f_boxgauss}
\end{equation}
where $R_{eq,G}$ and $R_{pol,G}$ indicate the $1 \sigma$ standard deviation among the equatorial and polar semi-axis of the 2D-Gaussian function, respectively, and $A_G$ is the amplitude of the 2D-Gaussian function.
$S$ indicates the sum $U_b + U_G$, where $U_b(x,y,R_{eq,G},R_{pol,G})$ is defined by Eq.~\ref{eq:ellipse_base}, and $U_G$ indicates the ellipse-shaped 2D-Gaussian function, defined as:
\begin{equation}
    U_G(x,y,R_{eq,G},R_{pol,G},A_G) = \left\{
    \begin{array}{l l}
        1 & {\rm if} \; H > 1 \\
        H & {\rm otherwise}
    \end{array} \right.
    \label{eq:f_gauss}
\end{equation}
where $H = A_G e^{-U_b/2}$.
This model is designed considering (1) both the 2D-Gaussian and the ECB-model centered at $x_C = y_C = 0$, and (2) the 2D-Gaussian function defined only outside the ECB-model region.
As we describe in Sect.~\ref{par:results} and discuss in Sect.~\ref{par:discussione_raggio}, the observed solar maps are well fitted with the 2D solar maps obtained through the 2D-model, using our beam pattern and the 2GECB-model as the solar signal.
\begin{figure*} 
\centering
{\includegraphics[width=89.5mm]{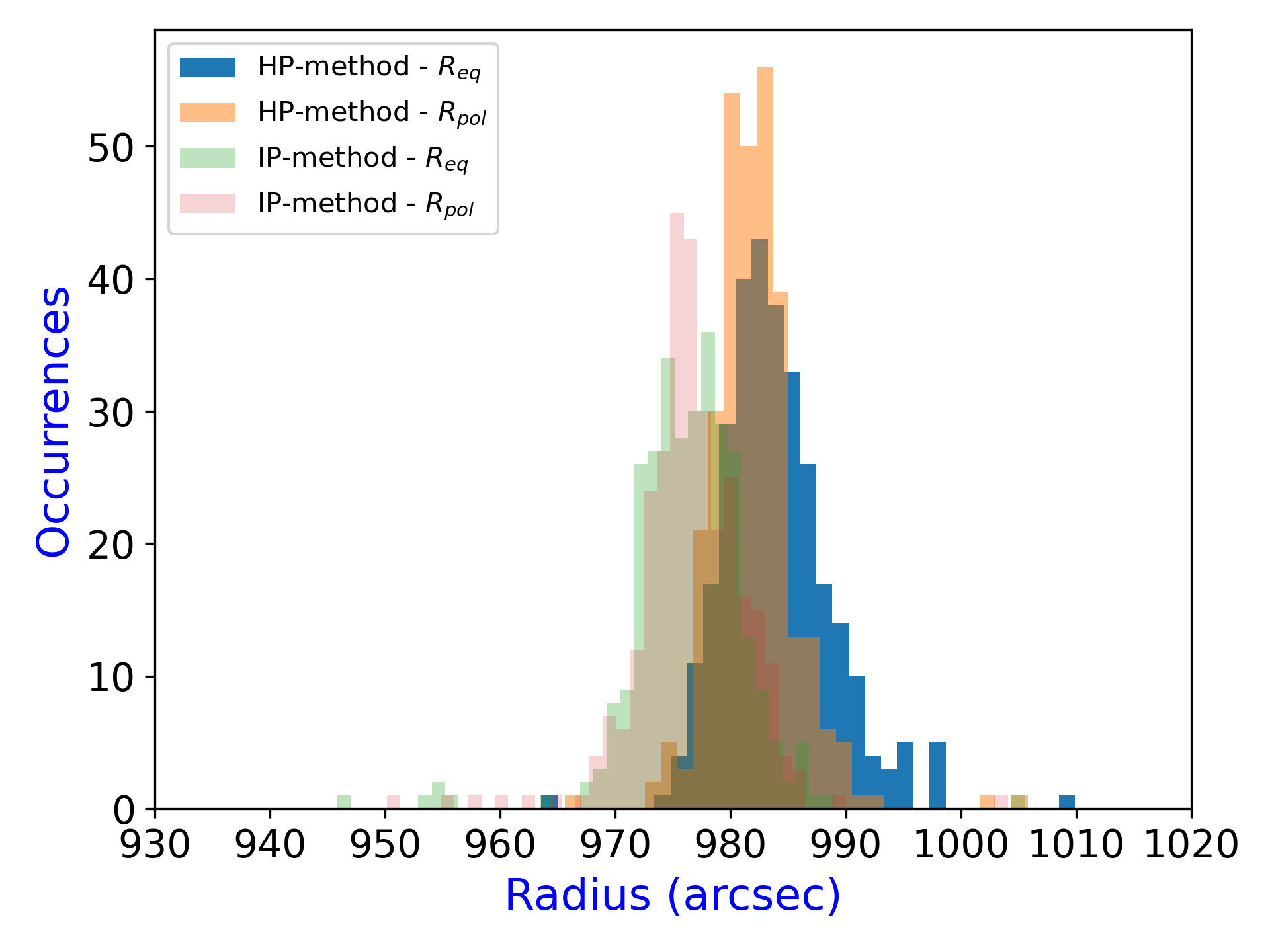}} \quad
{\includegraphics[width=89.5mm]{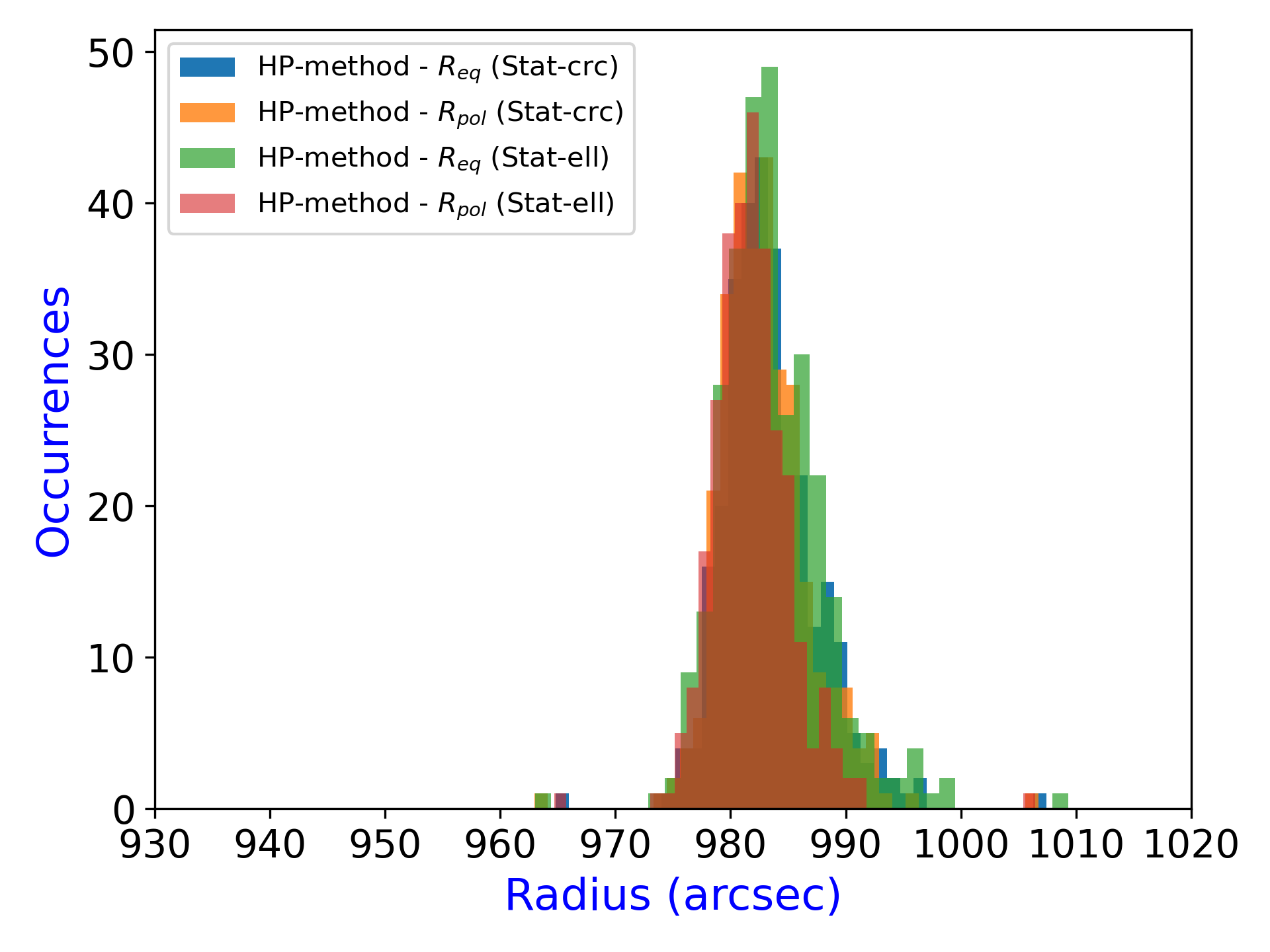}} \\
{\includegraphics[width=89.5mm]{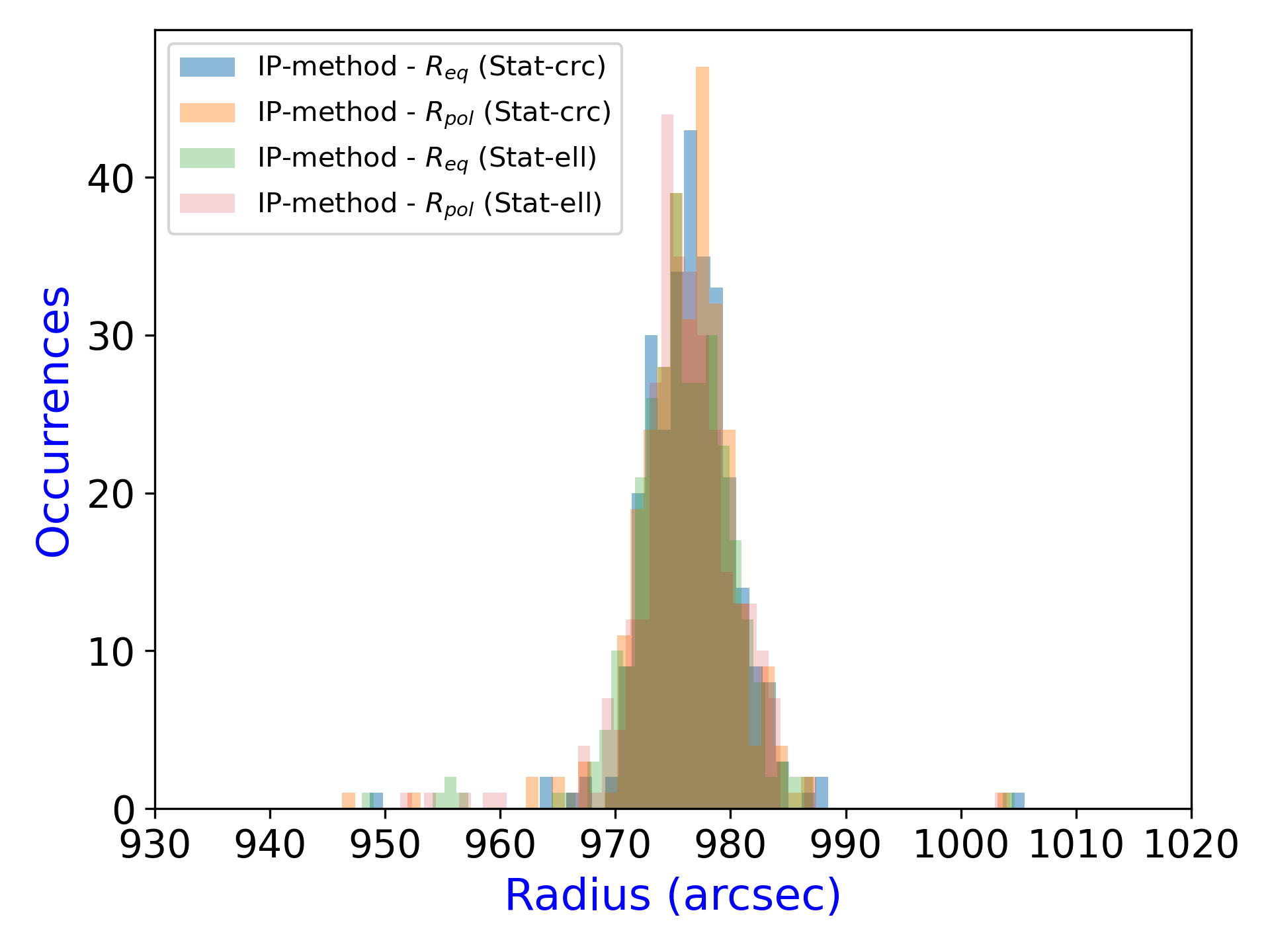}} \quad
{\includegraphics[width=89.5mm]{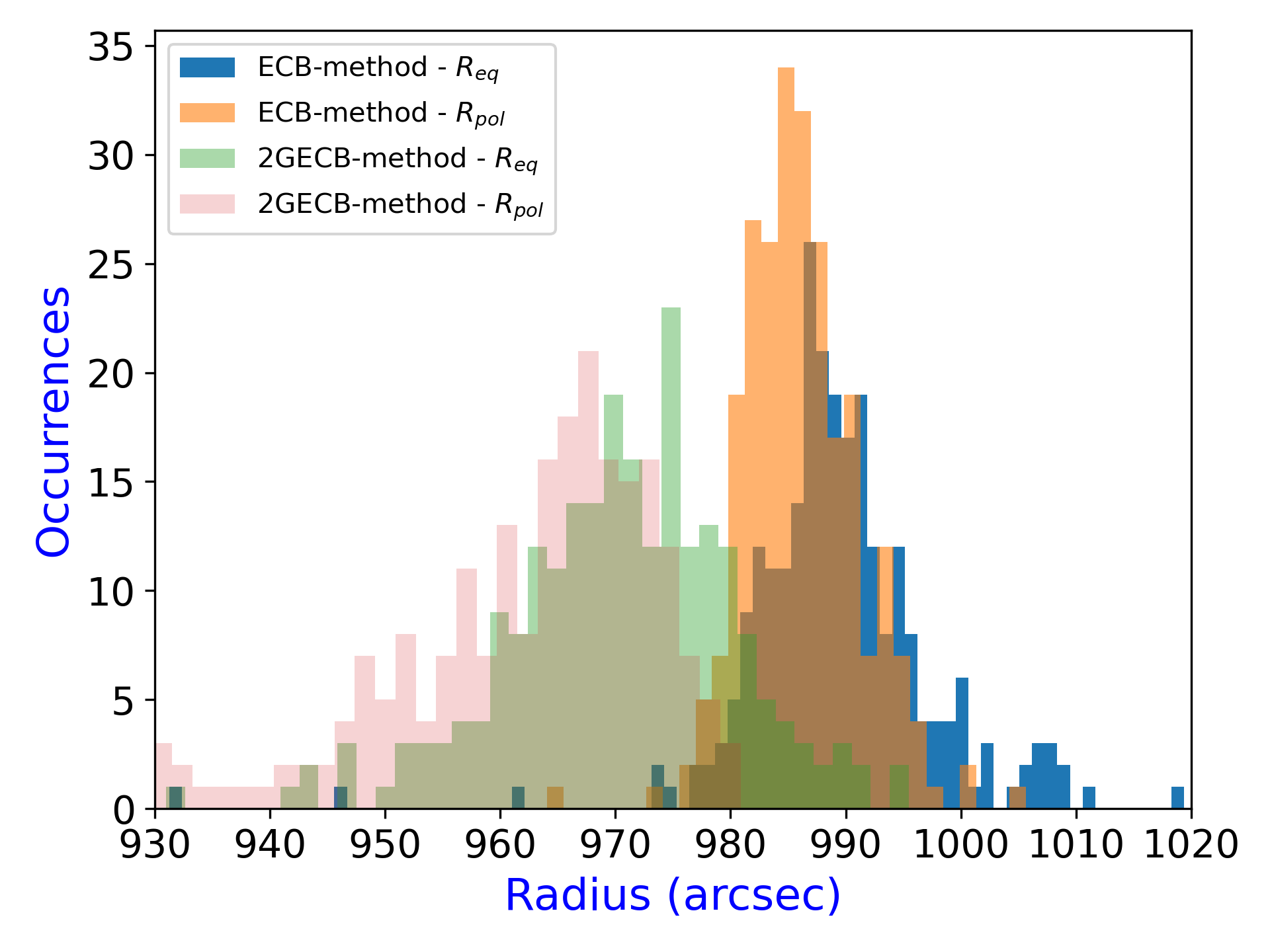}} \\
\caption{
Histograms depicting all the $R_{eq}$ and $R_{pol}$ measurements calculated from (top left) the HP- and IP-methods, (top right) the statistical approach using the HP-method, (bottom left) the statistical approach using the IP-method, and (bottom right) the ECB- and 2GECB-methods.
}
\label{fig:hist_resume}
\end{figure*}

These 2D-models were developed thanks to a Bayesian approach based on Markov Chain MonteCarlo (MCMC) simulations (e.g. \citealp{Sharma17}).
For this approach we used the Python {\fontfamily{pcr}\selectfont emcee} package\footnote{\url{https://emcee.readthedocs.io/en/stable/}} \citep{Foreman13}.
The model parameters, analysed with {\fontfamily{pcr}\selectfont emcee} to flush out degeneracies, are constrained through the definition of prior distributions (uniform in this work) that encode preliminary and general information.
The complete uniform prior distribution adopted in our analysis is listed in Table \ref{tab:priors}.
The parameters labelled with 0 indicate the initial best-fit parameters, calculated through the maximisation of the likelihood function, using the sequential least squares programming tools available in the Python {\fontfamily{pcr}\selectfont SciPy} package\footnote{\url{http://www.scipy.org/}} \citep{Jones01}.
\begin{table*}
\caption{Uniform prior distribution adopted for the Bayesian approach in the 2D-models of our analysis.
}
\label{tab:priors}
\centering
\begin{tabular}{l|ccc}
\hline
\hline
Model          & Parameter    & Range                                                 & Unit    \\
\hline
ECB-model      & $R_{eq}$     & $R_{eq,0} - 100 \leq R_{eq,0} \leq R_{eq,0} + 100$    & arcsec  \\
ECB-model      & $R_{pol}$    & $R_{pol,0} - 100 \leq R_{pol,0} \leq R_{pol,0} + 100$ & arcsec  \\
\hline
2GECB-model    & $R_{eq}$     & $R_{eq,0} - 100 \leq R_{eq,0} \leq R_{eq,0} + 100$    & arcsec  \\
2GECB-model    & $R_{pol}$    & $R_{pol,0} - 100 \leq R_{pol,0} \leq R_{pol,0} + 100$ & arcsec  \\
2GECB-model    & $R_{eq,G}$   & $200 \leq R_{eq,G,0} \leq 1200$                       & arcsec  \\
2GECB-model    & $R_{pol,G}$  & $200 \leq R_{pol,G,0} \leq 1200$                      & arcsec  \\
2GECB-model    & $A_G$        & $0.2 \leq A_{G,0} \leq 10$                            & -       \\
\hline
\end{tabular}
\end{table*}
In the MCMC analysis, the beginning of the ensemble sampler is characterised by an initial period -- called “burn-in”, discarded by the analysis -- where the convergence of the average likelihood across the chains is unstable (recommended chains: $300$).
The number of subsequent Markov chains are set up in $1000$ steps, with a number of $20$ walkers.
All the uncertainties are reported at 68\% ($1 \sigma$).

\section{Results}
\label{par:results}

We used $304$ solar maps ($287$ with Grueff and $17$ with SRT) to calculate the radii, and the correlation between the radii and the solar activity (and between $R_{eq}$ and $R_{pol}$).
We observed at seven central frequencies with Grueff and SRT: $18.1$, $18.3$, $18.8$, $23.6$, $24.7$, $25.8$, and $26.1$ GHz.
Among these frequencies, for our analysis we selected four frequencies, characterised by a uniform time coverage from 2018 to date: $18.3$ and $25.8$ GHz for Grueff, $18.8$ and $24.7$ GHz for SRT.
Moreover, for this analysis we also used two averaged solar maps at $18.3$ and $25.8$~GHz acquired in Medicina during the minimum solar activity ($2018$--$2020$).
Our data set is composed of solar maps normalised at 1 AU, and we considered only the medium/high-quality maps to avoid systematic and/or meteorological undesired effects.
These averaged maps are obtained only from the Grueff Radio Telescope, since the data set obtained with SRT does not uniformly cover our temporal range of observations between $2018$ and mid-$2023$.
The discussion of these results is treated in Sect. \ref{par:disc_concl}.
In this analysis the images were not corrected for center-to-limb variation (CLV) to enhance the visibility of the disk features and show prominences at the same time (e.g., \citealp{Alissandrakis17}).

Our results on the calculation of the median values of $R_{c}$, $R_{eq}$, and $R_{pol}$ are summarised in Tables~\ref{tab:raggio_resume} and \ref{tab:raggio_resume2} -- where the radii are listed by observing frequency, type, and method of calculation -- and in Fig.~\ref{fig:hist_resume}.
Following the methods described above we obtained values of $R_{\odot}$ for the various observing frequencies.
Such values range from $959$~arcsec (obtained with the 2GECB-method) and $994$~arcsec (obtained with the ECB-method), with typical errors of a few arcsec.

\begin{figure*} 
\centering
{\includegraphics[width=89mm]{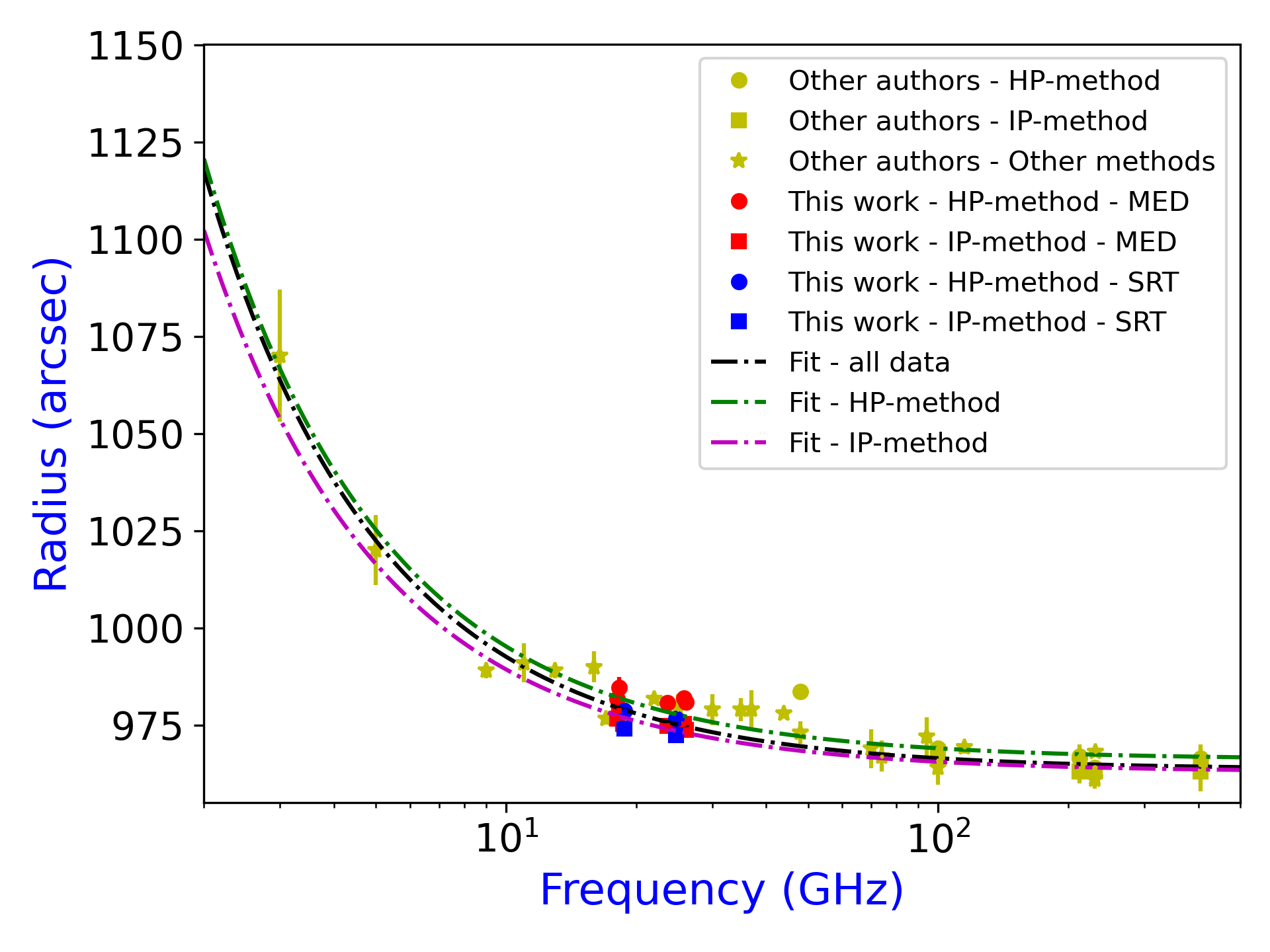}} \quad
{\includegraphics[width=89mm]{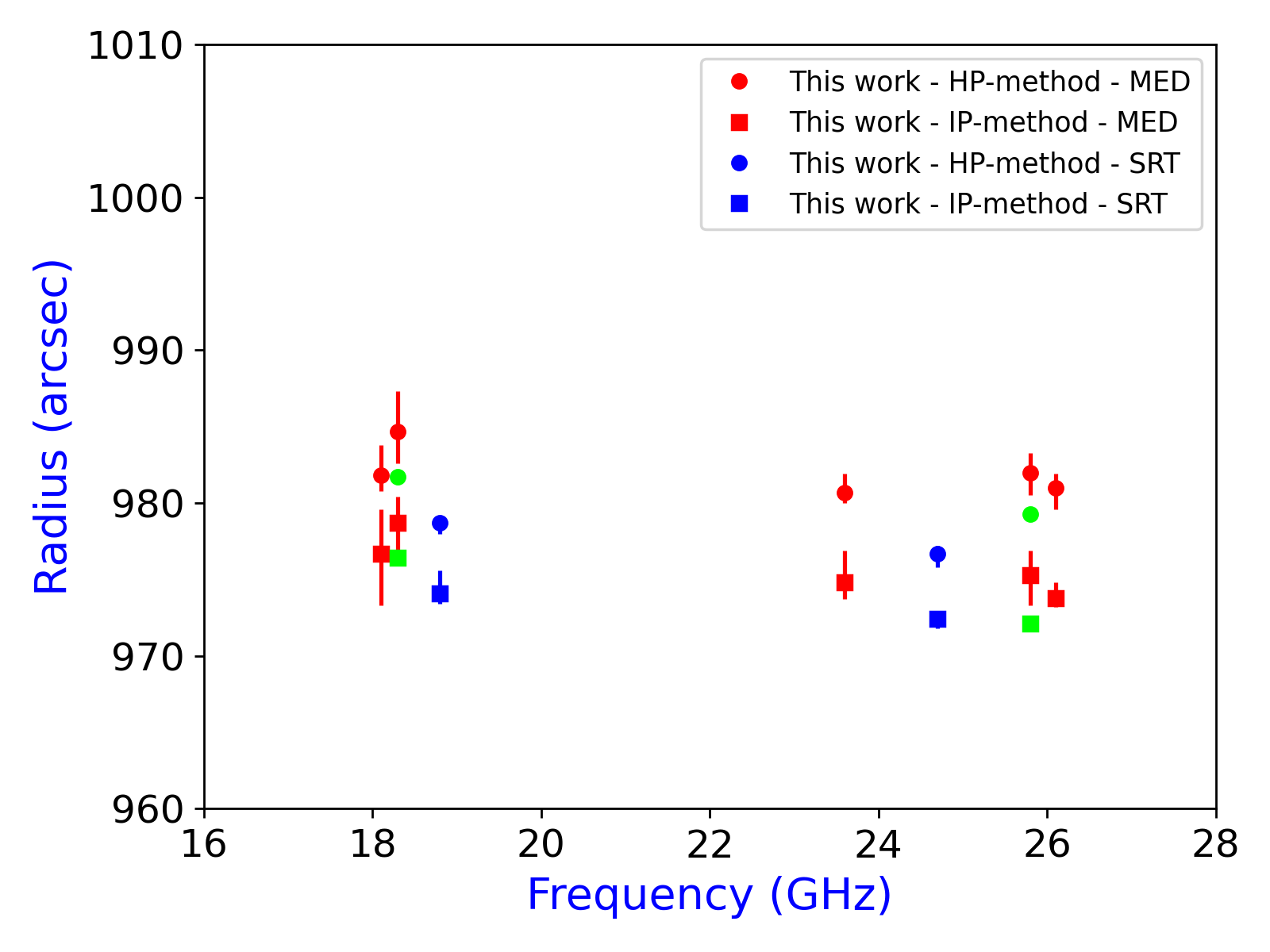}} \\
\caption{
Left: Solar radius as a function of frequency, with our measures of $R_{c}$ obtained through the modelling; the dashed lines represent the fitted exponential trend (black: all data; green: HP-method; magenta: IP-method); the yellow points are previous measurements from other authors (see Table 1 in \citealp{Menezes17,Menezes22}, and references therein), the circle and squared points are the present radius values derived from Grueff and SRT.
Right: Plot with our measures of $R_{c}$ obtained through the HP-method (circle points) and the IP-method (squared points).
The green points indicate the measures of $R_{c}$ obtained from the averaged solar maps of the Grueff Radio Telescope through the HP-method (circle points) and the IP-method (squared points).
}
\label{fig:radius_nu}
\end{figure*}

\begin{figure*} 
\centering
{\includegraphics[width=89mm]{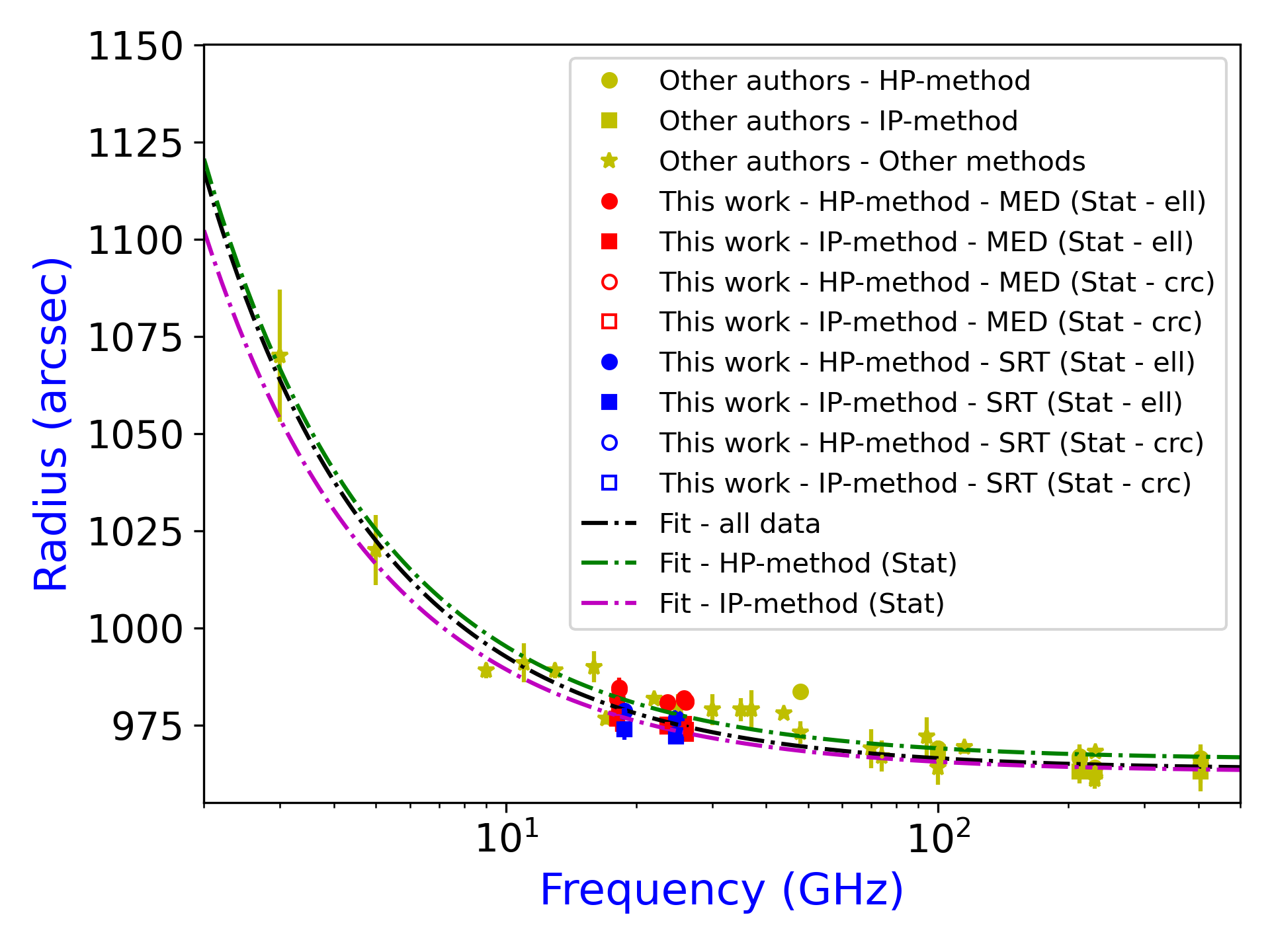}} \quad
{\includegraphics[width=89mm]{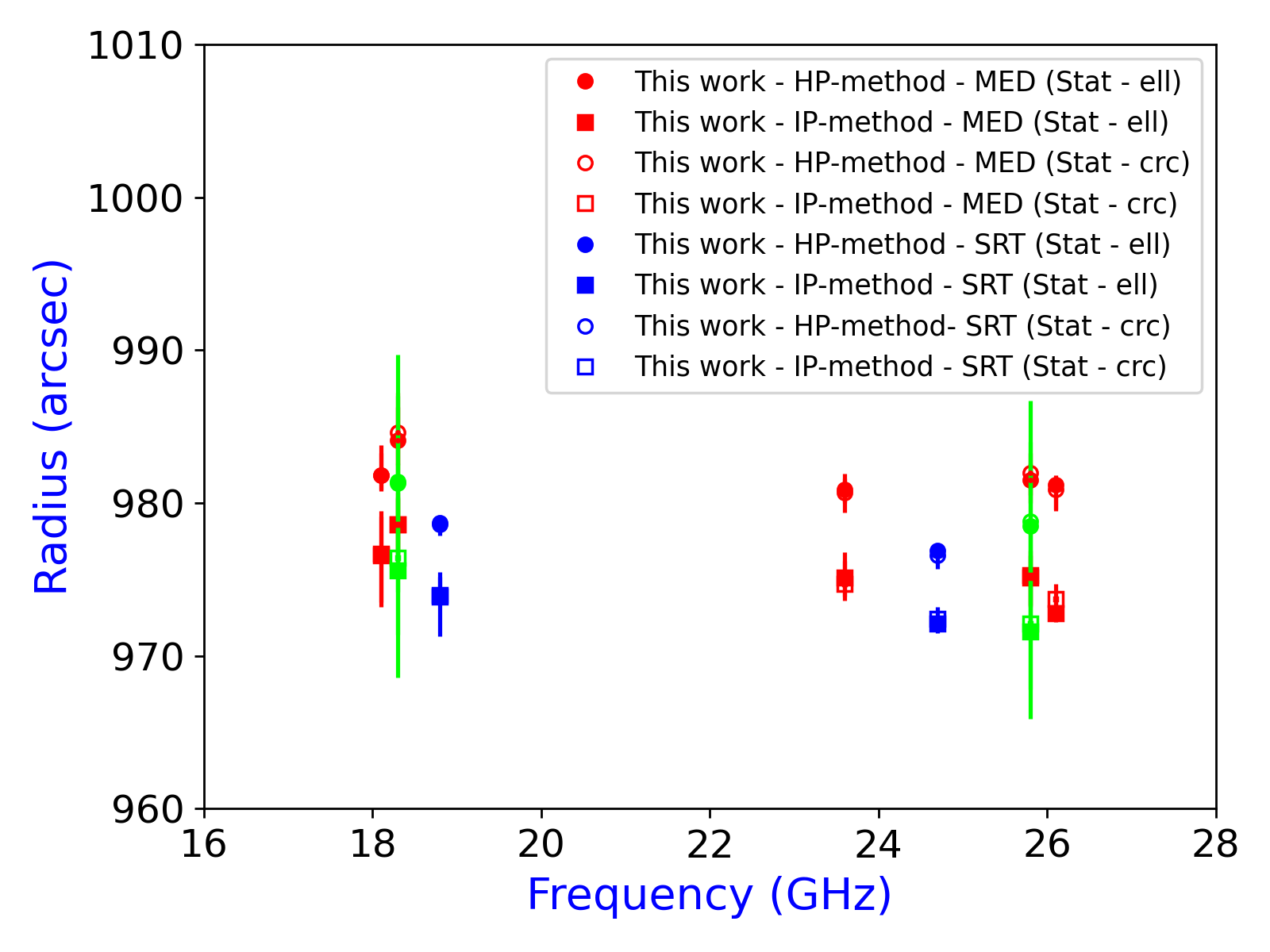}} \\
\caption{
Left: Solar radius as a function of frequency, with our measures of $R_{c}$ obtained through the statistical approach; see the caption of Figure~\ref{fig:radius_nu} for a full description of the colorbars and the symbols.
Right: Plot with our measures of $R_{c}$ obtained through the statistical approach.
}
\label{fig:radius_nu_stat}
\end{figure*}

To check for consistency, our results of $R_{c}$ are plotted in Figs.~\ref{fig:radius_nu} (modelling approach) and \ref{fig:radius_nu_stat} (statistical approach) with those from other authors at several radio frequencies and different measurement techniques (see \citealp{Menezes22}, and Table 1).
To guide the eye, an exponential curve (dashed line) is over-plotted to show the trend of the radius as a function of the observing frequency, indicating that the radius decreases exponentially at high radio frequencies.
Note that the trend curve is just a least-squares exponential fit, not a physical model.
Our results are shown by red (Grueff), green (averaged maps of Grueff), and blue points (SRT).
These solar radii seem to agree within the uncertainties with the trend, both for the modelling approach and the statistical approach (Figs.~\ref{fig:radius_nu} and \ref{fig:radius_nu_stat}).
The $R_{\odot}$ results (Tables \ref{tab:raggio_resume} and \ref{tab:raggio_resume2}) indicated that the HP-method yields larger radius values when compared to the IP-method, as already suggested by \citet{Menezes22} analysing the ALMA and SST data.
On average, the values derived from Grueff and SRT maps with the HP-method are $\sim 5$~arcsec larger than those derived with the IP-method.
We note that the results of the average $R_{\odot}$ measurements are comparable within $1\sigma$ error both for the modelling approach and the statistical approach.

Another way to check these radii was to compare them with the 2D-model, considering two different solar signal (ECB-model and 2GECB-model) and the beam patterns of the Grueff Radio Telescope and SRT (Sect.~\ref{par:ant_beam}).
Two examples of the application of the 2D-model are shown in Fig.~\ref{fig:fit_beam_med}.
In general, the radii obtained with the 2D-convolution model using the 2GECB-model are at least (1) $7$~arcsec smaller than those derived with the HP-method, and (2) $2$~arcsec smaller than those derived with the IP-method.
A specific comparison with other values obtained in the literature is described in Sect.~\ref{par:disc_concl}.
\begin{figure*} 
\centering
{\includegraphics[width=75mm]{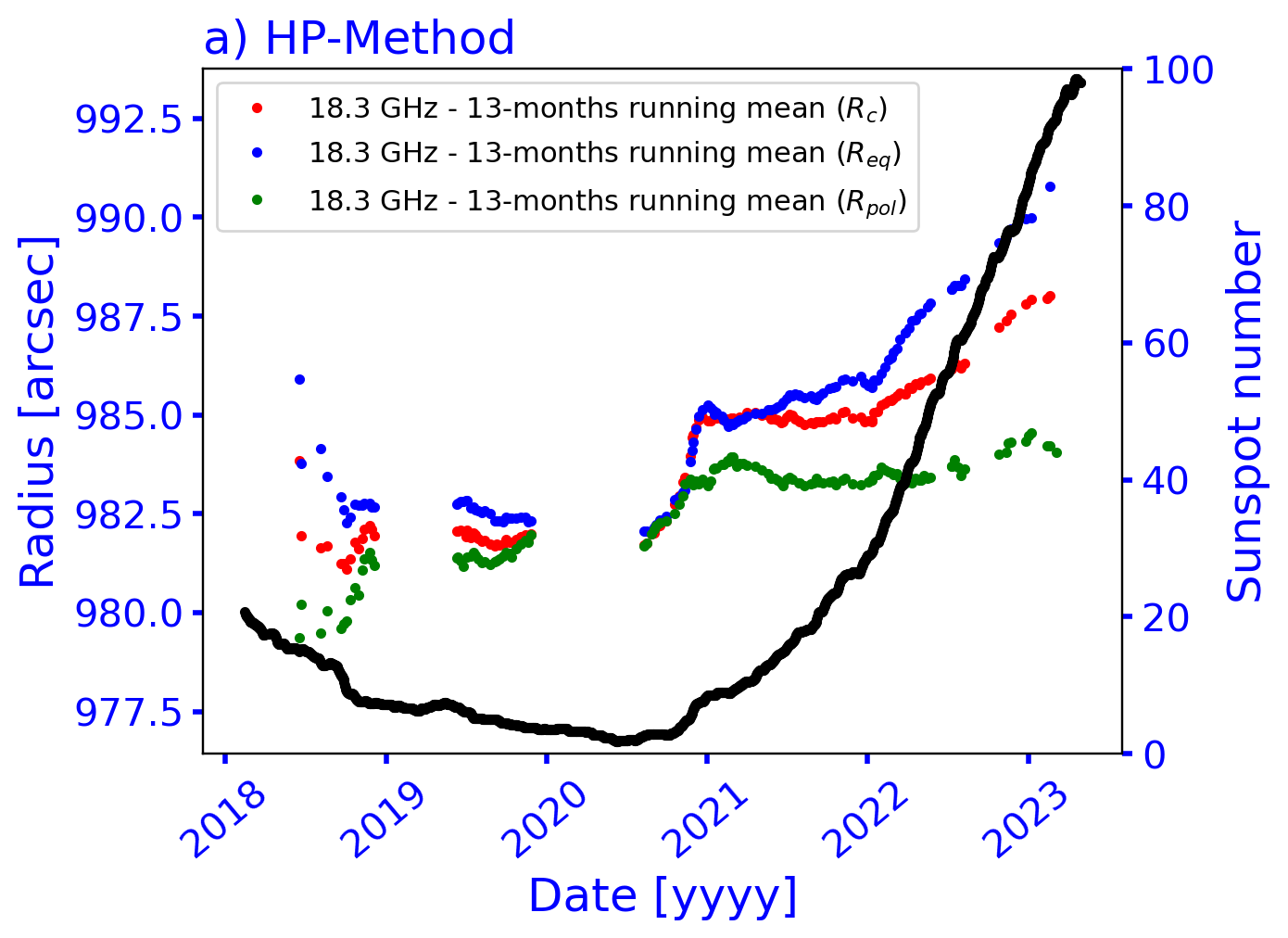}} \quad
{\includegraphics[width=75mm]{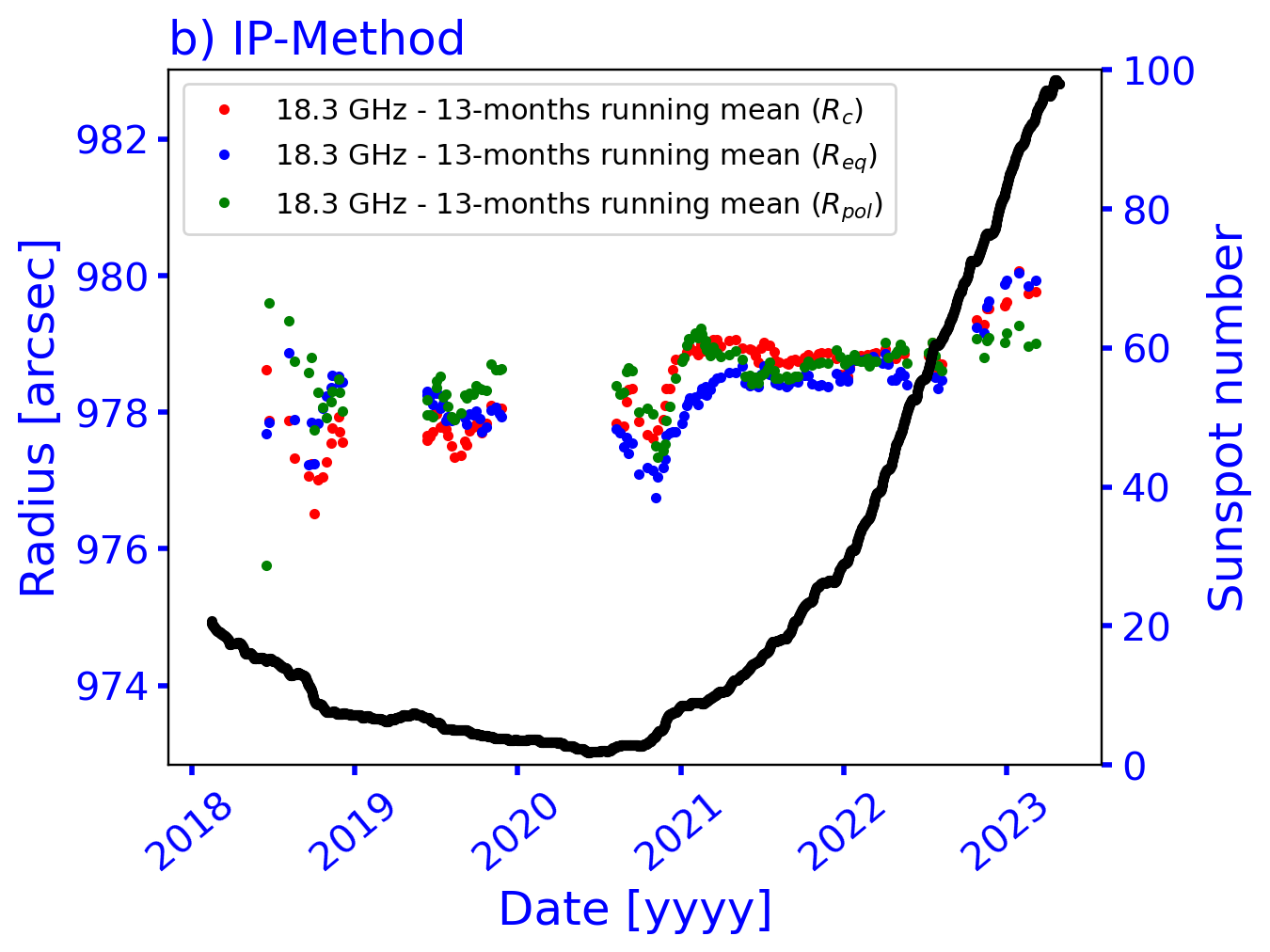}} \\
{\includegraphics[width=75mm]{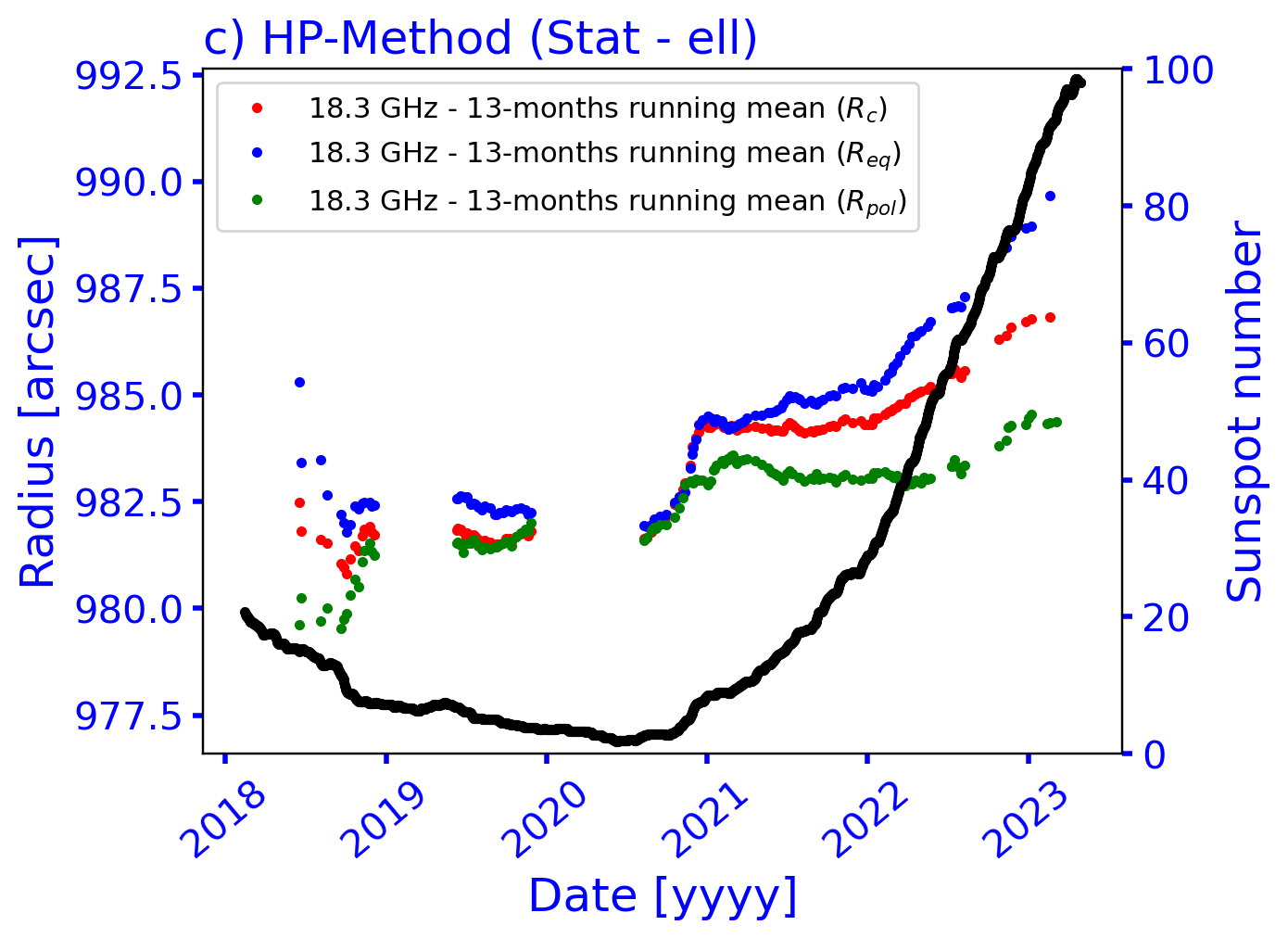}} \quad
{\includegraphics[width=75mm]{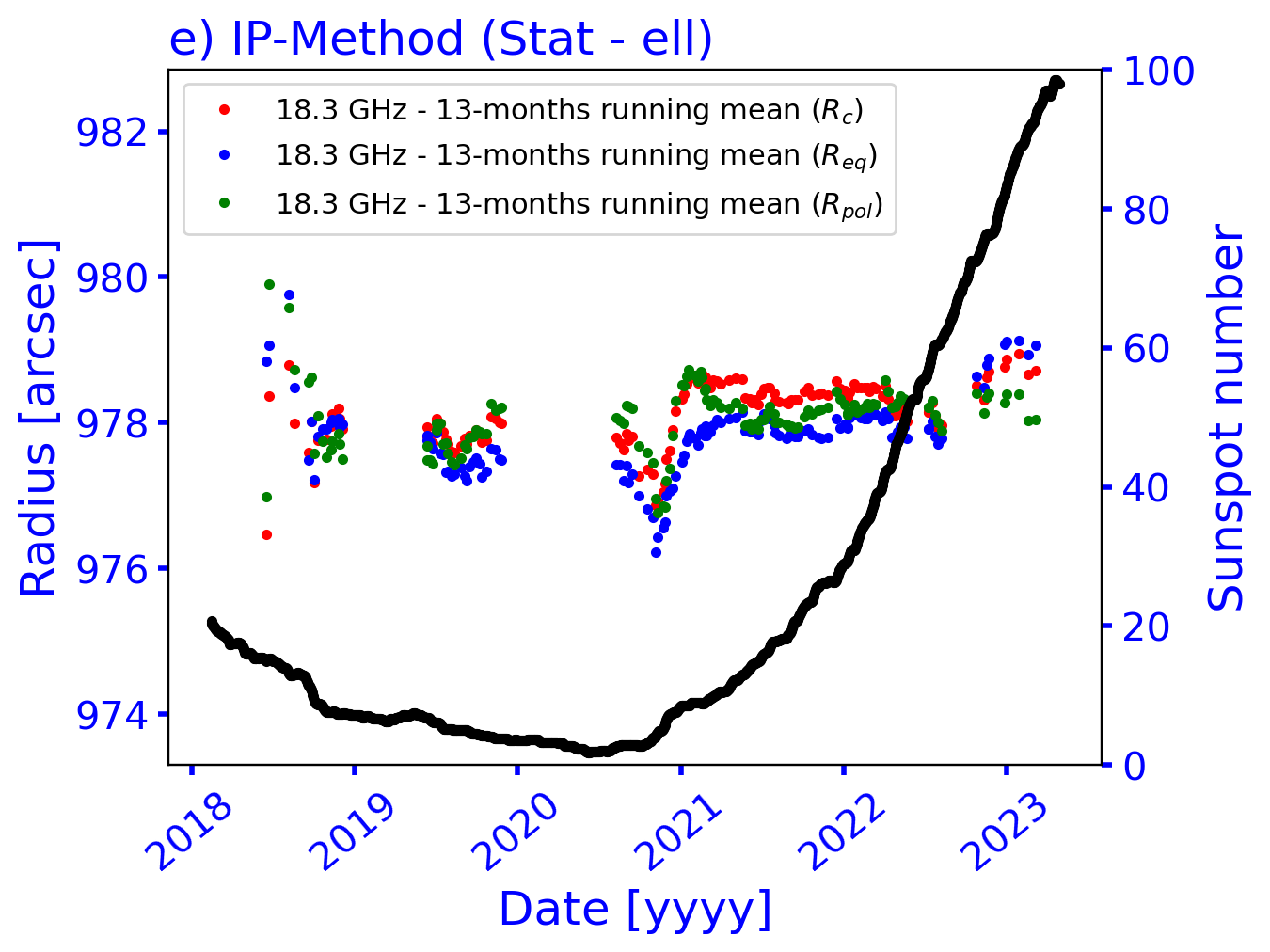}} \\
{\includegraphics[width=75mm]{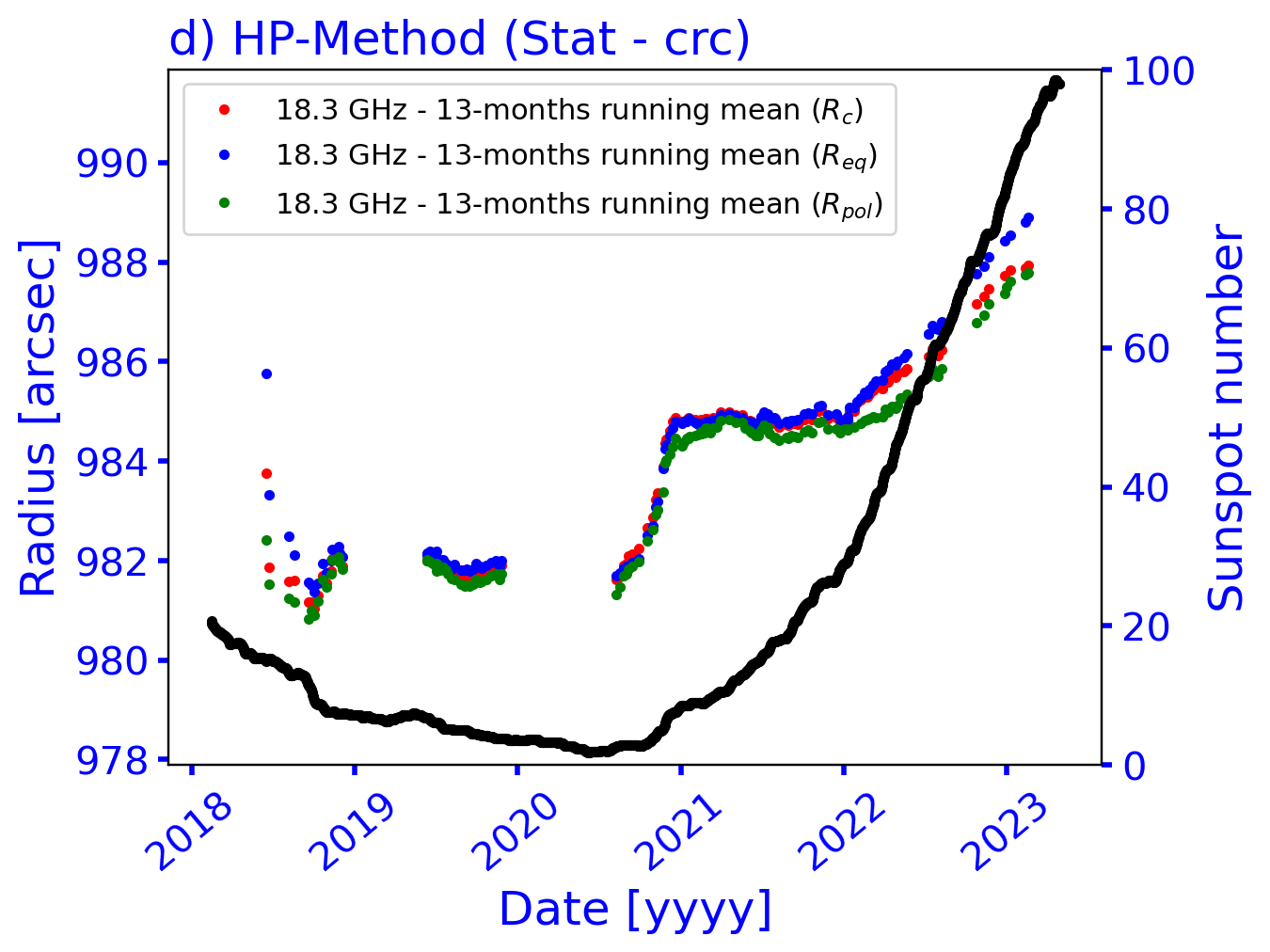}} \quad
{\includegraphics[width=75mm]{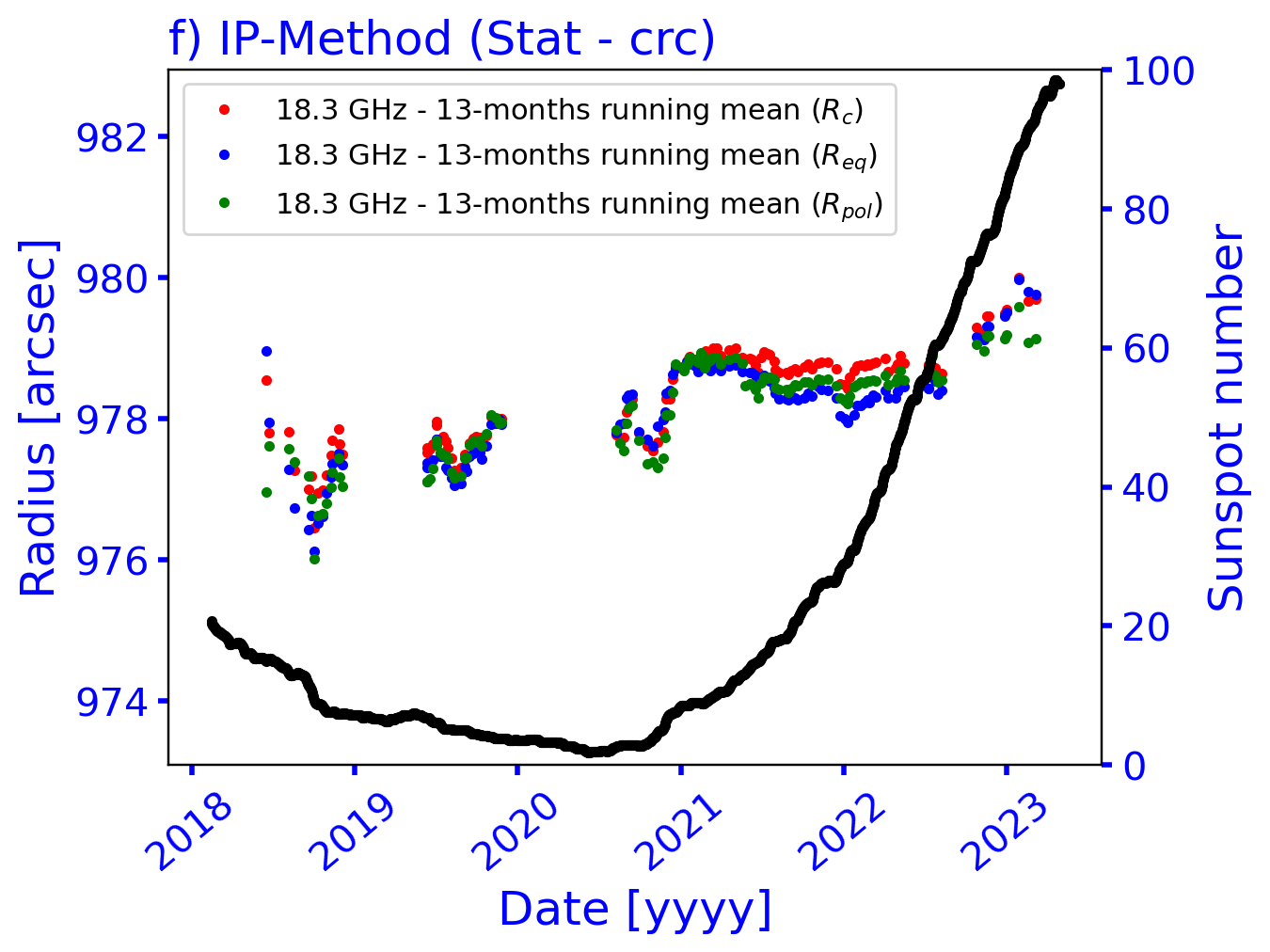}} \\
{\includegraphics[width=75mm]{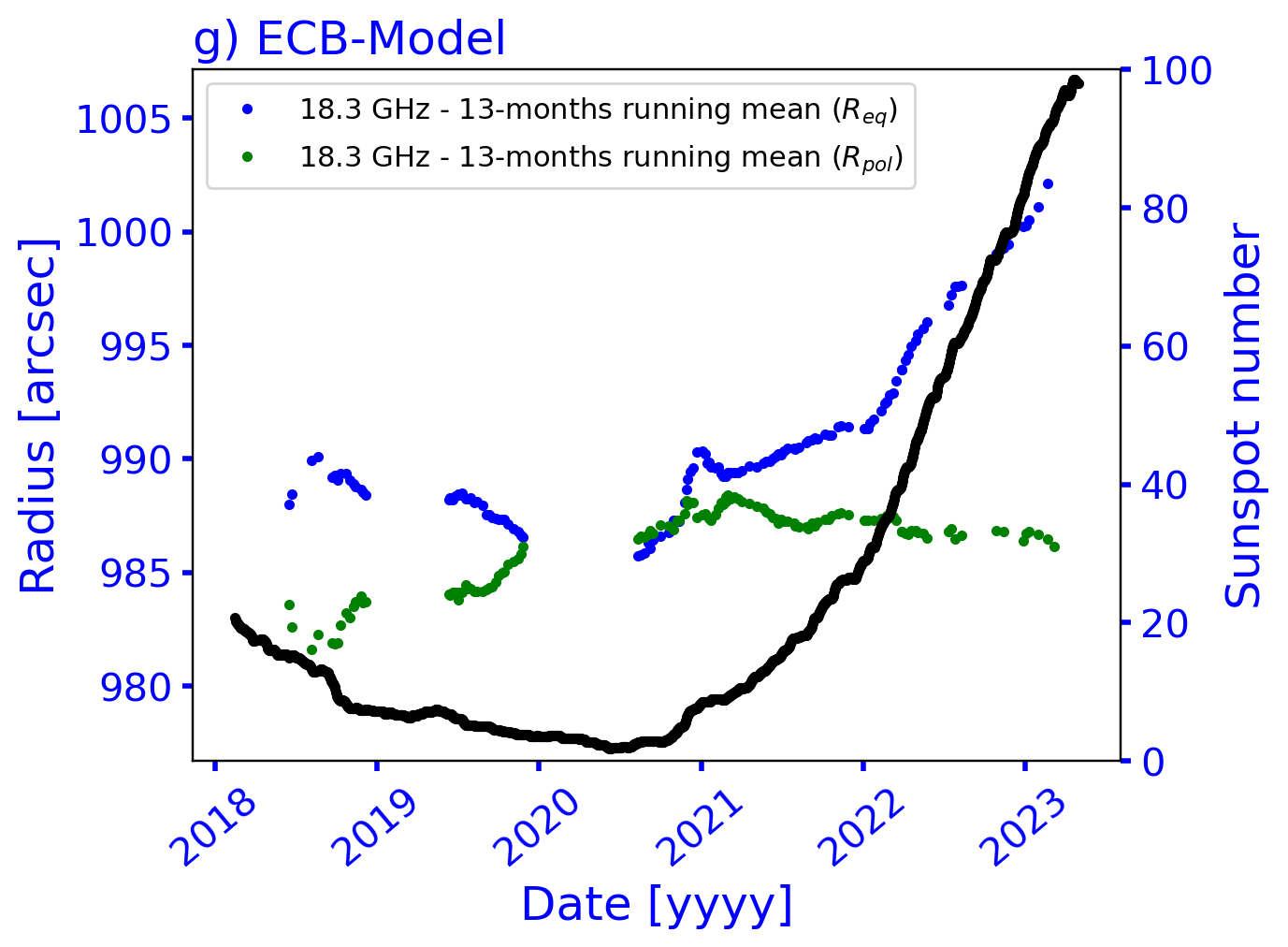}} \quad
{\includegraphics[width=75mm]{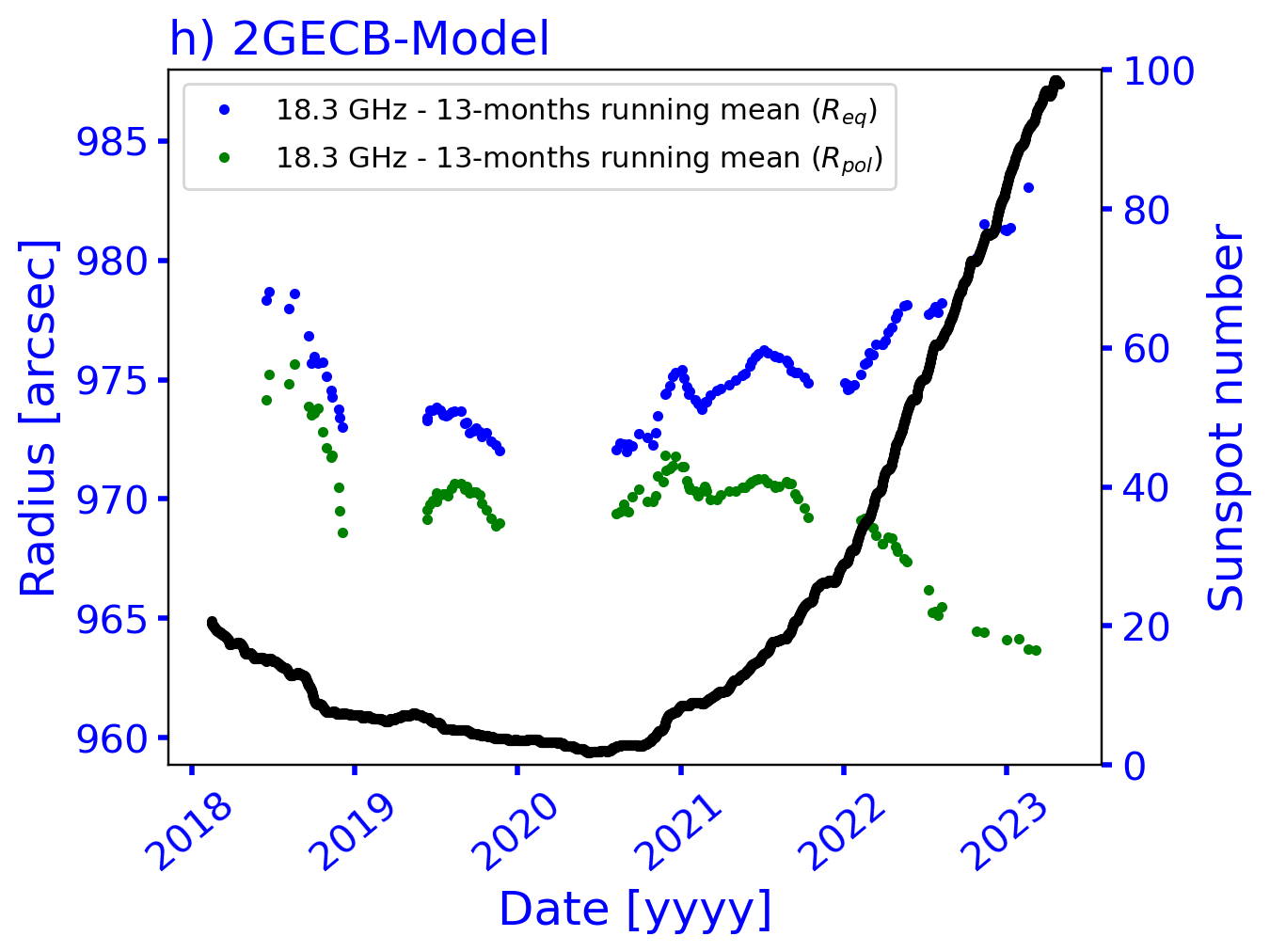}} \\
\caption{13-month running means applied at the solar radii obtained from the Grueff maps ($18.3$~GHz).
Red points indicate $R_{c}$, blue points indicate $R_{eq}$, green points indicate $R_{pol}$, and black points indicate the sunspot index number.
Plots (a) and (b) are referred to HP-method and IP-method, respectively; (c) and (d) are referred to HP-method and IP-method, respectively, for the elliptical-based statistical procedure; (e) and (f) are referred to HP-method and IP-method, respectively, for the circular-based statistical procedure; (g) and (h) are referred to ECB- and 2GECB-models, respectively.
}
\label{fig:raggio_evolution18}
\end{figure*}
\begin{figure*} 
\centering
{\includegraphics[width=75mm]{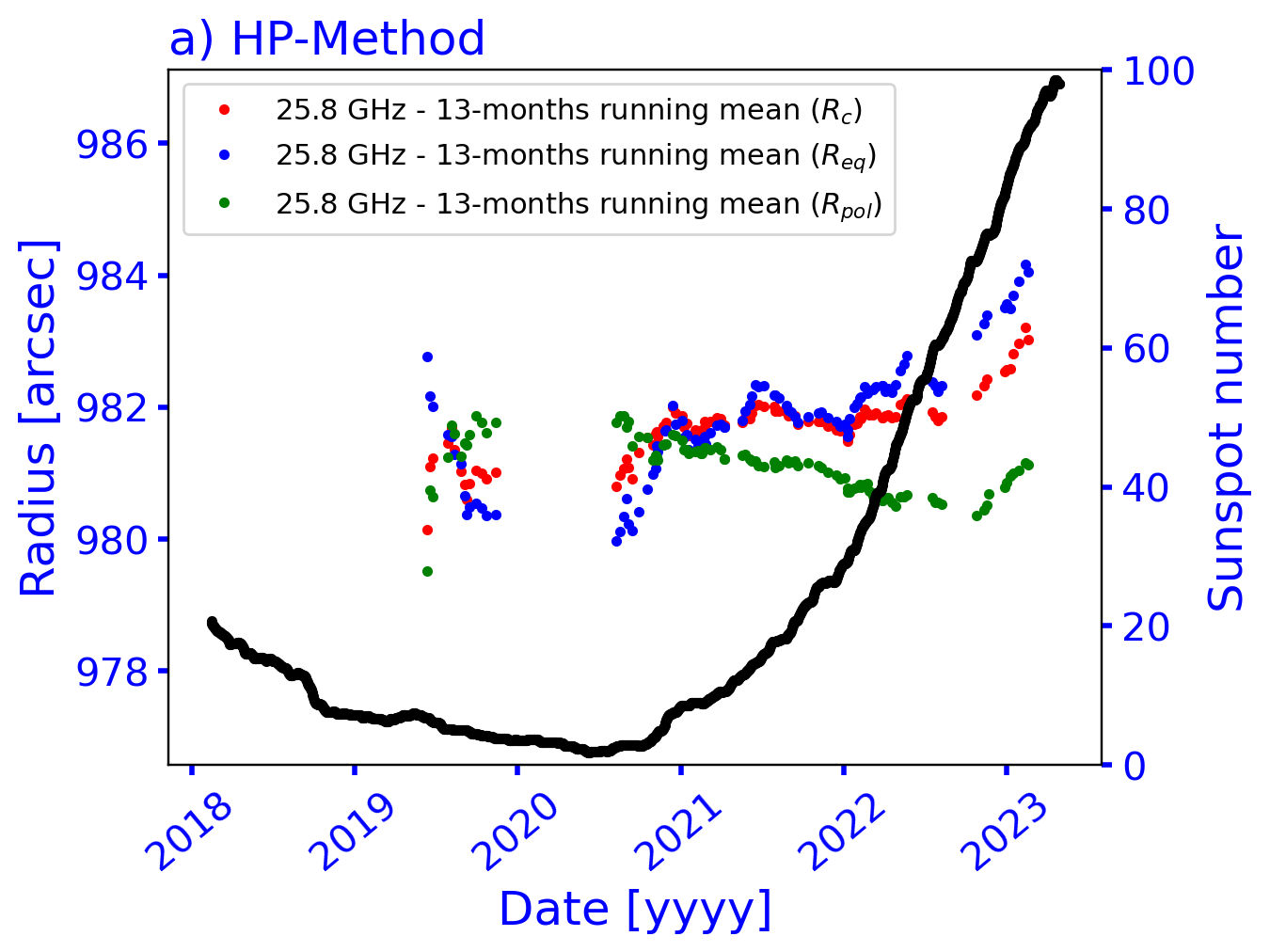}} \quad
{\includegraphics[width=75mm]{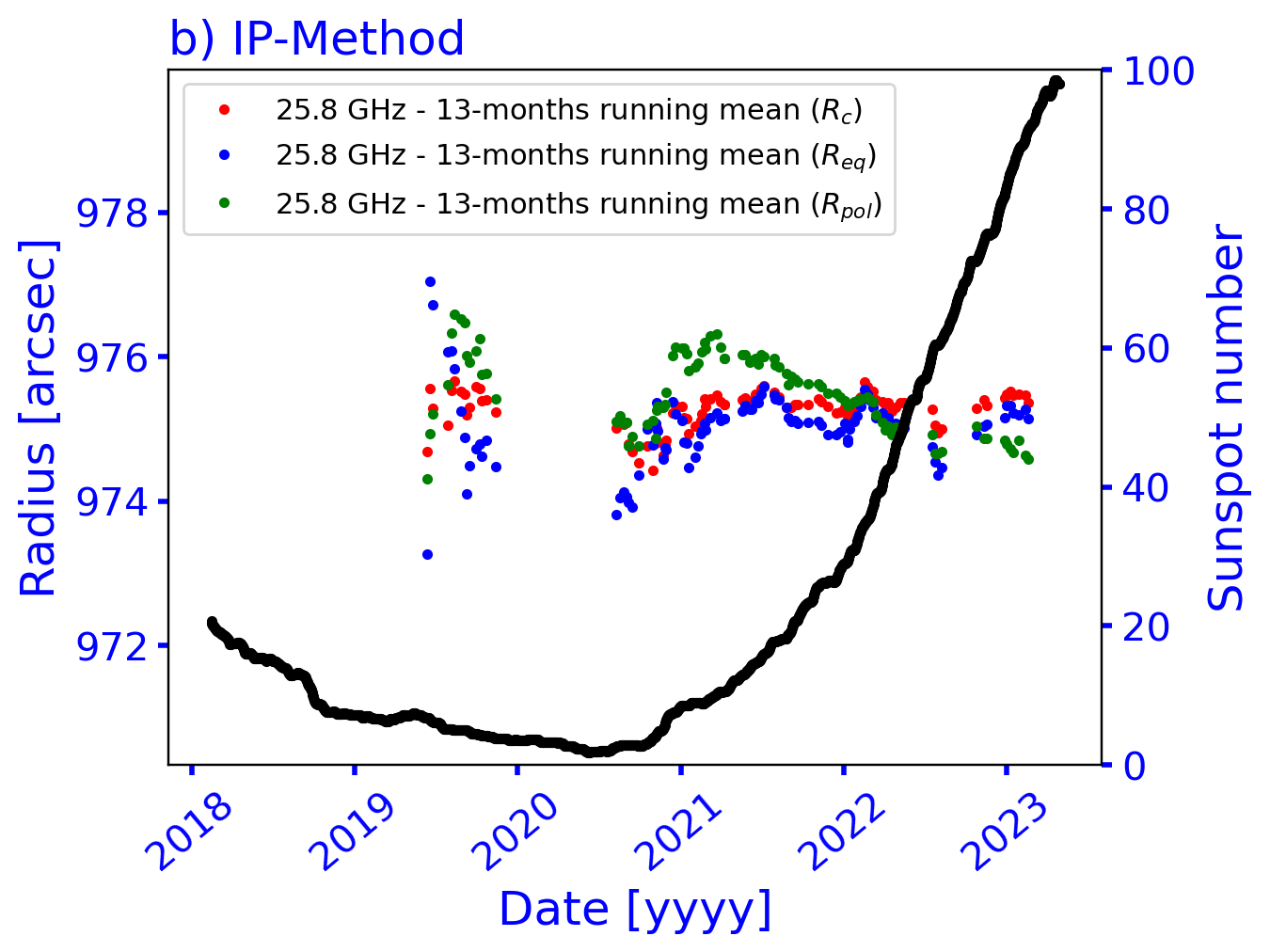}} \\
{\includegraphics[width=75mm]{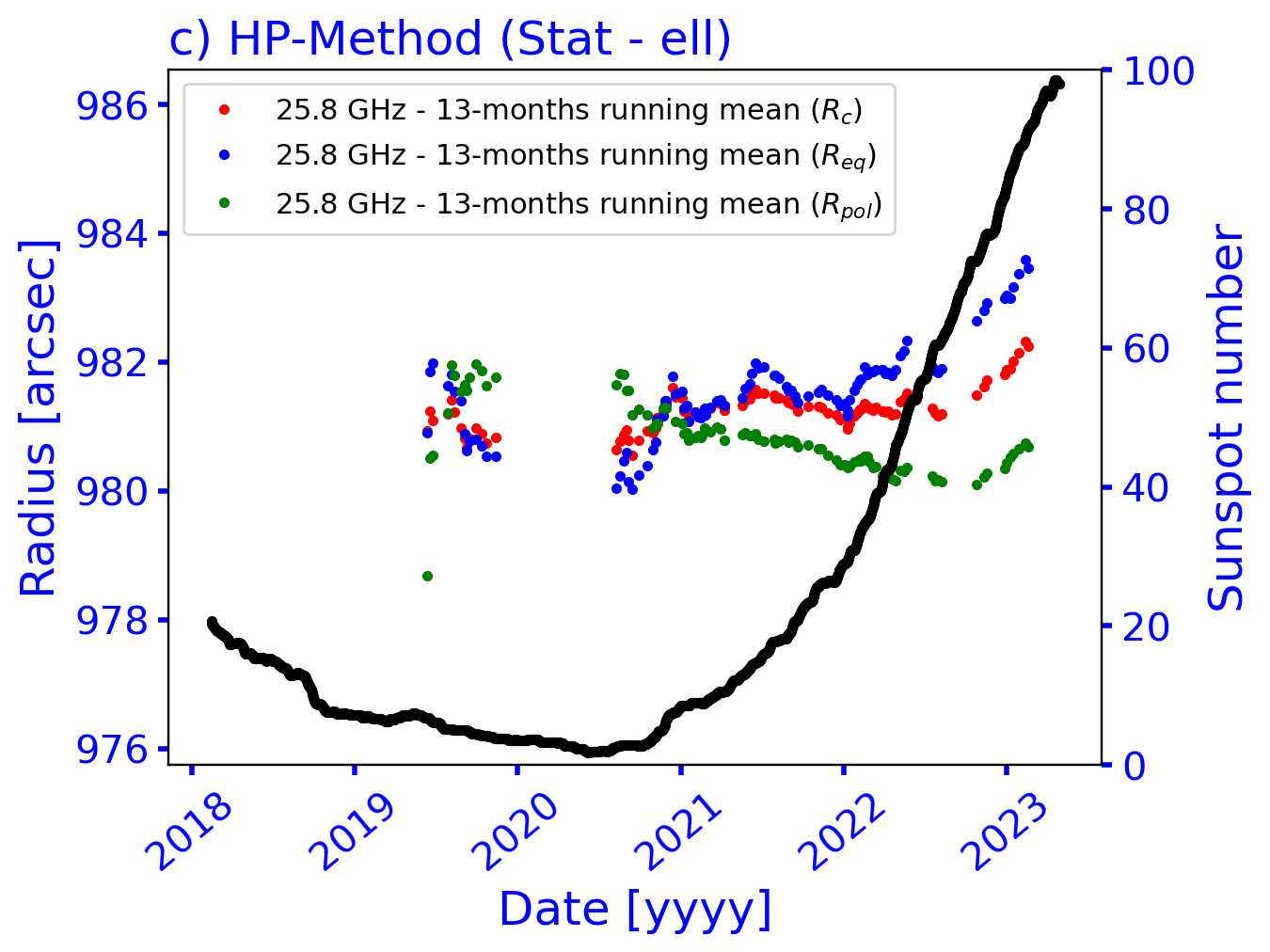}} \quad
{\includegraphics[width=75mm]{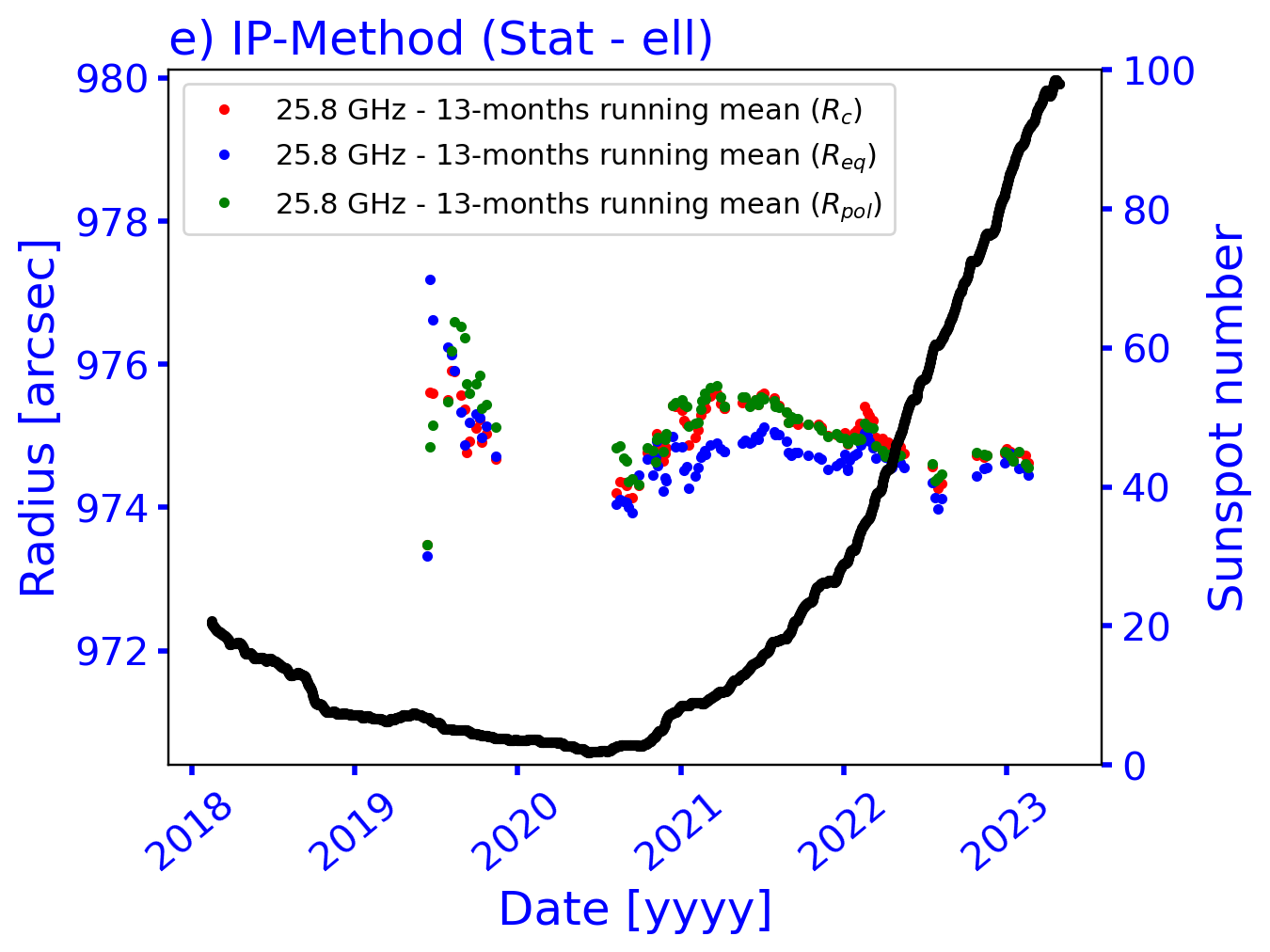}} \\
{\includegraphics[width=75mm]{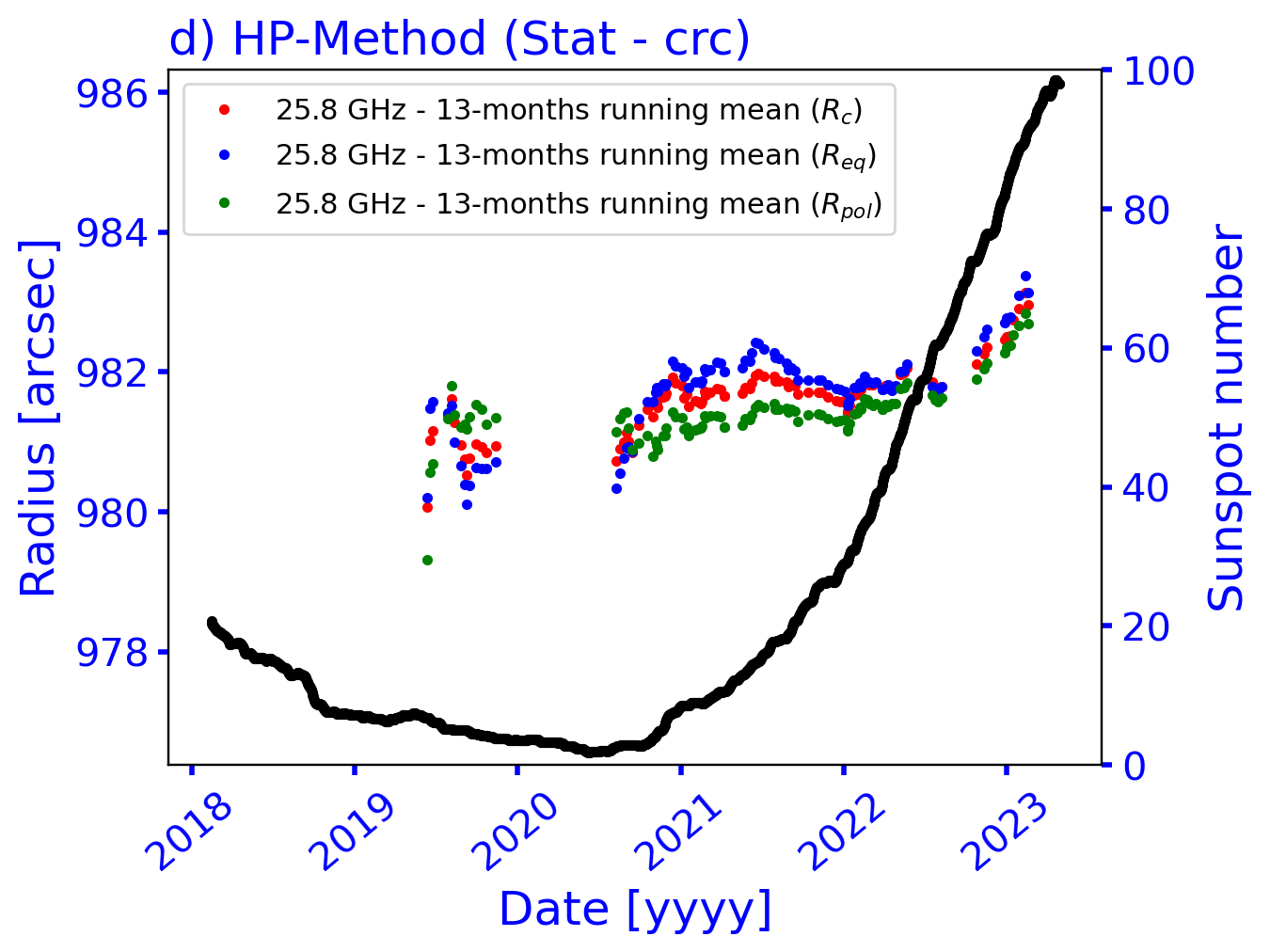}} \quad
{\includegraphics[width=75mm]{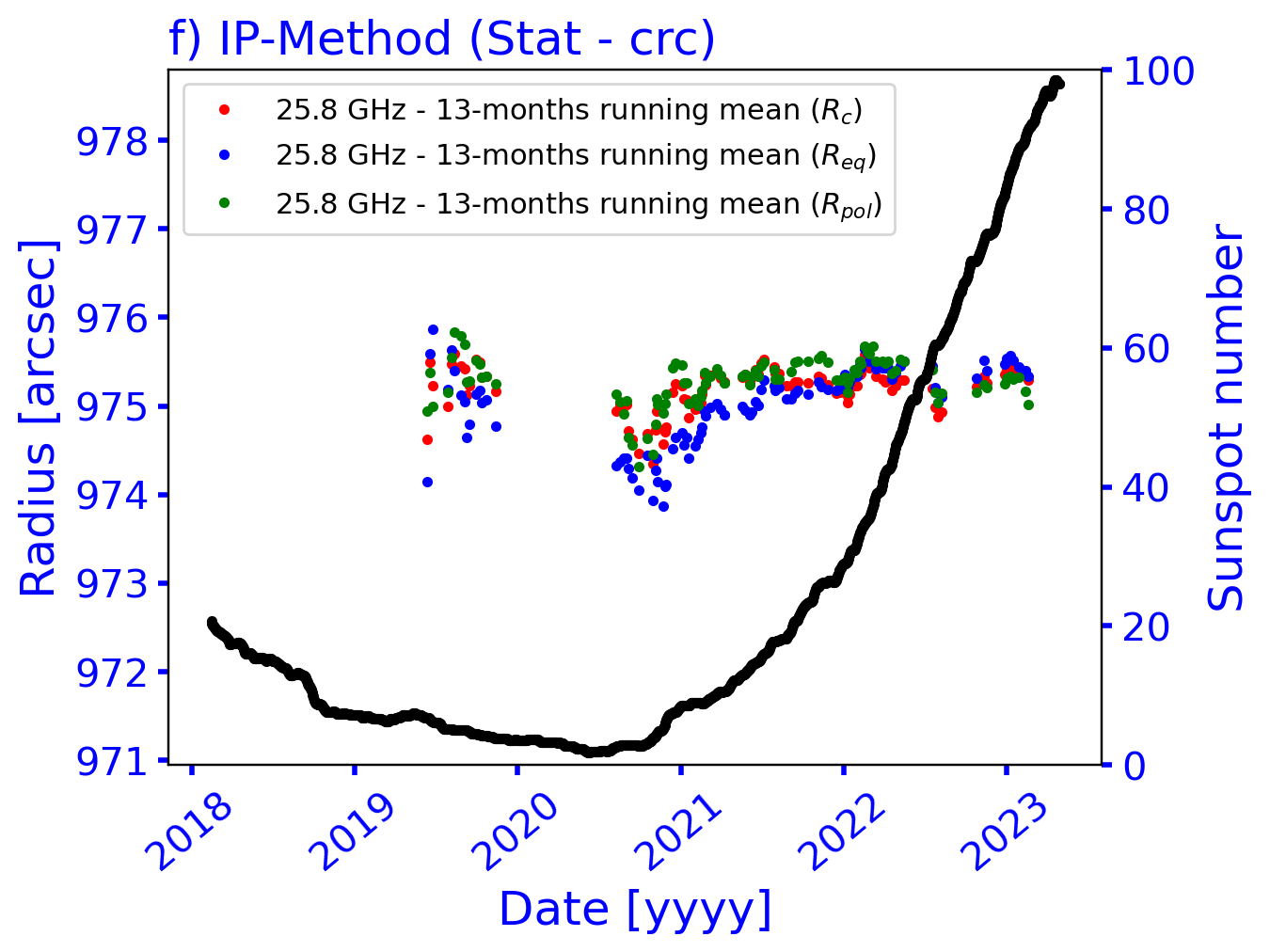}} \\
{\includegraphics[width=75mm]{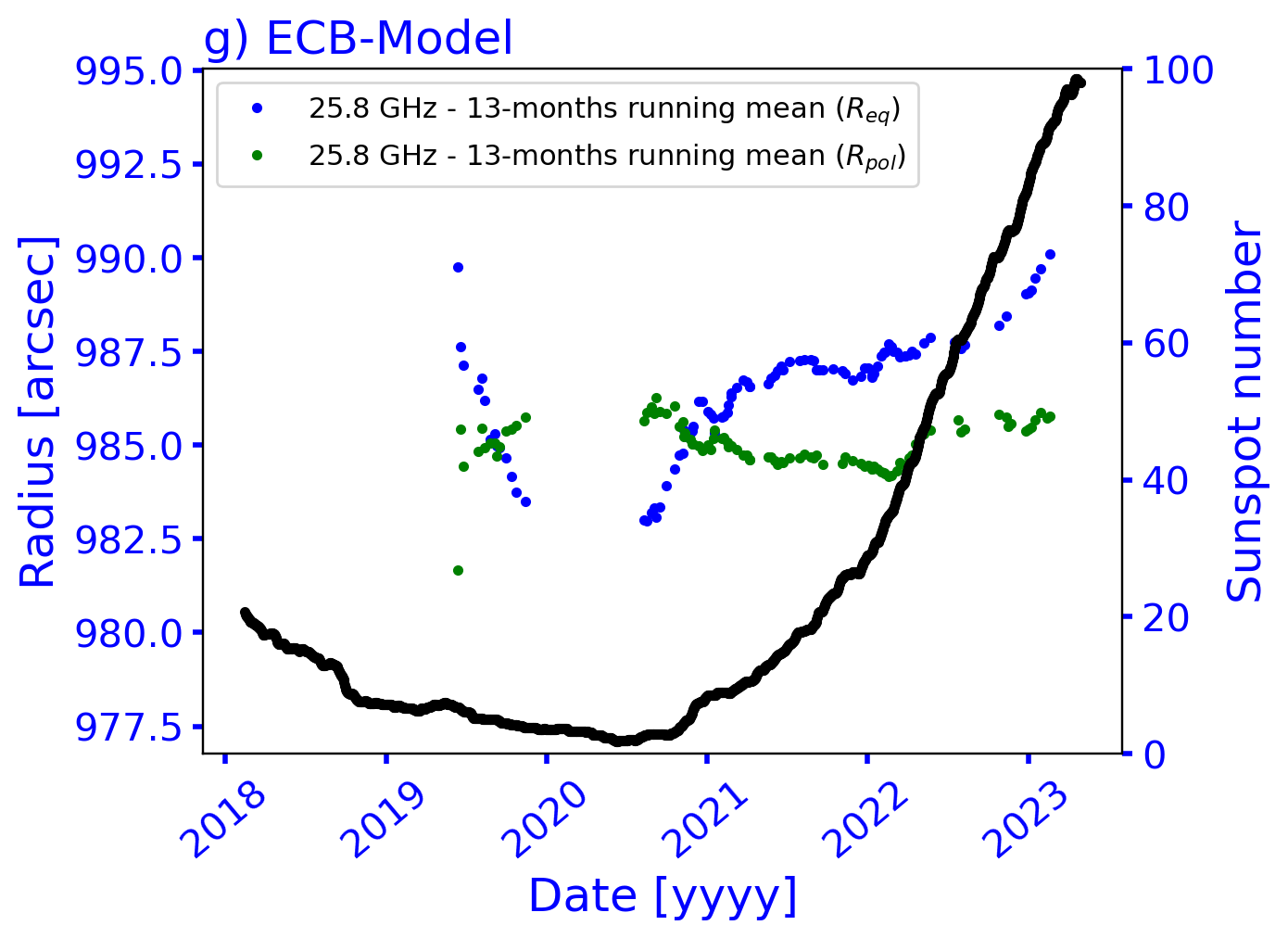}} \quad
{\includegraphics[width=75mm]{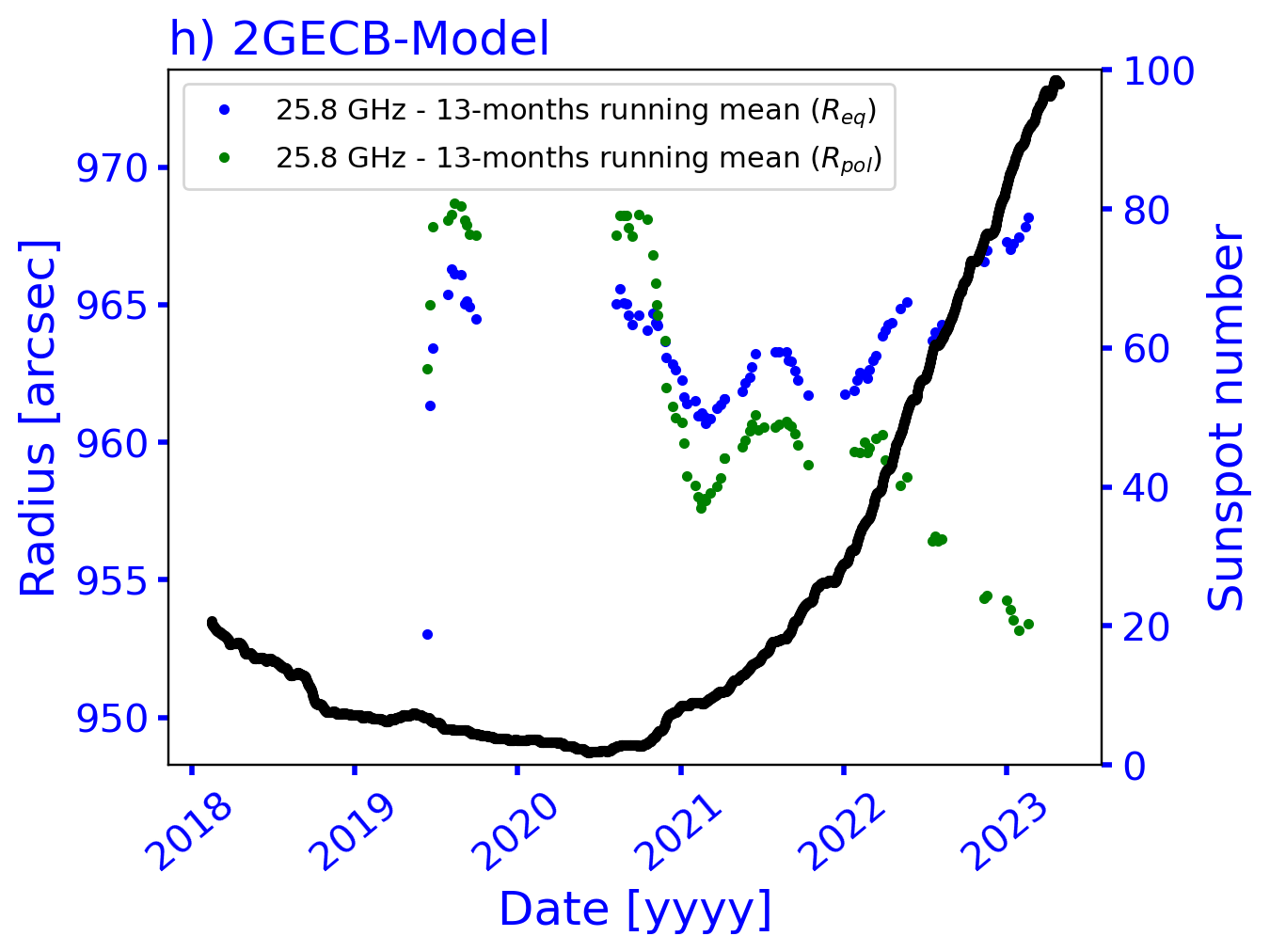}} \\
\caption{13-month running means applied at the solar radii obtained from the Grueff maps ($25.8$~GHz).
Red points indicate $R_{c}$, blue points indicate $R_{eq}$, green points indicate $R_{pol}$, and black points indicate the sunspot index number.
See the caption of Fig.~\ref{fig:raggio_evolution18} for a full description of the plots.
}
\label{fig:raggio_evolution25}
\end{figure*}

To date, the temporal range of our data set is limited with respect to the 11-years solar activity cycle.
However, we were able to investigate the temporal evolution of $R_{\odot}$ and its relationship with the solar activity.
This activity is described through the temporal variation of the sunspot index number, collected by the SILSO data of the Royal Observatory of Belgium (Brussels)\footnote{\url{https://www.sidc.be/silso/datafiles}}.
The strength of this relationship is checked through the Pearson's correlation coefficient (PCC) $\rho$.
Applying 13-month running means -- to avoid the influence of annual modulations \citep{Menezes17,Menezes21} -- at the solar radii obtained from our maps, we obtain the values of $\rho$ (Tables~\ref{tab:corrcoeff_resume_hp}, \ref{tab:corrcoeff_resume_ip}, and \ref{tab:corrcoeff_modelling}; Figs.~\ref{fig:raggio_evolution18} and \ref{fig:raggio_evolution25}).
We will discuss these results in Sect.~\ref{par:discussione_evo_raggio}.
Future observations with the Grueff Radio Telescope and SRT will expand our data set, resulting in a more exhaustive analysis of this kind of correlation.

\begin{table}
\caption{Values of the correlation coefficient $\rho$ between the solar radius (calculated with the Grueff radio telescope through the HP-method) and the solar activity at radio frequencies characterised by a good temporal coverage.
Green values in bold type indicate a strong correlation ($|\rho| \geq 0.7$), black values in bold type indicate a moderate correlation ($0.3 < |\rho| < 0.7$), and the rest indicates a weak correlation ($|\rho| \leq 0.3$).}
\label{tab:corrcoeff_resume_hp}
\scriptsize
\centering
\begin{tabular}{cc|ccc|c}
\hline
\hline
Frequency & Type correlation     & Modelling                        & Statistical                      & Statistical                      & Average $\rho$                   \\
(GHz)     &                      &                                  & (circle)                         & (elliptical)                     &                                  \\
\hline
18.3      & $R_c$ - sunspot      & \textcolor{teal}{\textbf{0.780}} & \textcolor{teal}{\textbf{0.779}} & \textcolor{teal}{\textbf{0.776}} & \textcolor{teal}{\textbf{0.778}} \\
18.3      & $R_{eq}$ - sunspot   & \textcolor{teal}{\textbf{0.913}} & \textcolor{teal}{\textbf{0.836}} & \textcolor{teal}{\textbf{0.913}} & \textcolor{teal}{\textbf{0.887}} \\
18.3      & $R_{pol}$ - sunspot  & \textbf{0.547}                   & \textcolor{teal}{\textbf{0.786}} & \textbf{0.588}                   & \textbf{0.640}                   \\
18.3      & $R_{eq}$ - $R_{pol}$ & \textcolor{teal}{\textbf{0.728}} & \textcolor{teal}{\textbf{0.979}} & \textcolor{teal}{\textbf{0.749}} & \textcolor{teal}{\textbf{0.819}} \\
\hline
25.8      & $R_c$ - sunspot      & \textcolor{teal}{\textbf{0.761}} & \textcolor{teal}{\textbf{0.760}} & \textbf{0.697}                   & \textcolor{teal}{\textbf{0.739}} \\
25.8      & $R_{eq}$ - sunspot   & \textcolor{teal}{\textbf{0.842}} & \textbf{0.606}                   & \textcolor{teal}{\textbf{0.848}} & \textcolor{teal}{\textbf{0.765}} \\
25.8      & $R_{pol}$ - sunspot  & \textbf{-0.612}                  & \textcolor{teal}{\textbf{0.784}} & \textbf{-0.586}                  & $-0.138$                         \\
25.8      & $R_{eq}$ - $R_{pol}$ & \textbf{-0.677}                  & \textbf{0.609}                   & \textbf{-0.534}                  & $-0.201$                         \\
\hline
\end{tabular}
\end{table}
\begin{table}
\caption{Values of the correlation coefficient $\rho$ between the solar radius (calculated with the Grueff Radio Telescope through the IP-method) and the solar activity at radio frequencies characterised by a good temporal coverage.
See the caption of Table~\ref{tab:corrcoeff_resume_hp} for a full description of the table.}
\label{tab:corrcoeff_resume_ip}
\scriptsize
\centering
\begin{tabular}{cc|ccc|c}
\hline
\hline
Frequency & Type correlation     & Modelling                        & Statistical                      & Statistical     & Average $\rho$                   \\
(GHz)     &                      &                                  & (circle)                         & (elliptical)    &                                  \\
\hline
18.3      & $R_c$ - sunspot      & \textbf{0.649}                   & \textbf{0.650}                   & \textbf{0.457}  & \textbf{0.585}                   \\
18.3      & $R_{eq}$ - sunspot   & \textcolor{teal}{\textbf{0.786}} & \textbf{0.584}                   & \textbf{0.622}  & \textbf{0.664}                   \\
18.3      & $R_{pol}$ - sunspot  & \textbf{0.444}                   & \textbf{0.576}                   & $0.263$         & \textbf{0.428}                   \\
18.3      & $R_{eq}$ - $R_{pol}$ & \textbf{0.591}                   & \textcolor{teal}{\textbf{0.912}} & \textbf{0.599}  & \textcolor{teal}{\textbf{0.701}} \\
\hline
25.8      & $R_c$ - sunspot      & $0.299$                          & $0.299$                          & $-0.273$        & $0.108$                          \\
25.8      & $R_{eq}$ - sunspot   & $0.121$                          & \textbf{0.599}                   & $-0.209$        & $0.170$                          \\
25.8      & $R_{pol}$ - sunspot  & \textbf{-0.596}                  & $0.161$                          & \textbf{-0.452} & $-0.296$                         \\
25.8      & $R_{eq}$ - $R_{pol}$ & $0.257$                          & \textbf{0.682}                   & \textbf{0.588}  & \textbf{0.509}                   \\
\hline
\end{tabular}
\end{table}
\begin{table}
\caption{Values of the correlation coefficient $\rho$ between the solar radius (calculated with the Grueff Radio Telescope) and the solar activity at radio frequencies characterised by a good temporal coverage.
$R_{\odot}$ is calculated through the ECB- and 2GECB-models described in Sect.~\ref{par:ant_beam}.
See the caption of Table~\ref{tab:corrcoeff_resume_hp} for a full description of the table.
}
\label{tab:corrcoeff_modelling}
\scriptsize
\centering
\begin{tabular}{cc|cc}
\hline
\hline
Frequency & Type correlation     & ECB                              & 2GECB                             \\
(GHz)     &                      &                                  &                                   \\
\hline
18.3      & $R_c$ - sunspot      & $-$                              & $-$                               \\
18.3      & $R_{eq}$ - sunspot   & \textcolor{teal}{\textbf{0.979}} & \textcolor{teal}{\textbf{0.862}}  \\
18.3      & $R_{pol}$ - sunspot  & $0.212$                          & \textcolor{teal}{\textbf{-0.785}} \\
18.3      & $R_{eq}$ - $R_{pol}$ & $0.253$                          & $-0.330$                          \\
\hline
25.8      & $R_c$ - sunspot      & $-$                              & $-$                               \\
25.8      & $R_{eq}$ - sunspot   & \textcolor{teal}{\textbf{0.763}} & \textbf{0.474}                    \\
25.8      & $R_{pol}$ - sunspot  & $0.167$                          & \textcolor{teal}{\textbf{-0.747}} \\
25.8      & $R_{eq}$ - $R_{pol}$ & \textbf{-0.419}                  & $0.082$                           \\
\hline
\end{tabular}
\end{table}

\section{Discussion}
\label{par:disc_concl}

\subsection{Radius calculation}
\label{par:discussione_raggio}

In this work we obtained an accurate measure of the solar radii ($R_{c}$, $R_{eq}$, and $R_{pol}$) in the centimetric range ($18.1$ -- $26.1$~GHz) through single-dish observations with the Grueff Radio Telescope and SRT.
Our results show a weak prolatness of the solar limb ($R_{eq} > R_{pol}$), although $R_{eq}$ and $R_{pol}$ are statistically comparable within $3\sigma$ errors (Fig.~\ref{fig:hist_resume}).
This trend, as discussed by \citet{Selhorst04}, can be explained by variations in the equatorial solar atmosphere during the period of maximum activity that cause an increase in the solar radius at lower latitudes.

We compared the calculation methods (HP, IP, statistical, ECB, 2GECB), showing a variation of the calculated solar radius based on the used method.
By looking at values in Tables~\ref{tab:raggio_resume} and \ref{tab:raggio_resume2}, the modelling and the statistical procedures are comparable for the calculation of these mean radii within $1\sigma$ error.
In general, the values of $R_c$ obtained with the HP-method in the range $18$ -- $26$~GHz are compatible with the other radii obtained in the literature with the same method within a similar frequency range (e.g., \citealp{Wrixon70,Fuerst79,Costa86,Selhorst04,Selhorst11}).
Small differences could be ascribed to the different angular resolutions of the instruments.
At the same observing frequency, a lower resolution translates in values of $R_{\odot}$ (and its uncertainty) larger than those obtained with instruments characterised by higher resolution.
For example, the value of $R_c = 976.6 \pm 1.5$~arcsec measured by \citet{Selhorst04} at $17$~GHz with NoRH (characterised by a spatial resolution of $10$~arcsec) is (1) compatible within $1\sigma$ error to that obtained with SRT ($\sim 979$~arcsec) at $18.8$~GHz, where the beamsize is $\sim 60$~arcsec, and (2) smaller than that obtained with Grueff ($\sim 982$~arcsec) at $18.3$~GHz, where the beamsize is $\sim 120$~arcsec, but still within $2\sigma$ error.
Moreover, there are no significant biases arising from image measurement sampling.
In our OTF technique, each pixel on solar images is oversampled, since our maps include at least $6$ measurements per beam for each sub-scan \citep{Pellizzoni22}.
A comparison of values obtained with the IP-method is not possible since there are no available measurements in the literature in a frequency range comparable to ours.
The available radii are calculated at observing frequencies higher than $100$~GHz (e.g. \citealp{Menezes21}).
The mean radii obtained with the HP-method and the ECB-model -- compatible with each other within $1\sigma$ error -- are larger than those obtained with the IP-method and the 2GECB-model.
This feature suggests that specific procedures to measure $R_{\odot}$, designed to also describe both the behaviour of the solar disk (such as ARs) and the coronal emission (such as the case of the IP-method and the 2GECB-model), result in a lower bias in solar radius determination \citep{Menezes22}, and hence in a smaller value of $R_{\odot}$.
In particular, the 2D solar maps modelled using our beam pattern and the 2GECB-model as the solar signal (Sect.~\ref{par:ant_beam}), are well fitted with the observed solar map ($\chi^2_r \sim 0.8$).
This modelling is better than that obtained using the ECB-model ($\chi^2_r \sim 0.6$).
Despite the statistically non-significant data set from SRT, from our results (Tables~\ref{tab:raggio_resume} and \ref{tab:raggio_resume2}) we preliminarily estimate the possible dependence of the HP- and the IP-methods on the antenna beam pattern, analysing the radii difference caused by these methods.
This radii difference follows the same trend (the percentage difference $\Delta M$ between the radii obtained with the HP-method and the IP-method is $\sim 0.6\%$) both at Grueff and SRT, suggesting that these methods could be independent of the antenna beam pattern.
On the other hand, this independence is not strictly defined, since at 25 GHz this trend in the IP-method seems to be less pronounced ($\Delta M \sim 0.2\%$), suggesting possible dependence on the antenna beam pattern.
Future solar sessions at SRT will allow us to clarify this aspect.
It is worth noting that approximately $30\%$ of the $R_{\odot}$ values ($\sim 20\%$ of $R_{eq}$ and $\sim 40\%$ of $R_{pol}$) obtained with the 2GECB-model -- built adopting a Bayesian statistics -- are smaller than the canonical and "average" optical solar photospheric radius $R_{\odot, opt}$ (see Fig.~\ref{fig:hist_resume}, bottom right).
However, the peaks of the histogram ($\sim 973$~arcsec for $R_{eq}$ and $\sim 967$~arcsec for $R_{pol}$) are above $R_{\odot, opt}$ and lower values are considered as statistical fluctuations.

From Fig.~\ref{fig:radius_nu}, we note that our radii are in agreement with the radius-versus-frequency trend of the literature points, where the curve seems to flatten at higher frequencies.
That could explain why we obtained similar average values of the radii at our observing frequencies.
In this context, it is important to take into account the analysis of \citet{Meftah18}, that claimed the extreme weakness of the correlation between the solar radius and the observing frequency in the visible and the near-infrared, in contrast to the claim made by \citet{Rozelot15}.
This weak correlation is even expected due to the sharpness of the solar density profile near the photosphere.
\citet{Meftah18} claimed that the analysis of the possible solar radius dependence on the frequency should be at least based on two different regions of the solar atmosphere (photosphere and chromosphere).
This aspect is crucial before fitting any polynomial functions to the measurements.

\subsection{Time evolution of the solar radius}
\label{par:discussione_evo_raggio}

Despite the irregular and limited coverage both in time and observing frequency of our data set, our investigation at $18.1$ -- $26.1$~GHz about the temporal evolution of the solar radius and its relationship with the solar activity suggests some interesting trends.
For this correlation analysis, the SRT data set obtained at $18.8$ and $24.7$ GHz does not show exhaustive information on the correlation with the solar activity due to the low statistics.
Therefore, we used only the data set from Grueff (at $18.3$ and $25.8$~GHz).

The 13-month running means applied to the solar radii measured at Medicina and the sunspot index number (Tables~\ref{tab:corrcoeff_resume_hp}, \ref{tab:corrcoeff_resume_ip}, and \ref{tab:corrcoeff_modelling}; Figs.~\ref{fig:raggio_evolution18} and \ref{fig:raggio_evolution25}), suggest a strong/moderate positive correlation between the 11-year solar activity cycle and the temporal variation of both $R_{c}$ and $R_{eq}$ at all observing frequencies.
Only at $25.8$~GHz there is a weak positive correlation obtained with the IP-method.
The other cases of correlation show different results according to the approaches for the $R_{\odot}$ calculation described in Sects.~\ref{par:det_radius} and \ref{par:ant_beam}.
In general, the correlation between $R_{pol}$ and the solar activity is:
\begin{enumerate}
    \item weakly/moderately positive at $18.3$~GHz with the HP-, IP-, and ECB-methods, and at $25.8$~GHz with the ECB-model;
    \item weakly negative at $25.8$~GHz with the HP- and IP-methods;
    \item strongly negative at $18.3$~GHz and $25.8$~GHz with the 2GECB-model. 
\end{enumerate}
Finally, the correlation between $R_{eq}$ and $R_{pol}$ is:
\begin{enumerate}
    \item strongly positive at $18.3$~GHz with the HP- and IP-methods;
    \item weakly/moderately positive at $25.8$~GHz with the IP-method and the 2GECB-model, and at $18.3$~GHz with the ECB-model;
    \item weakly negative at $18.3$~GHz with the 2GECB-model;
    \item weakly/moderately negative at $25.8$~GHz with the HP-method and the ECB-model.
\end{enumerate}

These results suggest that specific procedures to measure $R_{\odot}$, tailored to describe both the behaviour of the solar disk and the coronal emission (such as the case of the IP-method and the 2GECB-model), show a weak/moderate anti-correlation between the temporal variation of $R_{pol}$ and the solar activity, especially when the 2GECB-model is applied.
Among the $R_{\odot}$ prescriptions described in this work, the strength of the correlation is inversely proportional to the robustness/complexity of the prescription: for the case of the HP-method, based only on the $T_B$ level of the brightness profiles, the strength of the correlation is higher than those obtained with more complex prescriptions, such as the ECB-model, IP-method, and the 2GECB-model.
In particular, the 2GECB-model is also able to unearth anti-correlation between the temporal variation of $R_{pol}$ and the solar activity at $18.3$~GHz, where generally the thermal emissions from the solar activity -- a bias for the radius calculation -- are stronger than the counterpart at $25.8$~GHz.

Our results of the Medicina "Grueff" Radio Telescope seem to be in agreement with the analyses presented in the literature for similar radio frequencies \citep{Costa99, Selhorst04, Selhorst11, Selhorst19a}, which are expected to be correlated to the solar cycle positively for $R_c$ and negatively for $R_{pol}$.
These results are compatible with the presence of correlations between the radius variations and the solar activity, as suggested by \citet{Menezes21} monitoring the Sun for more than a solar cycle (from $2007$ to $2019$) at higher frequencies ($212$ and $405$~GHz) with SST and ALMA facilities.
In that work the authors suggest that the $R_{eq}$ variations are expected to be positively correlated to the solar activity, since the equatorial regions are more affected by the increase of the AR number during solar maxima, making the solar atmosphere warmer in these regions.
On the other hand, \citet{Menezes21} suggest that the anti-correlation between polar radius time series and the solar activity proxies could be explained by a possible increase of the polar limb brightening during solar minima, as also suggested by \citet{Selhorst04}.

Moreover, as shown in Fig.~\ref{fig:raggio_evolution18}, at $18.3$~GHz a bump appears at the end of 2020 both in the radii and in the solar activity.
This effect seems to corroborate a higher level of the positive correlation between $R_{\odot}$ and the solar activity at this observing frequency than $25.8$~GHz, in agreement with the conclusions of \citet{Menezes21} mentioned above about the stronger solar activity in the equatorial regions at the approach of the solar maxima.
As we can see in Fig.~\ref{fig:raggio_evolution25}, at $25.8$~GHz the time evolution of the 13-month running means applied to the measured solar radii obtained using the ECB- and 2GECB-methods, shows that this bump in the solar activity corresponds to the time when $R_{eq}$ becomes greater than $R_{pol}$, suggesting the starting point of the rising phase of the 11-year solar activity cycle.

\section{Conclusions and future developments}
\label{par:concl_svil}

In this paper, we mainly focus on the first measurements of the solar radius carried out using the Medicina "Gavril Grueff" Radio Telescope and SRT -- along with a comprehensive correlation analysis that also takes solar activity into account.
Spanning a period of five years, from $2018$ to mid-$2023$, we collected around $300$ single-dish observations.
This time frame covers approximately half of a solar cycle, and the observations were conducted in the radio K-band ranging between $18.1$~GHz and $26.1$~GHz.
From the seven observing frequencies available during our solar sessions, we have specifically chosen four frequencies for our analysis.
These selections were made based on their consistent time coverage.
Precisely, we employed frequencies of $18.3$~GHz and $25.8$~GHz for observations at Medicina, and frequencies of $18.8$~GHz and $24.7$~GHz for observations at SRT.
These chosen frequencies contribute significantly to a robust and meaningful analysis of our solar radius measurements.
To assess the quality of our radius determinations in our solar maps, we analysed the role of the antenna beam pattern on these maps employing two 2D-models for solar emission convolved with accurate beam models, built adopting a Bayesian approach (ECB- and 2GECB-models).

The mean radii ($R_c$, $R_{eq}$, and $R_{pol}$) calculated among all our solar maps (for each observing frequency and radio telescope) -- reported in Tables~\ref{tab:raggio_resume} and \ref{tab:raggio_resume2} -- show values compatible with the ones reported in literature.
Our measurements show a weak prolatness of the solar limb ($R_{eq}$ > $R_{pol}$), although $R_{eq}$ and $R_{pol}$ are compatible within $3\sigma$ errors.
In particular, the radii calculated with the HP-method and the ECB-model are larger than those measured through the IP-method and the 2GECB-model.
Moreover, while the HP-method seems to be independent of the antenna beam pattern, future observations of the Sun at SRT will allow us to clarify whether the IP-method is also independent of the antenna beam pattern.

As reported in Tables~\ref{tab:corrcoeff_resume_hp}, \ref{tab:corrcoeff_resume_ip} and \ref{tab:corrcoeff_modelling}, the 13-month running means applied to the solar radii measured at Medicina ($18.3$ and $25.8$~GHz) and the sunspot index number indicate (1) a positive correlation between the 11-year solar activity cycle and the temporal variation of both $R_{c}$ and $R_{eq}$ at all observing frequencies, and (2) an anti-correlation between the temporal variation of $R_{pol}$ and the solar activity, especially at $25.8$~GHz.
Moreover, the time variation of $R_{eq}$ and $R_{pol}$ for the solar data of the Grueff Radio Telescope shows a positive correlation, especially at $18.3$~GHz.
Our results about the correlation analysis may indicate an agreement with the analysis presented in literature for similar radio frequencies.

The bump observed at the end of 2020 both in the radii and in the solar activity, especially at $18.3$~GHz, constrains the positive correlation between $R_{\odot}$ and the solar activity at this observing frequency.
At $25.8$~GHz, the time evolution of $R_{\odot}$ obtained using the ECB- and 2GECB-models shows that the bump in the solar activity corresponds to the time when $R_{eq}$ becomes greater than $R_{pol}$, suggesting the starting point of the rising phase of the solar activity cycle.

Our results show that specific procedures for the $R_{\odot}$ measurement, suited for the modelling of both the behaviour of the solar disk and the coronal emission (such as the case of the IP-method and the 2GECB-model), result in a lower bias in solar radius determination (and hence in a smaller value of $R_{\odot}$) and in the detection of anti-correlation between the temporal variation of $R_{pol}$ and the solar activity.
This aspect suggests that the IP-method, and especially the 2GECB-model, could be among the most physically reliable methods for the calculation of $R_{\odot}$ and its variation over time.
Moreover, the radii obtained with the 2GECB-model range between the values obtained with the specific theoretical atmospheric models and the HP/IP-methods, suggesting that a non-negligible coronal emission level causes the decrease of the radius.
We will discuss this aspect in a separate paper \citep{Marongiu23b}.

Future observations and detailed theoretical analysis with the Grueff Radio Telescope and SRT -- for longer periods of time and in multi-frequency approach thanks also to new PON receivers at SRT operating up to $116$~GHz –- are crucial to better clarify several aspects, such as (1) the correlation between the solar activity and the solar size, (2) the polar and equatorial trends of the solar atmosphere, and (3) the question of the limb brightening (especially the presence of the polar brightening during the solar minima), which will be the subject of an upcoming work.


\begin{acknowledgements}


The Medicina radio telescope is funded by the Ministry of University and Research (MUR) and is operated as National Facility by the National Institute for Astrophysics (INAF).

The Sardinia Radio Telescope is funded by the Ministry of University and Research (MUR), Italian Space Agency (ASI), and the Autonomous Region of Sardinia (RAS) and is operated as National Facility by the National Institute for Astrophysics (INAF).

The Enhancement of the Sardinia Radio Telescope (SRT) for the study of the Universe at high radio frequencies is financially supported by the National Operative Program (Programma Operativo Nazionale - PON) of the Italian Ministry of University and Research "Research and Innovation 2014-2020", Notice D.D. 424 of 28/02/2018 for the granting of funding aimed at strengthening research infrastructures, in implementation of the Action II.1 – Project Proposal PIR01\_00010.

We acknowledge the Computing Centre at INAF - Istituto di Radioastronomia for providing resources and staff support during the processing of solar data presented in this paper.

\end{acknowledgements}

\section*{ORCID iDs}

M.~Marongiu: \url{https://orcid.org/0000-0002-5817-4009} \\
A.~Pellizzoni: \url{https://orcid.org/0000-0002-4590-0040} \\
S.~Mulas: \url{https://orcid.org/0000-0002-5455-1233} \\
S.~Righini: \url{https://orcid.org/0000-0001-7332-5138} \\
R.~Nesti: \url{https://orcid.org/0000-0003-0303-839X} \\
G.~Murtas: \url{https://orcid.org/0000-0002-7836-7078} \\
E.~Egron: \url{https://orcid.org/0000-0002-1532-4142} \\
M.~N.~Iacolina: \url{https://orcid.org/0000-0003-4564-3416} \\
A.~Melis: \url{https://orcid.org/0000-0002-6558-1315} \\
G.~Valente: \url{https://orcid.org/0000-0003-1197-9050} \\
G.~Serra: \url{https://orcid.org/0000-0003-0720-042X} \\
S.~L.~Guglielmino: \url{https://orcid.org/0000-0002-1837-2262} \\
A.~Zanichelli: \url{https://orcid.org/0000-0002-2893-023X} \\
P.~Romano: \url{https://orcid.org/0000-0001-7066-6674} \\
S.~Loru: \url{https://orcid.org/0000-0001-5126-1719} \\
M.~Bachetti: \url{https://orcid.org/0000-0002-4576-9337} \\
A.~Bemporad: \url{https://orcid.org/0000-0001-5796-5653} \\
F.~Buffa: \url{https://orcid.org/0000-0001-9256-4476} \\
R.~Concu: \url{https://orcid.org/0000-0003-3621-349X} \\
G.~L.~Deiana: \url{https://orcid.org/0000-0002-5404-5162} \\
C.~Karakotia: \url{https://orcid.org/0009-0002-9669-7692} \\
A.~Ladu: \url{https://orcid.org/0000-0003-1920-9560} \\
A.~Maccaferri: \url{https://orcid.org/0000-0001-7231-4007} \\
P.~Marongiu: \url{https://orcid.org/0000-0003-0314-7801} \\
M.~Messerotti: \url{https://orcid.org/0000-0002-5422-1963} \\
A.~Navarrini: \url{https://orcid.org/0000-0002-6191-6958} \\
A.~Orfei: \url{https://orcid.org/0000-0002-8723-5093} \\
P.~Ortu: \url{https://orcid.org/0000-0002-2644-2988} \\
M.~Pili: \url{https://orcid.org/0000-0003-3715-1091} \\
T.~Pisanu: \url{https://orcid.org/0000-0003-2510-7501} \\
G.~Pupillo: \url{https://orcid.org/0000-0003-2172-1336} \\
A.~Saba: \url{https://orcid.org/0000-0002-1607-5010} \\
L.~Schirru: \url{https://orcid.org/0000-0002-8199-6510} \\
C.~Tiburzi: \url{https://orcid.org/0000-0001-6651-4811} \\
P.~Zucca: \url{https://orcid.org/0000-0002-6760-797X} \\

\bibliographystyle{aa}            
\bibliography{aanda}          

\end{document}